\documentclass[11pt]{article}

\usepackage{pdfpages}
\usepackage{multirow,multicol,soul}
\usepackage[normalem]{ulem}
\usepackage{tabularx}
\usepackage{longtable,eurosym}
\usepackage{booktabs}
\usepackage{amsmath,bbm}
\usepackage{amssymb, epsfig}
\usepackage{float,graphicx,subfigure}
\usepackage[english]{babel}
\usepackage[utf8]{inputenc}
\usepackage[counterclockwise]{rotating}
\usepackage{lscape}
\usepackage[top=2cm, bottom=2cm, left=2cm , right=2cm]{geometry}
\usepackage{url,hyperref}
\usepackage{color}
\definecolor{darkgreen}{RGB}{51,153,55}

\usepackage[T1]{fontenc}

\newcommand{\1}{\mathbf{1}}

\def\R{\mathbb{R}}

\usepackage[natbibapa]{apacite}

\usepackage{setspace}
\usepackage{todonotes}

\date{}

\begin{document}

\title{Building up Cyber Resilience by Better Grasping Cyber Risk \\
Via a New Algorithm for Modelling Heavy-Tailed Data}
\vspace{-2ex}

\author{Michel Dacorogna\textsuperscript{a}, Nehla Debbabi\textsuperscript{b} and Marie Kratz\textsuperscript{c}\\[1ex]
\small
\textsuperscript{a}PRIME RE Solutions, Zug, Switzerland; Email: michel.dacorogna@prs-zug.com\\
\small
\textsuperscript{b}ESPRIT School of Engineering, Tunis, Tunisia; Email: nehla.debbabi@esprit.tn\\
\small
\textsuperscript{c} ESSEC Business School, CREAR, Cergy, France; Email: kratz@essec.edu
\vspace{-6ex}}

\maketitle

\begin{abstract}
\noindent
Cyber security and resilience are major challenges in our modern economies; this is why they are top priorities on the agenda of governments, security and defense forces, management of companies and organizations. Hence, the need of a deep understanding of cyber risks to improve resilience. We propose here an analysis of the database of the cyber complaints filed at the {\it Gendarmerie Nationale}. We perform this analysis with a new algorithm developed for non-negative asymmetric heavy-tailed data, which could become a handy tool in applied fields.
This method gives a good estimation of the full distribution including the tail. Our study confirms the finiteness of the loss expectation, necessary condition for insurability. Finally, we draw the consequences of this model for risk management, compare its results to other standard EVT models, and lay the ground for a classification of attacks based on the fatness of the tail.
\end{abstract}

\noindent {\emph 2010 AMS classification}: 
60G70; 
60E05; 
62P05;  
90B50; 
91G70; 
\\[1ex]

\vfill

\newpage
\tableofcontents
\newpage

\section{From operational risk to cyber risk}\label{sec:ORcyber}
\vspace{-2ex}
For a very long time, IT risks were classified in operational risk. Many databases recording operational failure events also contain cyber incidents, which have been modelled using traditional operational research methods. Recently, as cyber failures originate more from malicious attacks and may exhibit strong systemic effects, cyber risk has become a class of risks on its own, requiring specific approaches to study it. Besides the usual IT literature on cyber security and defense, the scientific literature on cyber concerns mainly two fields, Management and Operational Research (OR), for general models and study of properties, and Actuarial Research, when suggesting pricing models for insurance. These two paths clearly appear in the state-of-the-art review presented below.\\[1ex]
Extreme Value Theory (EVT) methods have already proven to be quite useful for operational risks (among recent papers, see e.g. \cite{DAS2021,Embrechts2018}) in the industrial and financial sectors, loss severity being even larger in manufacturing  than in finance. 
Relying on EVT becomes even more essential when studying cyber risk because it has moved, over the years, from a possible threat to an important emerging risk, which generally goes with a high probability of extreme events (due to immature management). Then, it also calls for the development or improvement of dynamic EVT approaches. This might be facilitated when using unsupervised EVT methods, where the threshold above which observations are considered as extremes is automatically detected. It is what we propose here,  extending  a recent method to skewed non-negative data, as those explored in this study. Our approach makes the empirical estimation of the whole heavy-tailed distribution more straightforward, and of easier use for non-specialists of EVT. It may become a handy tool in many applied research fields dealing with heavy-tailed data, in particular operational research.
\vspace{-3ex}

\paragraph*{Facing an emerging risk: the compromise between cyber security and cyber resilience -}
~\\
Cyber threats and crimes have increased exponentially in recent decades, due to a rapid diffusion of new and evolving Information and Communication Technologies such as Social Media, cloud computing, big data, Internet of Things (IoT) and smart cities (see e.g. \citep{Pasculli2020}). There are innumerable examples of cyber crimes having a huge impact in terms of human and financial costs, e.g. one of the latest attacks, attributed to 'Darkside' (cyber crime group), causing  in early May 2021 the shut down of {\it Colonial Pipeline} network and a shortage of oil supply in the North-East Cost of USA.  The frequency of cyber attacks increased even more since the beginning of the Coronavirus pandemic; we have seen a simultaneous surge in the use of Internet and in the cyber attacks targeted against individuals, hospitals, and small businesses (see e.g.~\citep{NCSC2020}).
Even though all actors in society are getting more and more aware of the growing importance of cyber risk, we are still far from having reached the same level of understanding and assessment for this specific risk, as we do for financial risk or natural disasters.\\[1ex]
From a societal point of view, we clearly need to develop ways of becoming more resilient as we increasingly depend on well functioning IT systems. Managing this risk does not only mean minimizing and preventing cyber-attacks, but also, if an attack is successful, ensuring that its consequences are not too severe for organizations or individuals, in other words, making society more resilient to them.
While the term `cyber security' is as old as computers themselves, the term `cyber resilience' has emerged recently and is gaining currency. Cyber security is focused on security alone, the term security referring to defense, protection, precaution. But organizations need a broader strategy that includes their ability to survive an attack, to recover with as little harm as possible, to continue to operate when experiencing a cyber attack, and finally to insure some unavoidable risks. This is what cyber resilience refers to. As already pointed out in the review by \cite{Aven2016} and in \cite{Aven2019}, integrating resilience principles and methods may participate to the development of  modern risk management. The future of this resilience will be multidimensional, combining prevention and protection measures like: users' education, security protocols, redundancies in IT systems, clear managerial attention for implementing an adequate strategy, insurance coverage to ensure the survival and functioning of the system.\\[1ex]
In order to fight against cyber-criminality in France, since the end of the 90’s, the Central Criminal Intelligence Service (SCRC\footnote{Service Central de Renseignement Criminel}) of the GN Judicial Pole (PJGN\footnote{Pôle Judiciaire de la Gendarmerie Nationale}) has developed various strategies, among which the setting up of a Cybercrime Fighting Center named C3N \footnote{Centre de lutte Contre les Criminalit\'es Num\'eriques}. SCRC, which consists of the C3N and of the Intelligence Division (DR\footnote{Division du Renseignement}), aims at improving prevention and protecting individuals and companies from cyber crimes, mainly small and medium businesses, which have less capacity to invest in cyber security\footnote{We describe here the GN organizational structure during the period covered by our dataset. The C3N has recently moved to the newly created COMCyberGEND (Commandement de la gendarmerie dans le cyberespace)}. In 2014, the GN started collecting centrally the complaints related to cyber attacks, of individuals or companies from rural and peri-urban areas in all metropolitan and overseas territories (it covers 95\% of the national territory and 55 \% of the French population). One task of C3N is to collect data and exploit criminal information with the DR, relying on the analysis of the thousands complaints that are received at GN and registered by the SCRC.\\[1ex]
From a management point of view, cyber security has to be weighed up against building resilience to IT attack or failure.
It is clear that nowadays, companies need to give access to their services through Internet. But systems connected to Internet cannot be 100\% safe. Hence, there will always be a certain amount of system failures due to cyber-crimes. 
The question is where to allocate resources: to security and/or to resilience? How much should be invested in security through fire-walls, multiple identifications, security officers, etc., against in building resilience  e.g. by  prevention measures, using multiple systems with backups, or/and by covering the remaining risk through an insurance policy whenever it concerns material damages (or any quantifiable damage)? The answer depends on the type of risk faced by a company or organization.\\[1ex]
%
This raises the question of the insurer's point of view, who needs to go further in her understanding of cyber risk.  Moreover, the targets of cyber attacks are largely affecting intangibles such as data theft or reputation damages, making losses difficult to quantify and predict. Due to this lack of knowledge and their fear of a strong systemic component, insurance companies generally offer inadequate cyber risk coverage to their customers, whether they are individuals, businesses or organizations; see~\citep{Advisen2018}~or~\citep{SwissRE2017} for discussions on these issues. The insurance market for cyber risk is in its infancy, although it is growing at a fast pace. The total gross premium was estimated to reach 8 billion US\$ in 2020, a very small proportion of the total gross insurance premium of 4,000 billion US\$. However, it has grown threefold since 2014; see \citep{MarshMicrosoft2019}.
Insurance companies do provide cyber covers but they are limited in their coverage and reimbursements. This is the major complaint from customers as stated in~\citep{Accenture2019}: {\it "There is a lack of capacity in the market and a willingness of insurance companies to take over this risk"}. It is probably why management still relies much more on security spending, expected to reach 124 billion US\$ compared to insurance premiums of 8 billion US\$.
Until recently, there were also many "silent" covers, {\it i.e.} exposures to loss or liability from a cyber-triggered event in other lines of business (Directors and Officers, Property Damages, Business interruption); see~\citep{CRO2016}. However, insurances companies are rapidly removing the silent covers from the contracts, leaving insured entities uncovered against cyber risks, if they do not buy specific cyber policies.\\[1ex]
From these various perspectives, it becomes clear that the analysis of cyber risk is a condition for building a resilient society against cyber crime. The first step towards modelling is to collect data reflecting accurately this risk.
Nevertheless, as it is an emerging risk, it is difficult to find relevant datasets that describe well the various types of attacks to the systems and measure their quantitative impact.
While most academic studies use available data breaches public datasets,  \cite{Eling2019} had the nice idea to extract 1,579 cyber risk incidents from an operational risk database, to consider a larger range of cyber risks.
In our case, we had the opportunity to get access to a large, unique and exceptionally rich database - the database of cyber complaints filed at the GN, covering many categories of cyber risk. Besides its sheer size (more than 200,000 data records), this database contains not only quantitative information on possible damages, but also qualitative one such as a text description of the complaints and additional information on the victims and actors.\\[1ex]
Having access to a new database, from a different source, is of much interest. Indeed, it gives the opportunity to compare results obtained from various sources, to look for some dependence between data coming from different databases, etc.
Hence, we proceed to a first analysis of the GN database, to shed light on some aspects of cyber risk, in particular, to the probability of extreme events and their frequency. 
For that, we introduce an algorithm based on a stochastic hybrid model, detecting automatically the threshold above which observations are considered as extremes, facilitating the estimation of the heaviness of the tail distribution. This method has been tested at different stages on various types of data (in engineering, finance, insurance) and could become part of the stochastic modelling techniques used in operations research (a software package is under construction, with two cases, one being the version developed in this study).
Our objective is to transform the cyber threats and its uncertainties into a measurable risk. Quantitative assessment of risks is the basis to design insurance covers for hedging the worst consequences of cyber crimes. Moreover, knowing the probability of occurrence of cyber attacks and their severity distribution allows management to find the right balance between investing into cyber security or/and in insurance protection. It is one of the many steps towards making society more resilient.\\[1ex]
In Section~\ref{sec:questions}, we review the state-of-the-art in cyber risk modelling and introduce the research questions we aim at answering with this study. We discuss the main issues related to data exploration and present basic statistics on the GN database in Section~\ref{sec:data}. In view of modelling heavy-tailed non-negative asymmetric data, a self-calibrating algorithm based on a general parametric model and on non linear optimization techniques is developed in Section~\ref{sec:model}.  Confidence intervals for the evaluated model parameters are introduced in this algorithm, revisiting the Jackknife technique because of its high execution speed. 
Application of our method to the GN cyber database follows in Section~\ref{sec:appliCyber}, turning to modelling damage severity (Section~\ref{ss:appliSeverity}), then to both severity and frequency of extreme damages (Section~\ref{ss:Poisson}).
Consequences for risk management are discussed in Section~\ref{ss:RM}, while investigating a possible classification of cyber attacks via their distribution heaviness is tackled in Section~\ref{ss:classes}.
We conclude the study in Section~\ref{sec:concl}, discussing management and research perspectives. 
In Appendix~\ref{App-test2and3components}, a series of experiments based on simulated data is conducted to challenge and prove the benefit of our method. Appendix~\ref{App-estim-comparison-EVTmethods} completes the study comparing the tail-threshold and the tail index estimations obtained with our algorithm and  various EVT methods included in the {\it tea R-package}.

\section{State-of-the-art in cyber modelling and our research questions}
\label{sec:questions}
\vspace{-2ex}
\paragraph*{State-of-the-art -} Modelling of cyber risk is an important issue, as it helps understand the underlying (dynamical) structure of this phenomenon through its realization (collected data). Moreover, it allows generalizing beyond the data, to obtain relevant  perspectives for future behaviors, in view of drawing possible scenarios. It is also a fundamental step for convincing insurance companies to take over this risk. As long as cyber risk is not widely understood, fears dominate the market and reluctance is the rule. The fast pace of information technologies makes this type of risk difficult to analyze, a challenge already taken up by researchers from different fields, as actuaries, data scientists, economists (in particular from game theory), IT system experts, probabilists and statisticians, etc (see e.g. \cite{Dacorogna2022}). Various toy/theoretical models (due to a lack of data), and models capturing some features of cyber risk have been suggested.

The field of research in cyber risk is very active, even though still in a fledging state leaving big holes in our understanding of it. Let us briefly review some of the directions of research.
We could refer to many papers; we do not give an exhaustive list, but examples of recent ones in which one can also find complementary bibliographies. \citep{Agrafiotis2018} provide a taxonomy of cyber harms and a study of their possible consequences, while \citep{Cohen2019} suggest their own taxonomy (with a few overlaps) and definition for the ever elusive cyber risk. Based on this and a database compiled by AON that contains 30,000 cases, they statistically describe financial cyber losses and suggest that the risk is very similar to operational risk. Efforts in the direction of a taxonomy can also be found in \citep{CRO2016} from the point of view of the insurance industry. Unfortunately, those various taxonomies are not fully compatible. Among the researchers working on the statistical modelling of empirical cyber risk data, the group around Eling pioneered this path (see ~\citep{Eling2016}). In one of their latest publications on the subject, \citep{Eling2019} apply a dynamic version of the standard peak-over-threshold (POT) method in EVT designed by \citep{Chavez2016} for operational risks, including covariates to analyze cyber losses compiled in an operational risk database.  Contrary to~\citep{Cohen2019}, they conclude that cyber risks differ from other operational risk categories. This debate illustrates the fact that data on the subject are sparse, which makes it difficult to come up with a consistent picture. It may differ from one dataset to another. That is why it is important to analyse as many datasets as possible, as soon as they become available.

Modelling goes in various directions: \citep{Baldwin2017} present a simple model to look at contagion of cyber attacks, while~\citep{Boehme2018} propose a review of cyber risk analysis from various disciplines and identify ways to improve cyber risk modelling. \citep{Fahrenwaldt2018} model, with an interacting Markov chain, the diffusion of cyber viruses or worms in a structured data network, while \citep{Farkas2021} apply Generalized Pareto regression trees to analyze cyber claims, identifying criteria for claim classification and evaluation on the same database as in~\citep{Eling2019}. \citep{Peng2018,Xu2017} argue that cyber risk should be tackled in a multivariate framework where the various risk factors are dependent on each other via copulas, otherwise, the risk would be underestimated. Nevertheless, they neglect the fact that treating the problem using EVT has already shown good estimation of the extreme risks. Other ways of tackling the question of cyber risk is to look at the amount of money that should be invested in cyber security versus buying insurance protection. It is what~\citep{Marotta2017,Wang2019} explore in their papers, while \citep{Nagurney2017} deal with the optimal investment in cyber security under budget constraints. Using a supply chain game theory network model, they study the vulnerability of the network to additional retailers or budget constraints.
Unfortunately, all these models are not backed by strong empirical evidences due to the lack of data of various sources. Along the same line of research,~\citep{Paul2021} propose a two-stage stochastic programming model to help decide on the optimal resource allocation strategies by governments and firms. They conclude that {\it "it is beneficial to spend more on intelligence given its increasing returns to the reduction of social costs related to cybersecurity"}. Intelligence means not only detection effectiveness, as defined by the authors, but also  a better understanding of the quantitative impact of the risk, hence identifying the likely attackers' targets. Modelling games between attackers and providers in an interdependent cyber-physical systems (CPS) is the chosen approach by \citep{He2020} to analyze the survival probability of CPS and Nash equilibrium strategies.

On the actuarial side, \citep{Romanosky2019} do a qualitative analysis of cyber insurance policies by examining those filed with state insurance US commissioners and found surprising variations {\it "in the sophistication (or lack thereof) of  the equations and metrics used to price premiums"}. This is another testimony of the lack of maturity of this market.~\citep{Carfora2019} propose an actuarial approach to compute insurance premium based on the publicly available dataset of the Privacy Rights Clearing House. \citep{Zeller2020} present a review of the actuarial literature on the topic and apply an approach based on marked point process for modelling cyber risk in view of determining the insurance premium. They identify and propose models for key co-variables required to model frequency and severity of cyber claims. Finally, let us refer to \citep{Bouveret2018}, who proposes to use a frequency severity model for the computation of the Value-at-Risk of the cyber risk of financial institutions.
This brief review of the various ways the insurance market and the academic literature tackle the problem, is far from being exhaustive, if ever possible. It shows, first of all, the lack of data, second the wide spectrum of approaches, and third, conclusions that seem not always congruent. More investigation, in particular empirical on diverse data sources, must take place to better understand this complex problem.\\[-7ex] 
\paragraph*{Our research questions -}
Our research program is one step in this direction and is structured around four questions; this study answers the first two, while setting up a methodology and giving hints for further investigating the last two.

The first question is whether the data collected by SCRC would add some value for studying cyber risk.
Of course, the very nature of this data, quite different from the sources studied so far, partakes in the originality of the study, as it can only help shed another light to the cyber risk landscape. 
It may also reveal general characteristics of this type of risk, if they are found in other databases too. It implies, as for any other source, checking first for reliability and relevance of the data.  
To do so, we explore the database and develop on it a statistical study. We conclude that, indeed, it is a unique source of information, confirming as well as complementing the picture given by databases studied so far. This is the object of Section~\ref{sec:data}.

Given that systemic risk appears in the literature as a crucial feature of cyber risk, it is natural to question its presence in our data. Hence our second query: do we observe and detect effects indicating systemic risk? It is an important issue, of high interest to the insurance market.  That is why we explore the existence of heavy-tailed distributions for variables characterizing the cyber risk. We do it on the damages reported in the complaints. It is intended to determine the threshold (denoted here, `tail-threshold'), above which observations are considered as extremes. Standard graphical methods of EVT could be used, but they are not always very robust. Here, we adapt to asymmetric non-negative heavy-tailed data, a model introduced in Debbabi {\it et al.}(see \citep{Debbabi2014}, \citep{Debbabi2017}) and developed the associated iterative algorithm, which identifies automatically the tail-threshold, is flexible and of easy use, whatever the size of the dataset. This unsupervised method gives the advantage of fitting the whole empirical distribution, whatever its heterogeneity, with a specific treatment for the tail where observations are scarce.  
The presence of very heavy tails is clearly assessed for the amounts at stakes in the complaints. However, a high probability of extreme event would not be enough to qualify the risk as systemic but it constitutes one piece of the puzzle.

Our focus on extremes is additionally motivated by the building of cyber resilience. We believe that cyber security may tackle the main attacks which probabilities fall in the main body of the distribution, while the tail distribution may concern the attacks they cannot handle or detect. Extreme attacks will be best fought through cyber resilience in two ways: on the operational side, increasing the redundancy of the system, on the other side, protecting the finance of the company through good insurance covers. In order for insurance to propose a good coverage to firms, the underwriters will perform a due diligence on their management of cyber security.  It is in the dialogue between insurance underwriters and risk managers that the company will be able to identify on the operational side, where resilience through redundancies of the IT system should be put in place in order to save on the insurance premiums. At the management level, this tradeoff between investments into securities and insurance covers has to be found. A good modelling of this risk will definitely help find the optimal tradeoff as well as determine the fair insurance premium. Moreover, by having a good model for the tail of the distribution, it becomes possible to estimate the capital needed for covering this risk, which is another piece of the puzzle to estimate the cost of bearing this risk.

The third question concerns the classification of cyber attacks. Our goal is to build an alternative classification of the complaints based on statistical regularities, in order to reduce the number of classes, whether it is those proposed by the ministry of justice, or those by the SCRC. It is too early to call an answer to this question, but hints about a varying tail index gives some hopes in this direction (see the last paragraph in Subsection~\ref{ss:classes}). Here again, we can improve cyber resilience by early detection and better characterization of the type of cyber attacks. Then, the security forces can focus their investigations and fight more efficiently the cyber attackers.

Finally, another aim we target is the dynamic modelling of this risk in a multivariate setting, since the size and content of the GN database offer the possibility of such investigation. Our algorithm should facilitate the processing of the extremes in this extended context. A standard argument in the literature is that cyber risk is characterized by its fast-changing environment. It is another study in itself, for which extra cleansing and treatment of the database are needed, in particular through the complaints description. As we manually double-checked the data on extremes, we concentrate, in this paper, on their time evolution, taking into account the frequency of the cyber attacks with extreme damages.  
This is tackled in Subsection~\ref{ss:Poisson}. It appears that the frequency of large damages in complaints does not vary significantly with time, at least on the observed period. Nevertheless, once we will have access again to the database, with an increasing amount of cyber attacks during the covid19 pandemics, we will proceed a second time to the same frequency study to check the current result.
\vspace{-2ex}
%
\section{Data Exploration} 
\label{sec:data}
\vspace{-2ex}
A scientific approach to study any phenomenon must rely on data, on one hand to inspire the modelling, on the other hand, to check the correctness of the model. We want to understand how our data have been generated. This may look trivial, but is, in fact, very important, as it is the foundation of  the study. For cyber risk, this is much more difficult as the risk is changing rapidly and the amount of representative data is scarce. Here, we have a unique opportunity to study data widely collected, and over few years, covering most of the French territories. 
A first phase of data cleansing is needed. 
In our case, the amount of data is too high to manually check its reliability. An automatic text recognition algorithm has been developed by C3N to ensure the congruence of the various fields. However, improvements need to be made to this fledgeling procedure. Therefore, we decided to also check manually the data presenting the largest damages, as we are particularly interested in their modelling. 

Note that the data presented here, have also been briefly described in a book chapter (see \citep{Ventre2020}) to illustrate the scientific approach when working on data. Nevertheless, since the aim here is quite different, we choose to come back on the descriptive statistics of the dataset under study and extend it, so that the paper is self-contained and more comprehensive.  Moreover, we have in the meantime double-checked the information given in the dataset and performed more detailed descriptive statistics for reaching a broader understanding, before going deeper in the analysis and modelling.

At the GN, we were explained the process of entering the data in the system. In every GN office in France, the officer (not a specialist in cyber security) receiving the complaint is in charge of writing a short report and filling the various fields of the database. 
From a geographical point of view, the data collection is spread over the entire country. However, the big cities (e.g. Paris, Lyon, Marseille) are not covered in this database as they are under the responsibility of the {\it Police Nationale}. The industry, however, is usually located in the peri-urban area surrounding big cities.
Nevertheless, given that cyberspace and geographical space are more or less independent, most of cyber crimes appear to be distributed demographically rather than geographically (except maybe in the case of  those of 'proximity' such as cyber-harassment). This seems to postulate that the GN and the police, having more or less half of the population living in their area of competence,  have to both know about cyber criminality in comparable proportions. The officer's first concern is to help the people who are filing complaints and try to find the culprit, rather than filling in a database. This may explain the existence of many errors due, either to wrong transcriptions, or typing errors. For this reason, we decided to review manually the extreme amounts. We compared, one by one in the database, the written description of the complaint to the other data fields (list given below) to assess consistency. We looked at 1,100 items containing the largest amounts in the declared 'dammages' field. They represent the 98.2\% quantile of the studied sample. We detected various problems in the dataset. Among them, the most frequent (90\% of the few errors found on these extremes) was a mistake in reporting the cents amount, where the dot marking the cents was missing (e.g. instead of 500.00 \euro{} a much larger amount of 50,000 was reported). 
In the rest of the section, we present results based on both filtering procedure, the automatic one developed by C3N and our own manual work.

\vspace{-2ex}
\subsection{The SCRC database}
\label{ss:database}
\vspace{-2ex}
The SCRC database starts from 2014, but is more reliable since July 2015. Thus, we consider in this paper the period  from 07-2015 to 04-2019, which includes 208,037 data.

The data to which we have access at SCRC, correspond to structured data and are presented under the {\it JavaScript Object Notation} (JSON) format. Each complaint has been first registered at SCRC according to 3 main fields: the cyber crime description, its victim, and its author. Then, it has been filtered and exported into a database presenting the following fields:\\[-6ex]
\begin{description}
\item [1- report\_date:] Reporting date of the cyber crime complaint \\[-5ex]
\item [2- damages:] Amount of the damage in Euro (\EUR)\\[-5ex]
\item [3- victims.dob:] Date of birth of the victim \\[-5ex]
\item [4- victims.sex:] Gender of the victim\\[-5ex]
\item [5- category:] Category of the crime (SCRC classification)\\[-5ex]
\item [6- natinf:] Nature/type of crime (Ministry of Justice classification)\\[-5ex]
\end{description} 
Note that the field 'category' corresponds to the classification of cyber crimes by SCRC into 10  groups subdivided into 52 subgroups (also subdivided into subsubgroups), while the field 'natinf', also referring to the type of crime (and represented by a code; see  \cite{natinf2014}), is defined by the `{\it p\^ole d'\'evaluation des politiques p\'enales de la direction des affaires criminelles et des gr\^aces du Minist\`ere de la Justice'}\footnote{cf. \url{http ://www.justice.gouv.fr/include_htm/pub/rap_cybercriminalite_annexes.pdf}} and includes 475 classes. In Table \ref{tab:dataPrej}, we present the composition of the cyber crimes sample communicated  by SCRC. 
About 70\% of the complaints do not mention any declared amount, or suggest a null amount. This class includes physical (e.g. child pornography) or moral (e.g. hate crimes, cyber-harassment) harm.
The other 30\%  complaints correspond to material damage recorded by the GN units on the declarations of victims of property crimes. We introduce there a new class defined with a threshold of $500$~\EUR, corresponding to the amount above which the judicial system authorizes generally for prosecution. 
\vspace{-2ex}
\begin{table}[H]
\centering
\caption{\label{tab:dataPrej}\sf\small Breakdown of the 208'037 data in terms of the damages amount. \vspace{0.7ex}}
{\small
\begin{tabular}{crr}\hline
&&\\[-1.5ex]
Amount $x$ (\EUR) & Sample Size $n$ & Percentage \\
&& (w.r.t. the total size)\\
&&\\[-1.5ex]
\hline
 \hline
&&\\[-1.5ex]
Not Declared (ND) or $x=0$ & 147,052 & 70.69\% \\
$0 < x < 500$ &  29,074 & 13.97\%\\
 $x \ge 500$ & 31,911 & 15.34\% \\[-2ex]
&&\\
 \hline
\end{tabular}
}
\vspace{-2ex}
\end{table}
%
It should be noted that besides the description of the cyber attack, no variable has yet been introduced in the GN database to discriminate between individuals and companies. We looked at it manually for the extreme damages above 40,000~\EUR, but, of course, it needs to be done automatically for every complaint; adding a dedicated field in the data entry form has been recommended to the GN.

{\bf $\triangleright$ Gender of the victims of cyber crimes -}
%
Looking at the type of population reporting to GN for a cyber attack (Table~\ref{tab:sexe}), we observe that 11.65\% do not contain a mention of gender (ND for "not defined"), leaving 183,801 data points instead of 208,037. The gender is not a discriminating parameter when considering the number of complaints, as can be seen in Table~\ref{tab:sexe}. Nevertheless, we could investigate if the type of complaints differs by gender. It will be considered in a further study.
\begin{table}[H]
\centering
\caption{\label{tab:sexe}\sf\small Gender classification of the sample with non-negative amounts. \vspace{0.7ex}}
\begin{tabular}{ccc}
\hline 
&&\\[-1.5ex]
Gender & Data number & Percentage \\
&&\\[-1.5ex]
\hline
 \hline
&&\\[-1.5ex]
F & 91,599	& 44.03\% \\
M & 92,202	 & 44.32\% \\
ND & 24,236 & 11.65\% \\[-2ex]
&&\\
 \hline
\end{tabular}
\end{table}
\vspace{-2ex}
The gender proportion is slightly obscured by the fact that it is not possible to distinguish between private complaints (made by individuals for attacks on their private systems) and complaints originating from companies. The gender portion would be better defined if this distinction could be made. Nevertheless, it could be retrieved, most of the time, through the complaint description.
Given the large amount of data, automatic text recognition is being developed to find out this (and other) information. 
Individuals and companies need to register their complaints at the National Police/Gendarmerie if they want to be covered by their insurance. We suspect that the majority of complaints comes from individuals. Nevertheless, we could check manually for the extreme observations the two following properties:
(i) Statistically, the amounts would be roughly the same for individuals and companies; (ii) the vast majority of complaints originates from individuals even for the very large amounts.

\begin{figure}[h]
  \centering
  \includegraphics[scale=0.6]{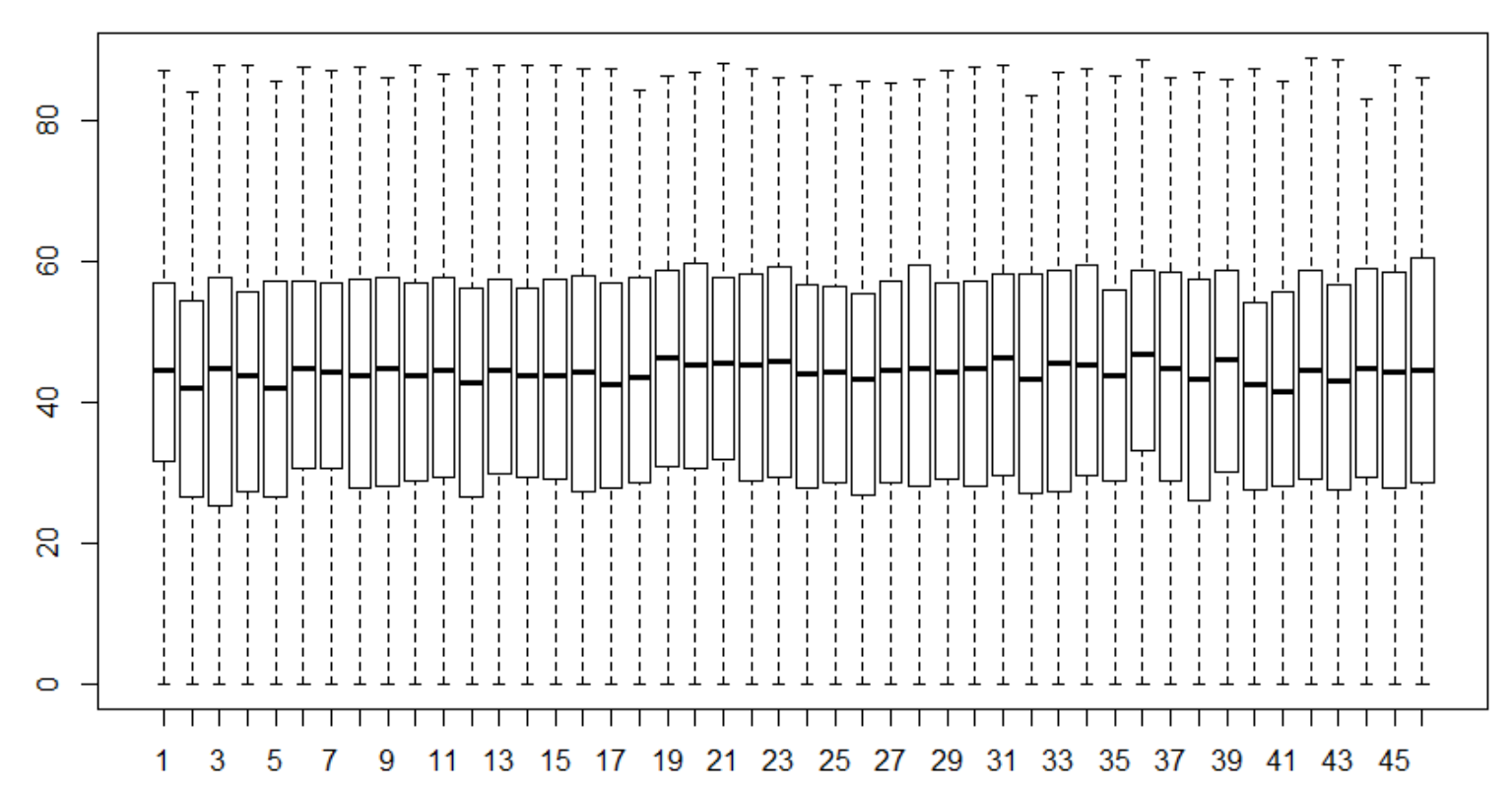}
  \parbox{282pt}{\caption{ \label{fig:boxplot-age} \sf\small Box Plot of the ages of the victims ($y$-axis) for every month of the period ($x$-axis) from July 2015 to April 2019.}}
\end{figure}

{\bf $\triangleright$ Age of the victims of cyber crimes -}
%
Let us turn to the age of the victims. We obtain the following box-plot given in Figure~\ref{fig:boxplot-age}.
We observe that the median remains more or less constant with time, with a value around $42.5$ years, which is also close to the average age ($43.4$ years) of the sample, and to the median age of the French population (around $40.5$).
The interquartile interval remains also more or less stable (interquartile interval [29.6; 54.4] years, on average). The lower and upper limits of this interval indicate a slight negative asymmetry. 

\begin{figure}[h]
  \centering
 \begin{minipage}{.33\textwidth}
\centering
  \includegraphics[scale=0.33]{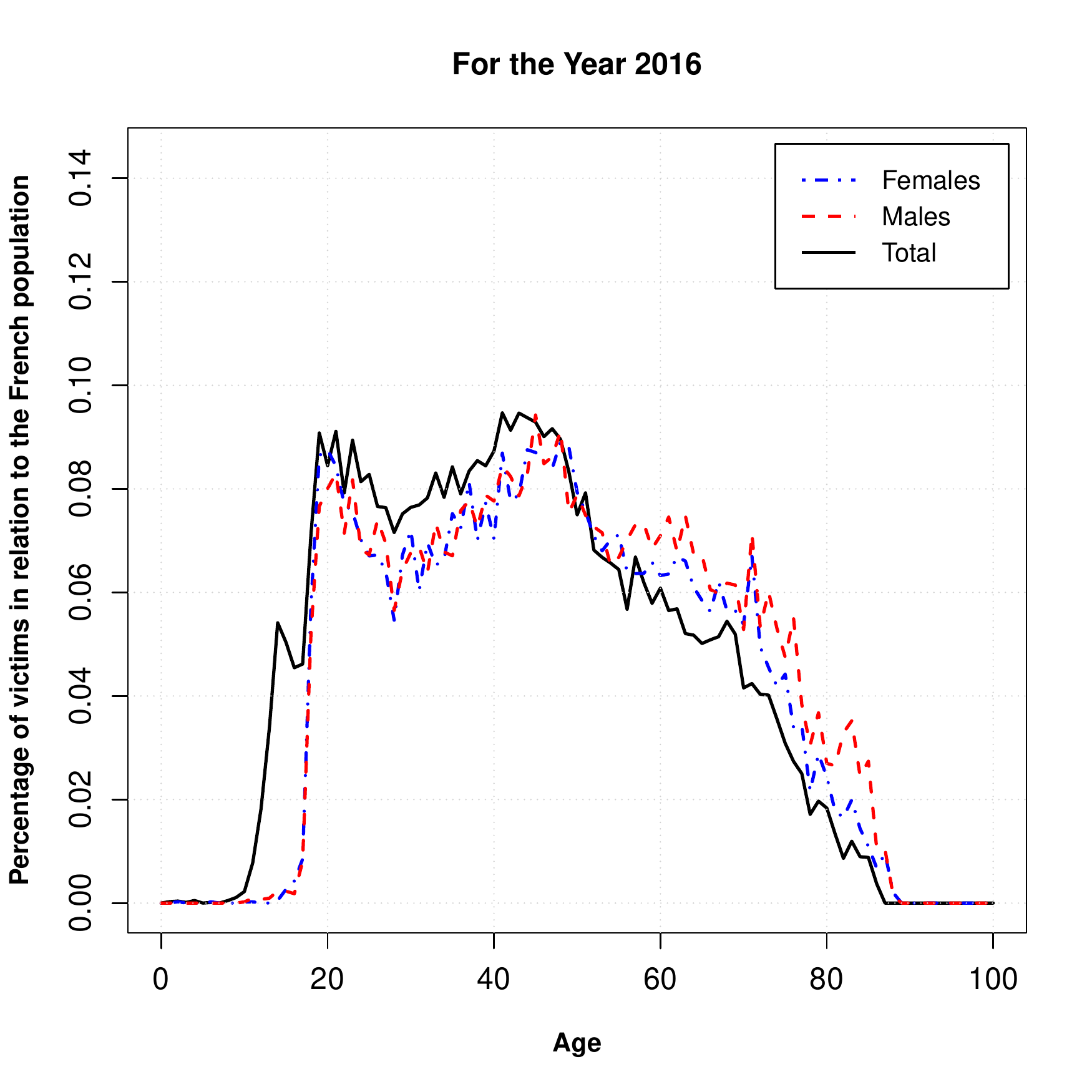}
\end{minipage}%
\begin{minipage}{.33\textwidth}
\centering
  \includegraphics[scale=0.33]{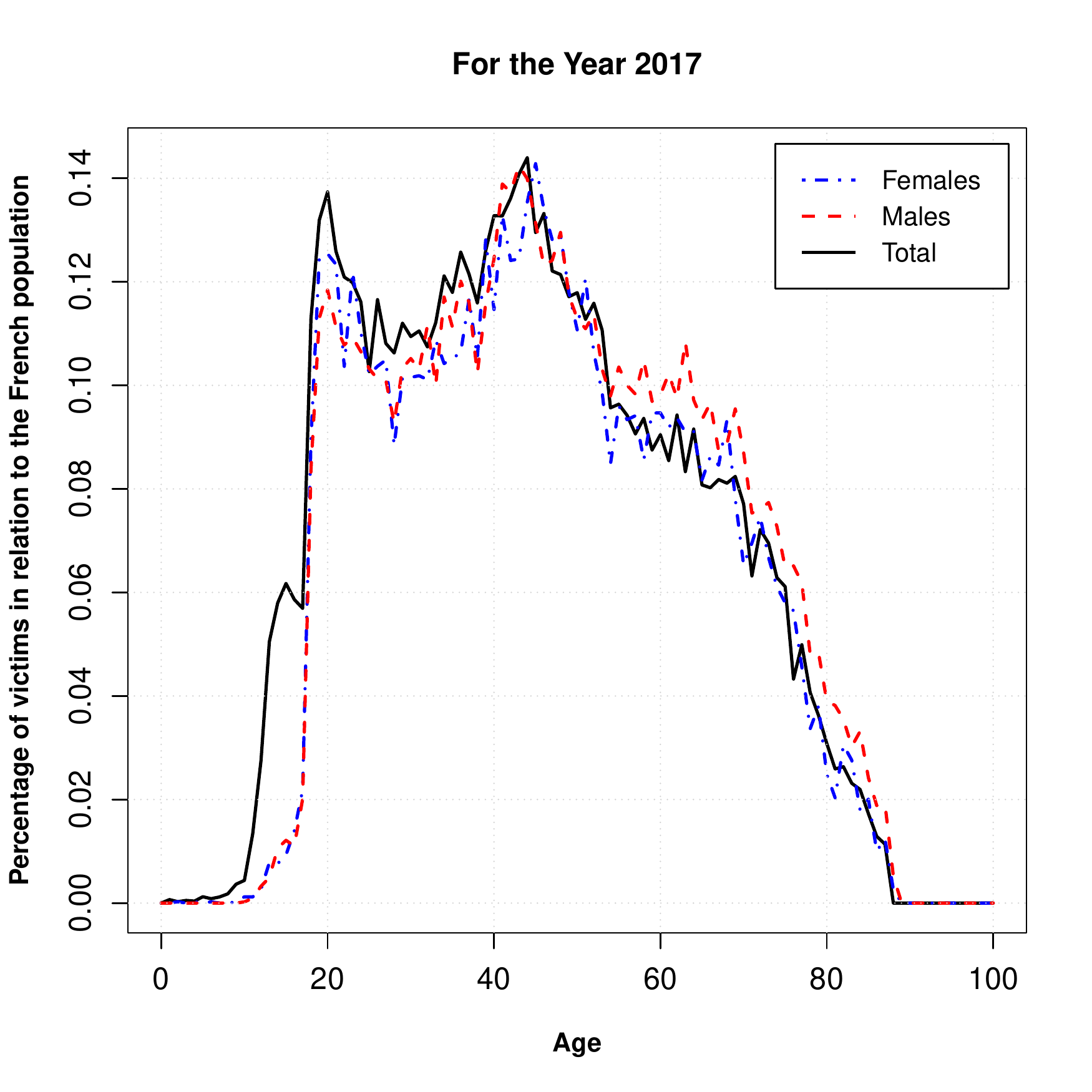}
\end{minipage}%
\begin{minipage}{.33\textwidth}
\centering
  \includegraphics[scale=0.33]{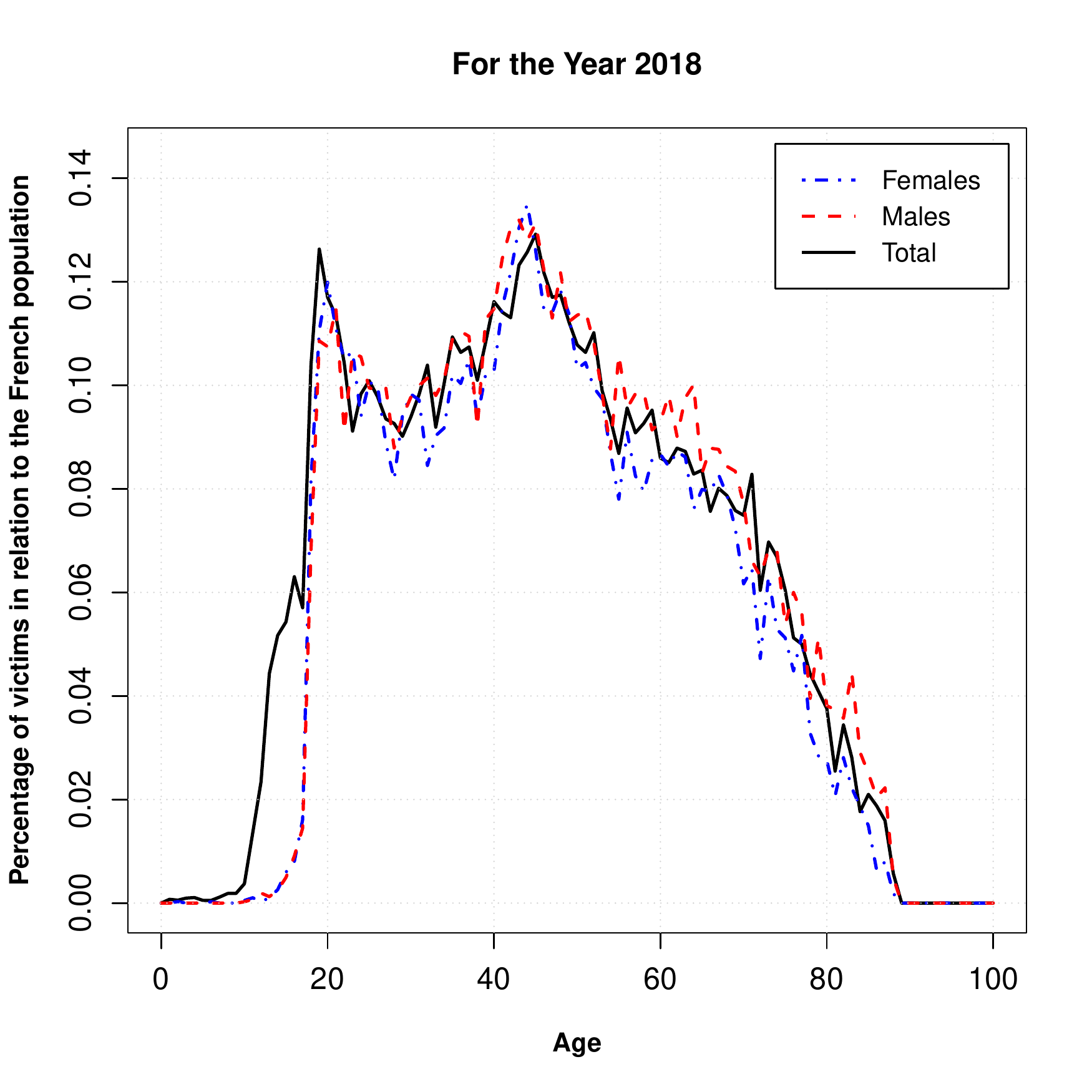}
\end{minipage}%
	\caption{\label{fig:age-pyr}\sf\small Comparison of the age of the victims to the ages pyramid of the French population for the years 2016, 2017, and 2018. The $x$-axis represents the age and the $y$-axis the percentage of victims w.r.t. the French population. Here, the label "Total" means the comparison with the sum of males and females.}  
\end{figure}
We study the distribution of the age of the victims for the three calendar years 2016, 2017, and 2018 (for which we have complete data that have been verified by SCRC), and compare it with the age pyramid of the French population, taking also the gender into account. For both years, the registered victims the most represented concern two classes, independently of the gender: young people around 20 years old and adults around 45 years old. The same criticism about the non-discrimination between individual complaints and company complaints could be made here as for the gender analysis. We are mixing the two sorts of origin. It is a limitation due to the fact that we do not have currently access to this information. Nevertheless, the small positive amounts are completely dominated by individual complaints and constitute the vast majority of the data. The proportion is accentuated to more than 95\% when considering the data with no declared amount (representing more than 70\% of the full dataset), which we also used in this age/gender analysis. This explains why we can afford comparison with the French population.

As illustrated in Figure \ref{fig:age-pyr}, the proportion of cyber crimes victims is, besides a small minimum around 30 years, of the same order between 20 and 45 years for both genders, then falls down rapidly below and above those ages. 
The proportion in Figure \ref{fig:age-pyr} is related to the pyramid of age as given by \cite{INSEE2018}.
Note that the value 100 on the $x$-axis collects all ages from 100 years old on. 
We looked at the cases where the ages of the victims were equal or above 100 years old. Those are generally cases of identity theft where the grand-child, or a relative, accesses bank accounts using the login and password of the owner. There is no case of ransomware or direct involvement of the elderly.
We observe in Figure~\ref{fig:age-pyr} that, in 2017, 0.10\% (0.06\% in 2016, and 0.09\%) to 0.14\% (0.11\% in 2016, and 0.13\% in 2018)\% of the French population aged from 20 to 60 years old have filed complaints as victims of cyber crimes. This represents already a large number of victims w.r.t. the 66,883,761 French inhabitants\footnote{\url{https://www.insee.fr/en/statistiques/2382601?sommaire=2382613}} (66,774,482 in 2016), given the fact that GN covers only 55\% of this population. All the more that it needs to be multiplied by a factor, according to the iceberg effect, estimated with various methods as roughly 250 on cyber complaints related to ransomwares in \cite{Dregoir2017}, using the GN database, but also external data (Google trends for the locky virus, which is of ransomware type)\footnote{For the study of general delinquency reportability rates, even if not so explicit for cyber aspects, see also \url{https://www.interieur.gouv.fr/Interstats/L-enquete-Cadre-de-vie-et-securite-CVS/Rapport-d-enquete-Cadre-de-vie-et-securite-2019}}.
Indeed, in cybercrime, a pronounced iceberg effect exists due to the absence of a complaint or, in the most serious cases, to the very absence of detection of the problem by the victims. Consequently, for security forces, the filing of complaints is only the visible part of a criminal phenomenon and does not grant access to the ground truth. To better understand the meaning of those percentages, one might compare them to the percentage of the (French) population victim of other types of attacks (non cyber ones). 

{\bf $\triangleright$ Cyber crimes by type -}
%
Now we turn to the type of cyber crimes and provide in Table~\ref{tab:X-natinf} the first 10 classes of the full sample of size 208'037, by decreasing order of class size. From the description registered at GN, it is not so easy to distinguish to which type a cyber crime belongs to, therefore how to classify it within a GN category and a Natinf one, especially given the large amount of those categories. We already know that, for insurance purpose,  the granularity must be much coarser: One future goal is to find a scientific way to regroup GN categories. One approach could be through the heaviness of the tail distribution, as discussed further when modelling the damages severity.  
\begin{table}[h]
\centering
\parbox{400pt}{\caption{\label{tab:X-natinf}{\sf\small Damages classified by type: the 10 classes the most represented among the full sample, identified by natinf code. It represents 78.1\% of the full sample of size 208,037.\vspace{0.7ex}}}
\footnotesize
\begin{tabular}{crlrr}
\hline 
&&&&\\[-1.5ex]
Class & Natinf code & Type & Complaints Number & Percentage \\
&&&&\\[-1.5ex]
\hline\hline
&&&&\\[-1.5ex]
{\bf 1} & {\bf 7,875} & {\bf Fraud}	& {\bf 123,536}	& {\bf 59.38\%} \\
{\bf 2} & {\bf 28,139}	& {\bf Identity theft} & {\bf 9,697} & {\bf 4.66\%}\\
{\bf 3} & {\bf 58}	& {\bf Breach of trust}	& {\bf 7,256}	& {\bf 3.49\%}\\
4 & 372	& Defamation & 4,888 & 2.35\%\\
{\bf 5} & {\bf 1,619}	& {\bf Violation to SADP\textsuperscript{a}} & {\bf 4,495} & {\bf 2.16\%}\\
{\bf 6} & {\bf 7,203} & {\bf Blackmail} & {\bf 3,295} & {\bf 1.58\%}\\
{\bf 7} & {\bf 7,151} 	& {\bf Theft} & {\bf 2,891} & {\bf 1.39\%}\\
8 & 10,765	& Invasion of privacy & 2,399 & 1.15\%\\
9 & 7,173 & Threat to individuals & 2,088 & 1.00\% \\
10 & 376 & Public abuse  & 1,997 & 1.00\%\\[-2ex] 
&&&&\\
\hline
\end{tabular}
\\
\textsuperscript{a}SADP: System of Automated Data Processing (STAD in French)~~~~~~~~~~~~~~~~~~~~~~~~~~~~~~~~~~~}
\end{table}
\vspace{-2ex}
When looking at the sample for which dammage amounts above 500~\EUR are provided, we obtain the following classification in Table \ref{tab:X-natinf-500}. Note that the classes common to Tables~\ref{tab:X-natinf} \& \ref{tab:X-natinf-500} are indicated in bold.

We observe that 'Fraud' is the most represented types of damages in both tables, with a proportion above 87\% of the 31,911 considered data in Table~\ref{tab:X-natinf-500}, and above 59\% (for the 208,037) in Table~\ref{tab:X-natinf}. This first (by size) class is far from the second class (Identity theft in Table~\ref{tab:X-natinf} and Breach of Trust in Table~\ref{tab:X-natinf-500}, respectively) which size is less than 5\%. Note the second gap in the size between 'Breach  of trust' and the other classes for damages above 500~\EUR,  going from 4.7\% to less than 1\%, whereas the percentage is regularly decreasing in Table~\ref{tab:X-natinf}.
\begin{table}[H]
\centering
\parbox{430pt}{\caption{\label{tab:X-natinf-500}\sf\small Damages above 500~\EUR classified by type: the 10 classes the most represented among the 31,911 data identified by natinf code. It represents 96.6\% of the sample of size 31,911. \vspace{0.7ex}}}
{\footnotesize
\begin{tabular}{crlrr}
\hline
&&&&\\[-1.5ex]
Class & Natinf code & Type & Complaints Number & Percentage \\
&&&&\\[-1.5ex]
\hline\hline
&&&&\\[-1.5ex]
{\bf 1} & {\bf 7,875}	& {\bf Fraud}		& {\bf 27,914}	& {\bf  87.5\%}\\
{\bf 2} & {\bf 58}		& {\bf Breach of trust}	& {\bf 1,497}	& {\bf 4.7\%}\\
{\bf 3} & {\bf 7,151} 	& {\bf Theft}		& {\bf 257}		& {\bf 0.8\%}\\
{\bf 4 }& {\bf 28,139}	& {\bf Identity theft} 	& {\bf 234} 	& {\bf 0.7\%}\\
5	  & 26012		& Fraud per legal entity & 214		& {\bf 0.7\%}\\
{\bf 6} & {\bf 1,619} 	& {\bf Violation to SADP} & {\bf 181}	& {\bf 0.6\%}\\
7 	& 94			& Harmful substances in educat. instit. & 153 & 0.5\%\\
{\bf 8} & {\bf 7,203} 	& {\bf Blackmail} 	& {\bf 152} 	& {\bf 0.5\%}\\
9 	& 560		& Use of falsified or forged cheque & 134 & 0.4\%\\
10 	& 7,881		& Violation to SADP at the prejudice & 94 & 0.3\%\\
&& of a vulnerable person && \\[-2ex] 
&&&&\\
\hline
\end{tabular}
}
\end{table}

\vspace{-2ex}
\subsection{Frequency} 
\label{ss:freq}
\vspace{-2ex}
Let us proceed to the inter-temporal analysis. Before doing it, we recall that the time stamp in our database is the reporting date, not the date of occurrence of the attack. This fact can of course bias the analysis. We are able to detect trends in the occurrences but not outbursts or seasonality of attacks.
\begin{figure}[h]
  \centering
  \includegraphics[scale=0.6]{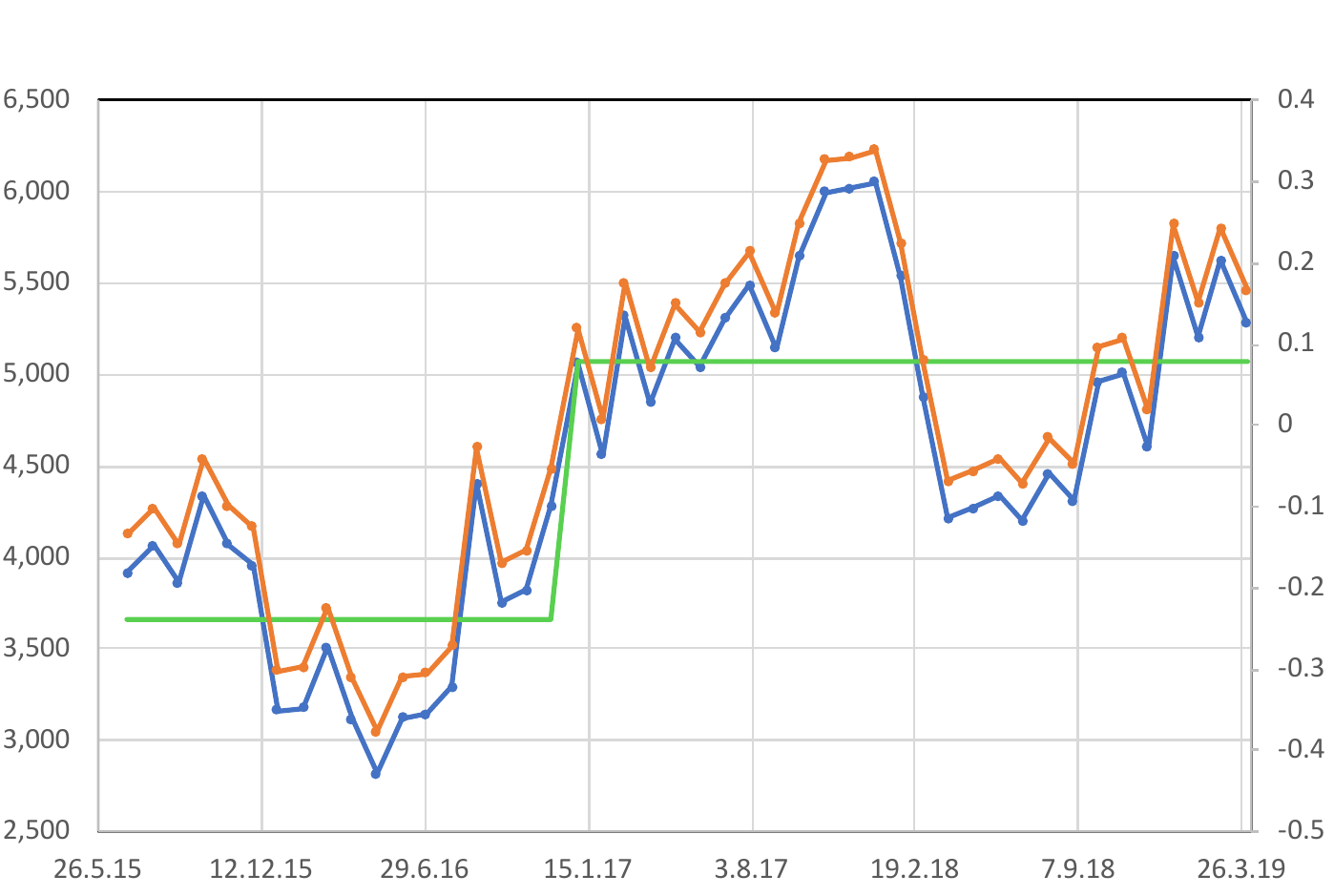}
\vspace{-2ex}
\parbox{346pt}{\caption{\label{fig:frm2} \small\sf Monthly frequency of the $N$ complaints. The $x$-axis represents the 46 successive months over the entire period. The left $y$-axis gives the monthly frequency of complaints, while the right one  gives the normalized number of complaints per month w.r.t. the monthly average on the full sample.}}
\end{figure}
In Figure~\ref{fig:frm2}, we present the variations of the monthly frequency of complaints on the sample from July 2015 to April 2019; see the blue (dark) curve. The mean of this quantity is slightly increasing over time, from slightly above 4,000 (4,194.5) for the two first years, to around 5,000 (5,032.8) for the last period; see the (green) horizontal lines in Figure~\ref{fig:frm2}. The orange (light) curve represents the variation of the monthly number of complaints against the average over the full sample. In the same figure, the values of the right vertical axis are normalized and defined as:
\begin{equation}\label{eq:moyNormal}
 \frac{m_i - \overline{M}_T}{\overline{M} _T} \quad \text{with}\quad \overline{M}_T= \frac1{K_T}\sum_{j=1}^{K_T} m_j ,\quad i=1,\cdots,K_T,
\end{equation} 
where $K_T=46$ is the total number of months in the sample, and $m_i$ is the monthly frequency for the $i$-th month with  $i \in [ \, 1,K_T \, ] \,$.

We see that the two curves (orange-light and blue-dark) are quite similar, as expected from the way they are computed. However, the scale displayed on the right is different and varies from negative (-0.45) to positive (0.35) values. It helps distinguish between two periods: The first 2015-2016 is negative and the second 2017 to 2018 is positive. Moreover, for the year 2017, a positive trend is clearly observable.

\begin{figure}[h]
  \centering
  \includegraphics[scale=0.6]{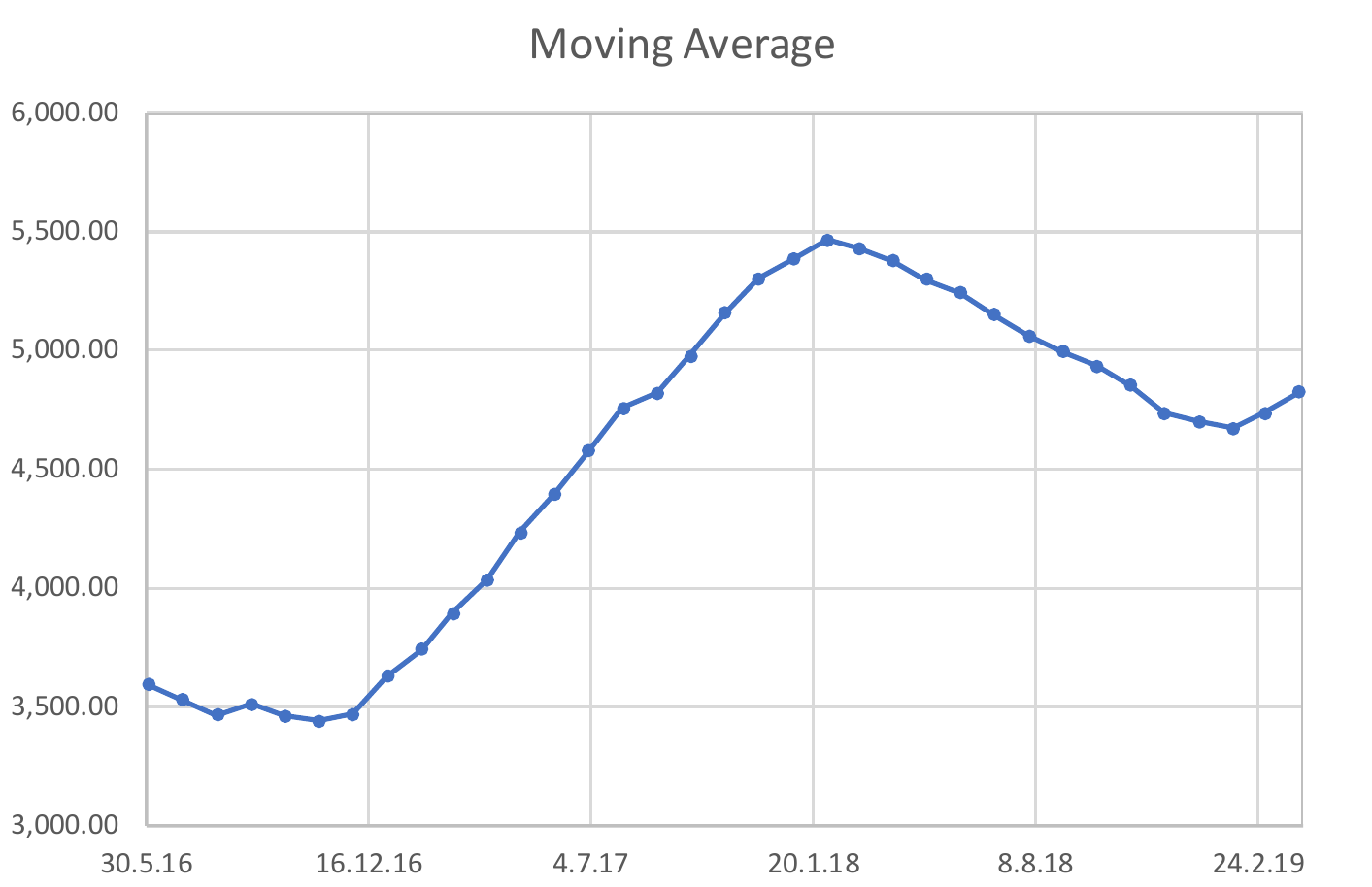}
  \parbox{300pt}{\caption{\label{fig:fra}\sf\small Annual moving average of the monthly frequency of complaints. The $y$-axis presents the average number of complaints, while the $x$-axis presents the dates of the moving average.}}
  \label{fig:Moy_nbr_ann}
\end{figure}

In Figure~\ref{fig:fra}, we draw the evolution from 2016 to 2019 of the annual moving average of the monthly frequency of the $n$ complaints, choosing a monthly rolling window. It means that for each month from June 2016 (as the yearly averaging  starts in July 2015) to April 2019, we use the data of the past year until the considered month, to compute the average of the complaints number.

Using the same notations as in Equation~\eqref{eq:moyNormal}, we compute the annual moving average during the whole period of $K_T=46$ months, as
$\displaystyle \overline{M}_{T,i}= \frac1{k_T}\sum_{j=i}^{i+k_T-1} m_j$, 
$\forall\, 1\le i \le K_T-k_T+1$, with $k_T=12$ months and $\overline{M}_{T,1}$ corresponding to the average on the month of June 2016.

We observe a strong increase of the number of complaints from the year 2017, going from 3500 to 5500 complaints, with a positive (linear) trend. In 2018, there is a slight decrease until the level 4700, then it seems to go up again in 2019. Those observations confirm those made on Figure~\ref{fig:frm2}.

\vspace{-2ex}
\subsection{Severity} 
\label{ss:sev}
\vspace{-2ex}
In this section, we proceed to the statistical study of the data for which damages amounts are given as positive, {\it i.e.} which correspond to material damage recorded by the GN on the declarations of victims of property crimes.
In Table~\ref{tab:statDesc}, we present the main descriptive statistics on this sample (of size 60,985; see Table~\ref{tab:dataPrej}), namely: mean, median, standard deviation,  dispersion index (DI) ({\it i.e.} ratio between the standard deviation and the mean: $DI=\sigma/\mu$), skewness and kurtosis (all quantities being, of course,  empirical). Note that those results are to be taken with caution, as the database for small or ND amounts still needs a lot of care and corrections. Once the corrections done, a specific study for the data with fields ND and $x=0$ (and, eventually, $0<x<500$) will be the object of another investigation. We pay a particular attention to the 31,911 amounts above $500$~\EUR ($x\ge 500$) for two reasons: The first is that $500$~\EUR  corresponds to the amount above which prosecution can be open,  the second is pragmatic, the database for small or ND amounts still needing some care, as already explained. We also give the main descriptive statistics for the two samples of positive amounts and of amounts above $500$~\EUR. 
\begin{table}[h]
\centering
\caption{\label{tab:statDesc} \sf Descriptive statistics for positive damages amounts. \vspace{0.7ex}}
\small
\begin{tabular}{lccccccc}
\hline 
&&&&&&&\\[-1.5ex]
 & Max & Mean & Median & Standard deviation & DI & Skewness & Kurtosis\\
&&&&&&&\\[-1.5ex]
\hline\hline
&&&&&&&\\
amounts >0  & 8,069,984 & 3,476.67 & 522.21 & 44,879.06  & 12.91  &  124.50 & 19,931.67   \\
(sample size: 60,985) &&&&&&&\\
&&&&&&&\\[-1.5ex]
\hline
&&&&&&&\\
amounts $\ge 500$ \EUR & 8,069,984 & 6,460.11 &  1,500.00  &  61,891.58  & 9.58 & 90.58 & 10,512.55 \\
(sample size: 31,911) &&&&&&&\\
&&&&&&&\\
\hline
\end{tabular}
\end{table}

\vspace{-2ex}
Besides the mean and median, which, of course, have different values between the two samples, the other descriptive statistics share the same characteristics.
We observe a strong skewness, whatever the sample considered, indicated by a very high value, but also by a strong difference between mean and median. For each sample, the variance exhibits a very high empirical value, pointing out that it can be interpreted as infinity. This is corroborated by the high values of DI and kurtosis. If the existence of the 2nd moment is questioned, a fortiori that of the higher moments, hence the very large value of the kurtosis. It already suggests the existence of heavy tails for the damages severity. 

\begin{figure}[h]
  \centering
  \includegraphics[scale=0.6]{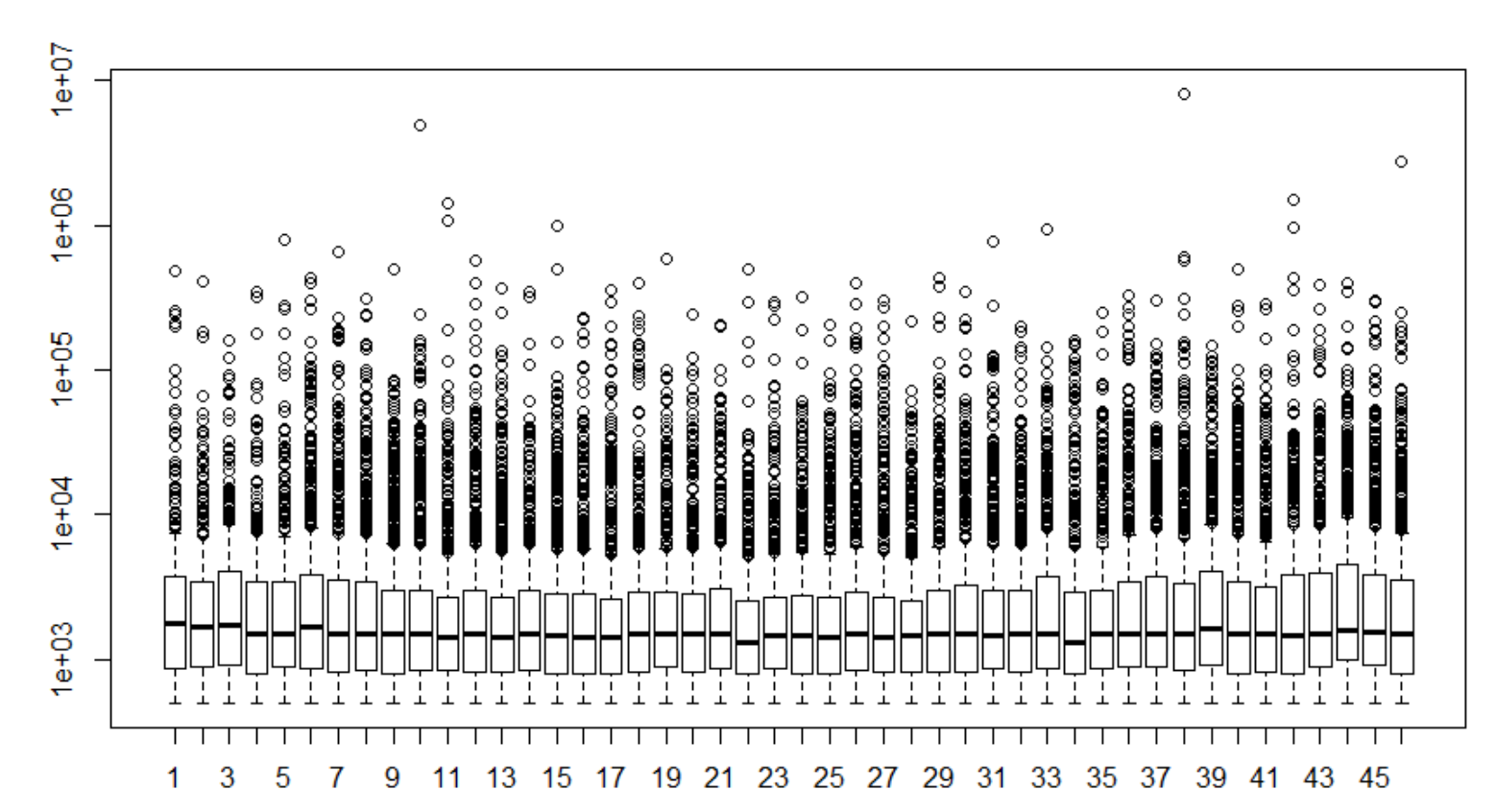}
\vspace{-2ex}
\parbox{360pt}{\caption{ \label{fig:boxplot-montant} \sf\small Box Plot of the monthly amounts $x\ge 500$ of the damages from July 2015 to April 2019. The $x$-axis corresponds to the 46 months. The $y$-axis is a logarithmic scale so that all values, extreme or not, can be seen on the plot. 
}}
\end{figure}

To conclude this section, we present the distribution of the positive amounts $\ge 500$\EUR for each month of the entire period, providing a box-plot for each month; see Figure~\ref{fig:boxplot-montant}.

We observe that the median is more or less constant for each month (with an average value of 1,493.6\EUR),  as well as 
the interquartile interval $[Q_1,Q_3]$, where $Q_1, Q_3$ denote the 1st and 3rd quartile, respectively (on average, $Q_1$ is 840.1\EUR and $Q_3$, 3,198.4\EUR). 
The position of the limits $Q_1-1.5\times (Q_3-Q_1)$ (4,378\EUR) and $Q_3+1.5\times (Q_3-Q_1)$ (-2,697\EUR; not visible here as we consider values larger than 500\EUR), respectively, indicate an important negative dissymetry. The mass of values beyond $Q_3+1.5\times (Q_3-Q_1)$ confirms the observation of a heavy tail distribution for this variable.

This temporal analysis validates the choice of a static approach chosen for our modelling, developed below.

\vspace{-2ex}
\section{Probabilistic modelling of heavy-tailed data}
\label{sec:model}
\vspace{-2ex}
As observed in the previous section, we are  clearly confronted to heavy-tailed and asymmetric (due to the strong skewness) data.
This characteristic is common to many fields (for instance on health data), among which OR where heavy-tailed data are often to be taken care of (see e.g. \cite{DAS2021}).

The presence of extreme risks induces specific risk management procedures and need for capital. Thus, it is essential to be able to quantify accurately the probability of extremes occurence for designing the appropriate hedging strategy.

In this context, building on the main ideas underlying the method developed in \cite{Debbabi2017}, we adapt its general hybrid model and draw a new algorithm allowing for a relevant fit of any heavy-tailed asymmetric non-negative data, thanks to the automatic detection of the tail-threshold. This new version completes the overall method, providing now 2 versions of the algorithm, one for symmetric data, the other here for asymmetric ones.

\vspace{-2ex}
\subsection{ Dealing with heavy-tailed data thanks to EVT}
\label{ss:EVT}
\vspace{-2ex}
In view of developing our model, we briefly recall some results from univariate EVT, namely the main asymptotic theorems for extremes,
as well as the founding ideas of the approach developed in \cite{Debbabi2014,Debbabi2017}.
This will be useful when considering the extreme loss severity associated with the registered complaints in Section~\ref{sec:appliCyber}. 
See also \citep{Ventre2020} and \cite{Kratz2019} for a recent overview of some standard (supervised) and new (unsupervised) methods in univariate EVT\footnote{Note that the wording used to describe well-known EVT notions in these references and in this paragraph (or, even, in any EVT (text)book), may be similar.}. For more details, we refer the reader to standard books on the EVT literature, e.g. \cite{Leadbetter2011}(1st ed. 1983), \cite{Resnick2008} (1st ed. 1987), \cite{Embrechts2011} (1st ed. 1997), \cite{Reiss2007} (1st ed. 1997), \cite{beirlant2004}, \cite{deHaan2006}, \cite{Resnick2007}.


While the (Generalized) Central Limit theorem provides a limiting distribution for the distribution bulk, thus describing the mean behavior of the phenomenon studied through an observed (iid) sample from this (unknown) distribution, the Extreme Value theorems consider the extreme behavior of such phenomenon.
There are two EVT pillar theorems, one (named the three-types theorem, unified into a single three-parameter family) proving that the renormalized maximum of a sample is asymptotically distributed as a Generalized Extreme Value (GEV) distribution, defined by
\begin{equation}
G_{\mu,\beta,\xi}\left(x\right)= \exp\left[ -\left(1+\xi \,\frac{x-\mu}{\beta}\right)^{-1/\xi}\right],~\text{for}~x~\text{such that}~~ 1+\xi~\frac{x-\mu}{\beta}>0~,
\end{equation}
 with location parameter $\mu$, scale parameter $\beta$, and tail index $\xi$, this latter parameter being our focus in EVT, as it determines the nature of the tail distribution.

Extracting more information in the tail of the distribution for a better estimation of $\xi$, we turn to thresholds methods and the second EVT pillar theorem, the Pickands-Balkema-de Haan theorem, which proves that, for a sufficiently high threshold $u$, a very good approximation to the excess distribution function ({\it i.e.} the distribution of the exceedances above $u$) is the Generalized Pareto distribution (GPD) $G_{\xi,\beta(u)}$ defined by
\begin{equation}\label{eq:defGPD}
G_{\xi,\beta(u)}(y)=\left\{
\begin{array}{ll}
1-\left(1+\xi\frac{y}{\beta(u)}\right)^{-1/\xi} &\mathrm{if}~ \xi \neq 0 \\ 
1-\exp\left(-\frac{y}{\beta(u)}\right) &\mathrm{if}~ \xi = 0
\end{array}\right.
\end{equation}
where $y\geq 0$ if $\xi\geq 0$, and $\displaystyle 0\leq y \leq -\frac{\beta(u)}{\xi}$ if $\xi<0$.

The distribution tail can be of three types, according to the sign of $\xi$, with a heavy (or fat) tail if $\xi>0$ (Fr\'echet domain of attraction), a light tail if $\xi=0$ (Gumbel domain of attraction), and a finite upper endpoint if $\xi<0$ (Weibull domain of attraction).

The main problem when applying the Pickands-Balkema-de Haan theorem, comes back to the identification of the threshold above which observations are considered as extremes, so that they can be fitted with a GPD.  Various methods have been developed for this purpose and for estimating the tail index, among which supervised methods (as the standard ones of the EVT literature) and unsupervised ones (see e.g. references in \cite{Debbabi2017} and \cite{Tencaliec2020}). 

A final reminder, useful for our analysis, concerns the relation between the value of the tail index and the existence of moments of the GEV and GPD distributions. 
The $\displaystyle k^{\text{th}}$ moment exits if $\displaystyle \xi < 1/k$ or, equivalently, the so-called shape parameter $\alpha=1/\xi$ satisfies $\alpha > k$. It means that the smaller the shape parameter (or, equivalently, the larger the tail index), the heavier is the tail.
For instance, if $1< \alpha \le 2$, the second moment of the distribution does not exist (i.e. infinite variance) but the first moment (the expectation) is finite.
In insurance companies, the severity of risk is often classified according to the range of the shape parameter $\alpha$. For instance, pandemics and natural catastrophes like windstorms or floods have $1.2< \alpha <2$, while earthquakes have fatter tails with $0.9<\alpha\le 1.1$ (if $\alpha\le 1$, the reinsurer will only give limited covers in order to force the loss distribution to have a finite expectation, hence an $\alpha$ back above 1); financial risks exhibit $2 < \alpha \le 4$. This is why, in the following, we may privilege discussing the range of $\alpha$ rather than that of $\xi$.
\vspace{-2ex}
\subsection{Towards an unsupervised modelling method}
\vspace{-2ex}

Let us describe succintely the main idea of the unsupervised method by Debbabi et al. (see \cite{Debbabi2014}, \cite{Debbabi2017}, and, for an overview, \cite{Kratz2019}, \cite{Ventre2020}). It has been developed for fitting multi-component data that exhibit heavy tails, determining in an automatic way the threshold $u_2$ above which the GPD fits the tail distribution, as well as the tail index $\xi$. Note that to be able to determine 'automatically' the  threshold $u_2$, we need all the information contained in the data, contrarily to standard EVT approaches, where only the information contained in the tail of the distribution is used, explaining in this latter case why defining $u_2$ is less straightforward. Using all information also means providing a model for the whole data at once, which may be seen as an advantage. 
Unlike existing statistical methods for density parameters estimation such as maximum (log) likelihood or moments ones, to name a few, this iterative algorithm is built on the solving of a set of  non-linear least squares problems by the Levenberg-Marquardt (L-M) technique  \citep{Levenberg1944,Marquardt1963}, which combines  Gauss–Newton and gradient descent methods to reach the desired minimum.

Our method has been developed in successive stages since 2014, testing it in terms of goodness-of-fit on simulated data, but also in many applications that help improve it and extend it as done here on cyber data, to better catch the data complexity. 
It is based on an algorithm calibrating on data a general hybrid model composed of  two main components, built on asymptotic theorems, splitting mean and extreme behaviors to use for each behavior a general limiting parametric distribution.  One of the key ideas to improve an earlier version of the method has been to introduce a bridge  between the two main components, chosen as an exponential distribution, to have a continuous hybrid distribution and to allow for a better determination of the extreme behavior (described by a GPD).
Indeed, we could always try to link directly the main and extreme distributions (without a bridge), but then, the GPD might have to go towards intermediate behavior (intermediate order statistics) rather than describing the largest order statistics. With the bridge's introduction, we do not face this issue, providing at the same time (thanks to the algorithm) a way to determine automatically the threshold above which the observations are considered as extremes and are fitted with a GPD (via the Pickands-Balkema-de Haan theorem).  This general model can then be calibrated on any type of heavy-tailed non-negative data (rather than finding a specific model).

The algorithmic method comes into two versions, depending on the nature of the data: symmetric versus asymmetric (skewed). While the first version has been built for symmetric data, approximating the mean behavior with a Gaussian distribution (using the CLT) (see \cite{Debbabi2017}),  the second version is  developed here for asymmetric non-negative data replacing the Gaussian behavior with a lognormal one, as it is well known that the CLT suffers of a slow speed of convergence for skewed data, as could be experimented on the cyber dataset of the GN; see Figure~\ref{fig:cdf-positiveDamages}. We observe on the blue curve (G-E-GPD  model) of this figure that the slow convergence has a negative impact on the whole fit. 
\vspace{-2ex}
\subsection{An algorithm for asymmetric data}
\vspace{-2ex}

If the main idea of the approach still holds when changing the Gaussian component into a lognormal one, we need of course to define the new model and relations between parameters, and adapt the self-calibrating algorithm accordingly. 


The model we consider here, denoted by  LN-E-GPD (Lognormal-Exponential-Generalized Pareto Distribution), is characterized by its probability density function (pdf)  $h$ expressed as: 
\begin{equation}\label{eq:LNEGPD}
h(x;\boldsymbol{\theta})= \gamma_1\, f(x;\mu,\sigma)\,\1_{(x\leq u_1)} 
                 			+ \gamma_2 \, e(x;\lambda)\, \1_{(u_1\leq x \leq u_2)}  
			          + \gamma_3\, g(x-u_2;\xi,\beta) \,\1_{( x \geq u_2)},  
\end{equation}
where $f(\cdot;\mu,\sigma)$ denotes the lognormal (LN) pdf with mean $\mu\in\R$ and standard deviation $\sigma >0$, defined for all $x>0$ by $\displaystyle f(x;\mu,\sigma)=\frac{1}{x\sigma\sqrt{2\pi}} e^{-\frac{(\log x-\mu)^2}{2\sigma^2}}$, $e$: the exponential pdf with intensity $\lambda$ (defined on  $\R^+$ by $\displaystyle e(x;\lambda)=\lambda\,\exp^{-\lambda\, x}$), 
$g$: the pdf of the GPD defined in \eqref{eq:defGPD} with tail index $\xi>0$ (heavy-tail condition) and scale parameter $\beta>0$, 
while $\gamma_i$, $i\in\{1,2,3\}$, are the non-negative weights (for $h$ to be a pdf) with $\gamma_1+\gamma_2+\gamma_3 \ge 1$, and $u_1$ and $u_2$ are the two junction points between the components, with $u_1\le u_2$.

 Let us define the relations between the parameters of the model, using the heavy-tailed framework and the $C^1$ assumption on the pdf that imposes smooth transitions from one component to another.
For heavy-tailed data, we have $\xi>0$ and the asymptotic (for high threshold) relation $\beta=\xi \, u_2$, which we are going to use in the algorithm. The $C^1$ assumption is translated by $\gamma_1\, f(u_1;\mu,\sigma)=\gamma_2 \, e(u_1;\lambda)$,  $\gamma_2 \, e(u_2;\lambda)=\gamma_3\, g(0;\xi,\beta)$, and similar equalities when considering the derivative of $f,e$ and $g$, respectively.
Therefore, after some computation, we obtain:
\begin{equation}\label{eq:relations-Param}
\left\{
  \begin{array}{ll}
   \beta=\xi \, u_2; \quad \lambda=\frac{1+\xi}{\beta}; & \gamma_2=\Big[ \xi \, e^{-\lambda\,u_2} +  \left(1+\lambda\,\displaystyle \frac{ F(u_1;\mu,\sigma)}{f(u_1;\mu,\sigma)} \right) e^{-\lambda\,u_1}\Big]^{-1}; \\
\lambda \sigma^2 u_1-\log u_1=\sigma^2-\mu; &\quad \gamma_1=\gamma_2 \,\frac{e(u_1;\lambda)}{f(u_1;\mu,\sigma)}; 
 \quad
\gamma_3=\beta\,\gamma_2 \, e(u_2;\lambda).
  \end{array}
\right.
\end{equation}
Those relations help reduce the size of the vector of parameters to be  estimated from 10 to 4,  namely $\big[\mu,\sigma,u_2,\xi]$. Then, we run the iterative algorithm for the LN-E-GPD model to estimate those 4 parameters (the other 6 being deduced via \eqref{eq:relations-Param}).

The model parameters are estimated via an iterative algorithm adapted from that developed for the G-E-GPD model, described in details in Debbabi et al. and which convergence has been studied analytically and numerically. We recall here the main principle (see the pseudo-code in \cite{Kratz2019}, Section 2.4.1): 
We initialize the parameters of the distribution body and the threshold $u_2$ to estimate in a first iteration the tail index $\xi$. Then, we fix the latter with this first value and estimate the other parameters. This back-and-forth process (between body and tail) is iterated by minimizing together, with the L-M technique, two distances between empirical and model distributions, one for the whole distribution and the other for the tail, until convergence.

If the main principle holds for the new model, it is also worth noticing that a critical point, which highlights the difference between modeling the bulk data by a Gaussian or a lognormal distribution, lies on the choice of initial parameters conducting the algorithm to convergence. Indeed,  we recall that for  G-E-GPD, the Gaussian mean corresponds to the distribution mode, which gives a nice strategy to initialize the Gaussian parameters. Unfortunately, it is no longer the case for the lognormal distribution for which we have tested several techniques to obtain initialization that holds up.

 Moreover, this algorithm provides an additional flexibility compared to a two components model:
If an observed phenomenon under study would be well explained by two components only, then the two thresholds $u_1$ (junction point between the body and the exponential bridge) and $u_2$ (junction point between the exponential bridge and the GPD) will collapse into one during the calibration. This has been shown for the G-E-GPD model in \cite{Debbabi2017}), providing the earlier G-GPD model introduced in \cite{Debbabi2014}. Here, we also show the same property for the LN-E-GPD model,  conducting a series of experiments based on Monte Carlo simulations (see Appendix~\ref{App-test2and3components}), leading to a LN-GPD model (with non uniform weights for each component), which is a generalized version of the Cszeledin distribution (a LN-Pareto distribution with specific weights) introduced in \cite{Knecht2003} and used in the cyber case by \cite{Eling2019}.
It demonstrates the outperformance of the three components model\footnote{Note that this general method and its two algorithmic versions will be part of a statistical software package. Meantime, the R code is available upon request.}.

Hence, the two components model should not be a purposedly chosen one, but come as a specific subcase of a general model that has been calibrated, moreover in an automatic way, without resorting to either costly computational techniques or via standard EVT techniques, recognized as oversensitive to the threshold above which observations are considered as extremes. This points out the outperformance of our model, which lies in its generality, simplicity, and self-calibrating property.

\vspace{-2ex}
\subsection{Assessing the parameters estimation via a re-sampling technique}
\label{ss:jackknife}
\vspace{-2ex}
Another important input in the extension of this method is the construction of confidence intervals for the estimated parameters via a re-sampling technique, as well as  the introduction of a better visualizing tool for the tail fit (see Section~\ref{ss:appliSeverity} and  the right plot of Figure~\ref{fig:cdf-positiveDamages}) (those elements have been introduced in the software package under construction).

To provide confidence intervals for the estimation of the model parameters, we revisit the Jackknife method (see~\cite{Kunsch1989}) that measures the variability of the estimation across sub-samples. This is one of the earliest re-sampling techniques, which is more suited for a large number of observations than the standard bootstrap (which we also ran to check that we would obtain the same results; it was the case, but it took a few days of computation to obtain the results, confirming the advantage of the Jackknife use in this context). Our main focus is on the tail of the distribution since it is more difficult to estimate than the features in the body such as mean and variance. Thus, we consider the three parameters of the GPD component, namely the tail index $\xi$, the scale parameter $\beta$ and  the exceedance threshold $u_2$.
To define a numerical confidence range, we build randomly  $m=10$ subsamples and run on each one the algorithm for calibrating the hybrid model. 
Each subsample is constructed in the following way: we omit some randomly selected data points that amount to $10$\% of the original dataset of size $n=60985$, making sure that each of those selected observations is omitted only once, while used in the 9 other computations (note that it means that each selected observation in the whole sample will be removed in 1 of the  10 subsamples). The estimation results obtained with the Jackknife method are based on the average of the 10 estimates obtained for each subsample. Taking the example of the tail index, its estimator (through this method) denoted by $\bar{\xi}^J$ is defined by $\displaystyle \bar{\xi}^J = \frac{1}{m}\sum_{i=1}^m \hat{\xi}_i^J $, where $\hat{\xi}_i^J$ is the estimator of $\xi$ obtained on the $i$-th subsample  and $m$ is the number of subsamples (here, $m=10$).
Similarly, we can compute the standard deviation of these estimated (via the Jackknife method) parameters
(see~\cite{Yang1986}) to obtain an estimation of the standard deviation of the estimator over the whole sample of size $n$. Taking back the example of the parameter $\xi$, the estimated standard deviation of the estimator $\hat{\xi}$ (over the whole sample), is defined by

\begin{equation}
 \widehat{\sigma(\hat\xi)} = \sqrt{(1-1/m)\sum_{i=1}^m (\hat{\xi}_i^J- \bar{\xi}^J)^2 }.
\end{equation}
As $\widehat{\sigma(\hat\xi)}/ \sigma(\hat\xi)\, \rightarrow 1$, 
we define the 95\% variability $a_{95\%}^J=\Phi^{-1}(0.975)\, \widehat{\sigma(\hat\xi)}$ (assuming asymptotic normality) and the confidence range displayed in Table~\ref{tab:Jack} is expressed as
\begin{equation}
 \hat{\xi} - a_{95\%}^J ~~\le~~\xi ~~\le~~ \hat{\xi} + a_{95\%}^J.
\end{equation}
A similar procedure can be applied  for all parameters.

This last step completes the algorithm that provides parameters calibrated on a general model with their CI, in a fast and reliable way. 

\vspace{-2ex}
\section{Application to Cyber Data}
\label{sec:appliCyber}
\vspace{-2ex}
We start, in Section~\ref{ss:appliSeverity}, considering the data field `damages', 
which provides the severity of the cyber attack.
This focus on the financial consequences of cyber attacks, i.e. the amounts, and not on causes of the attacks, is due to three reasons: First, the financial risk must be well understood for providing good insurance covers. Second, as the amounts have stable statistical properties over time (see Figure~\ref{fig:boxplot-montant}) a static univariate distribution will give a good picture of the severity variable. Third, the amounts constitute a large enough set of observations so that the extremes can be well modelled. For this latter reason, we mix as well all types of cyber attacks, and not only a specific one. The investigation of the causes will be the object of future studies.

Recall also that we were able to double check manually the information given on the 'amounts' variable in the GN database for the tail (for the amounts above 40,000\EUR, corresponding to the quantile of order 98.2\%). This is why our study concerns the modelling of the extreme amounts, including the frequency of the occurence of extreme damages (above a high threshold); see Section~\ref{ss:Poisson}.
We will investigate further the multivariate modelling, once the information given on all variables will have been carefully checked. 

Once having a calibrated model for the damages, we can then use it for risk management purposes, as for instance evaluating how much capital is required to cover cyber risk. An illustration is given in Section~\ref{ss:RM}.

We are also interested in the tail modelling of the various types of attacks listed in the GN database. 
The idea is to check if the tail index could be used as a discriminating criterion between various forms of cyber complaints. 
A first exploratory attempt is developed in Section~\ref{ss:classes}, 
considering the three most frequent types. The first class is preponderant compared to the other two. Nevertheless, the robustness of the parameters estimation we observed with our method makes it possible to apply the model for those latter classes of small size (but still of larger size than of most samples considered so far in the cyber literature).

\vspace{-2ex}
\subsection{Application to the damage severity}
\label{ss:appliSeverity}
\vspace{-2ex}
Based on the empirical results obtained for the damage severity in Table~\ref{tab:statDesc} with characteristics specific to heavy-tailed phenomena, we naturally look for a model able to capture such a feature and consider our flexible hybrid model, defined in \eqref{eq:LNEGPD} and \eqref{eq:relations-Param}, which we calibrate on the positive damages using the iterative algorithm discussed in Section~\ref{sec:model}.

The obtained estimates are given in Table~\ref{tab:param-PositiveDamages}, where we observe that the LN-E-GPD model reduces into two components, as the 2 thresholds on either side of the exponential bridge collapse to $u_1=u_2$.  It is interesting as this 2 components model has already been suggested for cyber data (see e.g. \cite{Zeller2020}). 
The threshold $u_2$,  automatically evaluated via the hybrid model, corresponds to a quantile of order $96.6$\%. 
The GPD fitted above $u_2$ exhibits a shape parameter $\alpha=1/\xi=1.23$, indicating a heavy tail with a finite first moment but no finite variance.
%
\begin{table}[htbp]
  \centering
  \caption{\label{tab:param-PositiveDamages} \sf Evaluated parameters of the hybrid LN-E-GPD model for positive damages. \vspace{0.7ex}}
{\small
    \begin{tabular}{lcccccccccc}
    \hline
    Model & $\mu$ & $\sigma$ & $\gamma_1$ & $u_1$ & $\lambda$ & $\gamma_2$ & $\xi$ & $u_2$ & $\beta$ & $\gamma_3$ \\
\hline
\hline
&&&&&&&&&&\\
    LN-E-GPD & 6.27  & 1.54  & 99.4\% & 9,999.34 & 0.0002 & 17.5\% & 0.81  & 9,999.34 & 8,087.11 & 3.4\% \\
&&&&&&&& q(96.6\%) &&\\
\hline
    \end{tabular}
		}
\end{table}

\begin{figure}[h]
{\resizebox*{8cm}{!}{\includegraphics[width=7cm,height=7cm]{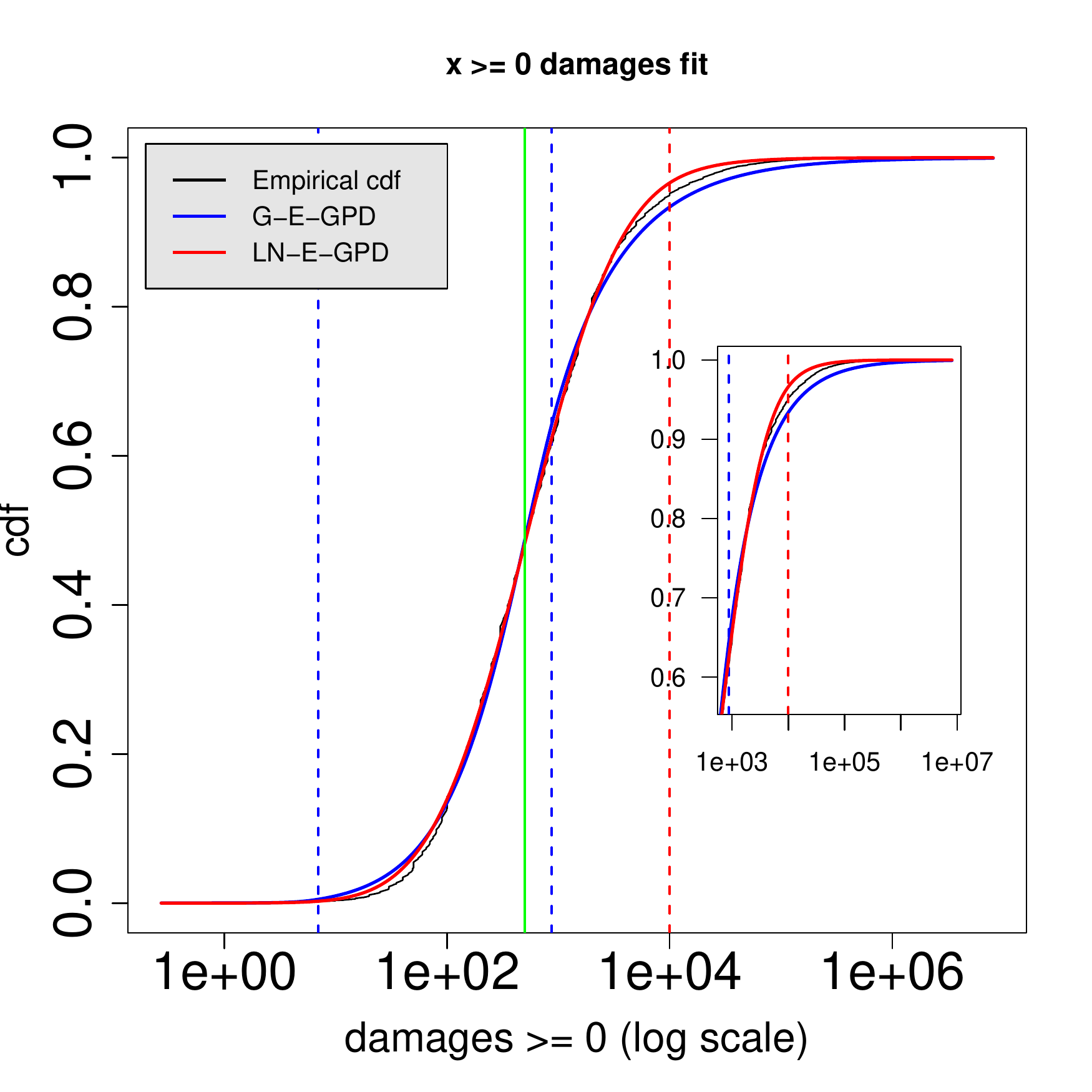}}}
\hfill
{\resizebox*{8cm}{!}{\includegraphics[width=7cm,height=7cm]{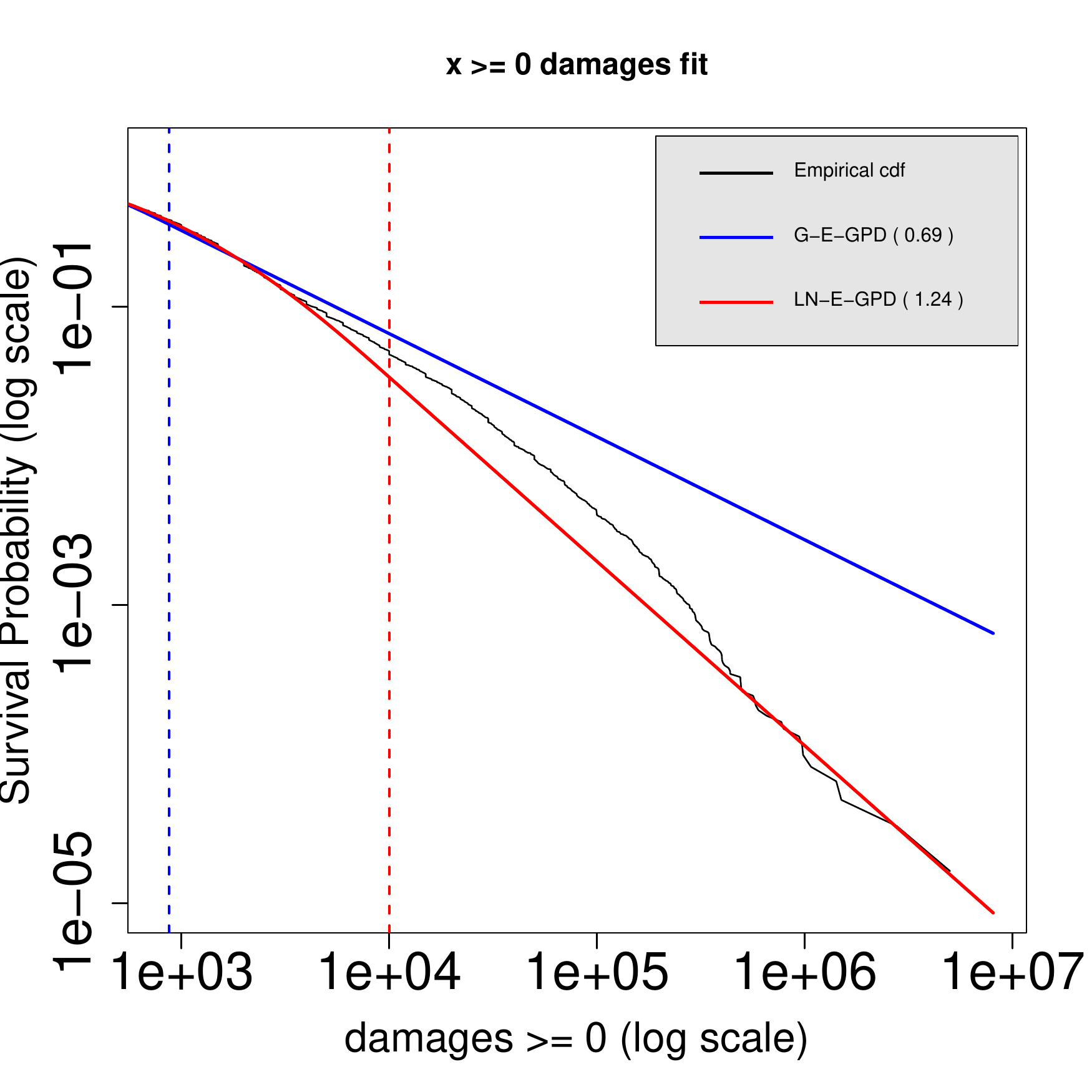}}}
\parbox{500pt}{\caption{\label{fig:cdf-positiveDamages} \sf\small Cdf (left plot, with a log scale for the $x$-axis) and survival cdf (right plot, with a log scale for both the $x$ and $y$ axes) of the positive damages. The empirical cdf is represented in black, the LN-E-GPD in light (red) and the G-E-GPD in dark (blue). The dashed vertical lines (with the same color code) correspond to the thresholds between the components of the hybrid model considered, while the continuous vertical (green) line points out the 500\euro -threshold.}}
\end{figure}

Fitting the  LN-E-GPD model on the damage severity data, we obtain the following output of the algorithm, given in Figure~\ref{fig:cdf-positiveDamages} and Table~\ref{tab:fit-PositiveDamages}. We also exhibit the fit of the G-E-GPD model, replacing the Lognormal component with a Gaussian (G) one, to illustrate its inadequacy to account for the asymmetry of the data due to the slow convergence of the CLT. It impacts, as expected, the whole fit, including the tail one, as can be observed, even if this impact is mitigated by the bridge component. 

In Figure~\ref{fig:cdf-positiveDamages}, we provide two types of graphs, one (left) giving the empirical cdf and the two fitted distributions (with a logarithmic scale on the $x$-axis), the other (right) displaying the corresponding survival distributions ($1-F$) on a double logarithmic scale (i.e. for the $x$- and $y$-axis).
\vspace{-2ex}

The interest of taking a double logarithmic scale is that, for such a representation, a GPD becomes a linear decreasing function, which slope is the negative value of the shape parameter, i.e. $-\alpha$. 
Indeed, considering the survival GPD for $x \,(\ge u_2)$, namely $1-G_{\xi,\beta}(x)= \left(1+\frac{\xi}{\beta}\, x\right)^{-1/\xi}$, and taking its logarithm gives 
$$
\log\left(1-G(x;\xi,\beta)\right)\,=\, -\frac1{\xi} \log\left(1+\frac{\xi}{\beta}\,x\right) 
\underset{x\,\text{large enough}}{\sim}  -\frac1{\xi}\log\left(\frac{\xi}{\beta}\right)-\frac1{\xi}\log x \,=:\,  y(\log(x)).
$$
The function $y(\cdot)$ is then a linear function in $\log x$, with slope $-1/\xi= -\alpha$ and intercept $-\frac1{\xi}\log\left(\frac{\xi}{\beta}\right)$.
This facilitates the comparison between empirical and fitted tail distributions, as the mismatch will appear clearly. It is an interesting alternative representation to QQ-plots.

Looking at Figure~\ref{fig:cdf-positiveDamages}, we observe that the LN-EXP-GPD model fits better the empirical data than the G-EXP-GPD, as expected. This is confirmed by the errors reported in Table~\ref{tab:fit-PositiveDamages}, where the total error of the latter is almost twice as big as the LN-EXP-GPD for both root mean square errors (RMSE) and mean absolute errors (MAE) (1.51\% versus 0.80\% for the RMSE, and 1.33\% versus 0.66\% for the MAE). 
\begin{table}[htbp]
\centering
\parbox{360pt}{\caption{\label{tab:fit-PositiveDamages} \sf\small Measuring the goodness of fit for the 2 considered hybrid models on the    positive damages. The total and tail errors are computed using respectively the root mean squared error (RMSE) and the mean absolute error (MAE). \vspace{0.7ex}}}\\
\begin{tabular}{lccccc}
		\hline
		&&&&&\\[-1.5ex]
    Model & \multicolumn{2}{c}{Total error in \%} & \multicolumn{2}{c}{Tail error in \%} & BIC criterion \\
& RMSE & MAE & RMSE & MAE &  \\[1ex]
\hline\hline
&&&&&\\
    LN-E-GPD & 0.80 & 0.66 & 0.94 & 0.79 &  -255,927\\
&&&&&\\
\hline
&&&&&\\
    G-E-GPD & 1.51 & 1.33 & 1.60 & 1.56 &   -222,225\\[2ex]
		\hline
    \end{tabular}
\end{table}
Focusing on the distribution tail, the right plot of~Figure~\ref{fig:cdf-positiveDamages} clearly depicts a better fit for the LN model than for the Gaussian model. We see that the Gaussian model overestimates the heaviness of the tail, while the LN model fits well the linear representation of the empirical tail (lower right part). 
The superiority of the LN hybrid model over the Gaussian one, is also confirmed by the Bayesian Information Criterion (BIC) which is 15\% lower for the first model than for the second one. The shape parameter of the tail distribution ($1/\xi$), estimated as 1.24, indicates a rather heavy tail, as already commented. 
\begin{table}[h]
\centering
\caption{\label{tab:Jack} \sf\small Variability of the GPD parameters estimation using the Jackknife method. \vspace{0.7ex}}
\begin{tabular}{lccc}
		\hline
		&&&\\[-1.5ex]
     & $\alpha$ & $\beta$ & $u_2$ \\[0.5ex]
\hline\hline
&&&\\[-0.5ex]
Estimation & 1.236 & 8,087 & 9,999\\
95\% Confidence Range (CR) & [1.213 ; 1.260] & [7,929 ; 8,245] & [9,980 ; 10,018] \\[1.5ex]
\hline
\end{tabular}
\end{table}
We observe in Table~\ref{tab:Jack} that the results of the fit are robust towards the samples. The choice of $u_2$ is very stable (0.2\% variation), while the scale index $\beta$ is in range of $\pm 2$\% and the shape parameter $\alpha$ in range $\pm 1.8$\%.

Further, to stress the stability of the result and to be closer to a realistic frame (given that positive values below 20\euro\, do not really make sense, given the context), we also ran the iterative algorithm to fit the LN-E-GPD model on a second sample obtained when removing the 583 damages below 20\euro\, (from the sample of size 60,985). Note that those removed data correspond, most probably, to data badly reported in the database due to a wrongly placed decimal point. Once corrected, they would be above 20 (implying that the second sample cannot be considered as a censored sample). Given the method, such a sample should not change the tail of the damages distribution. Indeed, we found an estimate of 1.26 for the shape parameter (compared with 1.24 when considering the positive damages), hence well within the uncertainty range.

%
For comparing the evaluation of the tail heaviness, we also introduce the classical Hill estimator (see \cite{Hill1975}) for the tail index $\xi$, defined as
\begin{equation}
\hat\xi=H_{k,n} =\frac 1k \sum_{i=0}^{k-1} \log \left(\frac{X_{n-i,n}}{X_{n-k,n}}\right)
\end{equation} 
where $\displaystyle X_{n,n}=\max_{1\le i\le n} X_i \ge X_{n-1,n}\ge\cdots \ge  X_{n-k+1,n}\ge  X_{n-k,n}$ are the largest order statistics of the heavy-tailed observations (here, the damage severity) $(X_1,\cdots,X_n)$, with $k$ such that $X_{n-k,n}=u_2$. The main problem faced when using this type of tail index estimators, is to evaluate $u_2$, {\it i.e.} to select the number $k$ of largest order statistics; here we use the threshold $u_2$ determined automatically by our algorithm, instead of going through the standard EVT graphical (supervised) methods. The Hill estimator is weakly consistent for heavy-tailed data and satisfies, under some second order property, $\displaystyle\sqrt{k}\, \left( H_{k,n} - \xi\right)\ \underset{n\to\infty}{\stackrel{d}{\longrightarrow}} N\left(0,\, \xi^{2}\right)$,  
 from which we build an asymptotic confidence interval (CI) for $\hat\xi= H_{k,n}$.

Considering $u_2$ as the 96.6\% quantile (see Table~\ref{tab:param-PositiveDamages}), we obtain as Hill estimate: $\widehat H_{k,n} =0.962$ with 95\% asymptotic confidence interval $[0.55; 1.37]$. Note that the estimate obtained via the algorithmic method (0.81) lies within this confidence interval.

\vspace{-2ex}
\subsection{A Poisson-GPD model for the Severity and Frequency of Extreme Damages}
\label{ss:Poisson}
\begin{figure}[h]
\centering
{\resizebox*{7cm}{!}{\includegraphics{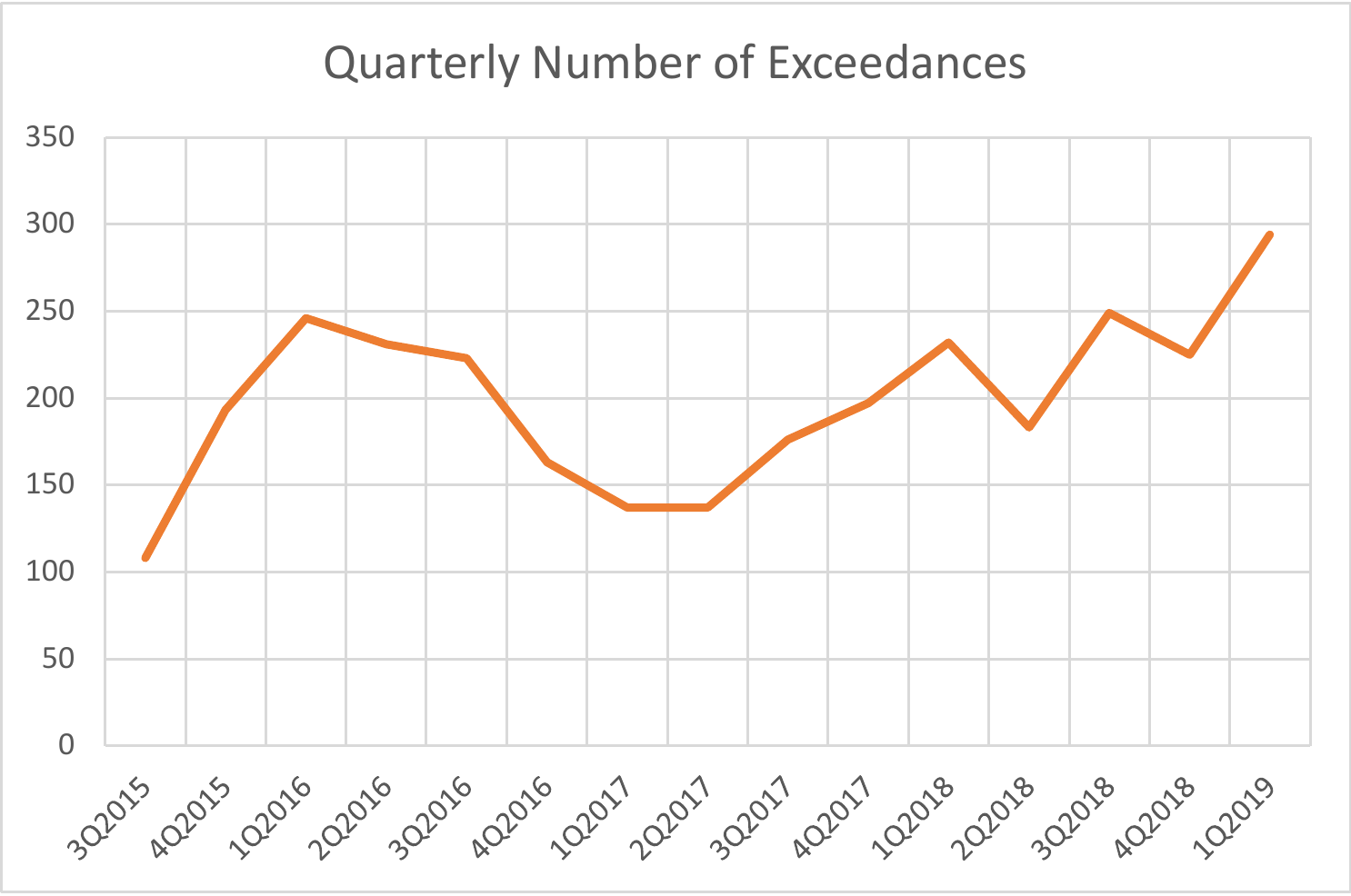}}}
\hspace{12pt}
{\resizebox*{7cm}{!}{\includegraphics{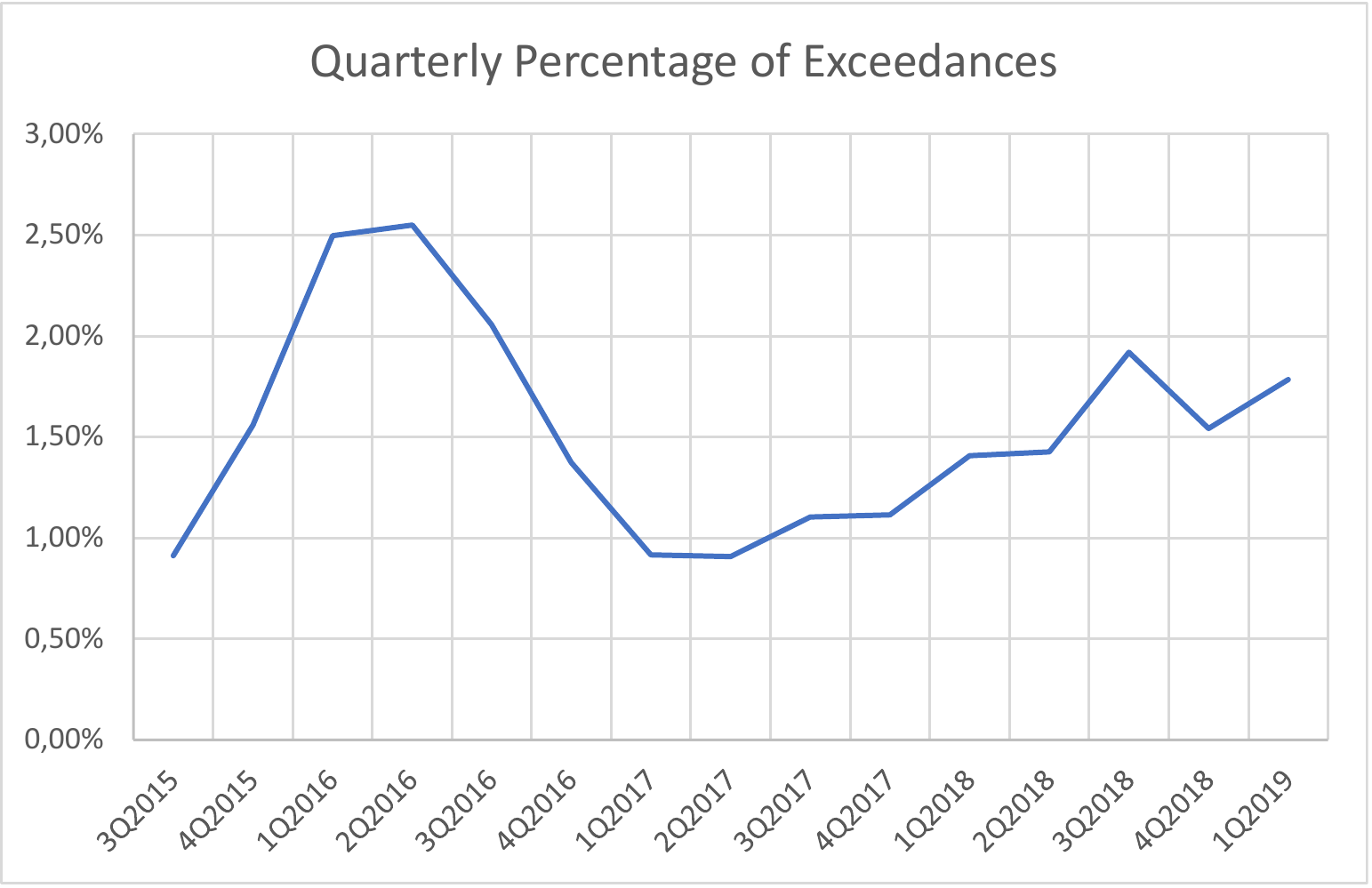}}}
\parbox{450pt}{\caption{\label{fig:freq-extremes} \sf\small Quarterly frequency of damages (left $y$-axis) with magnitude larger than $u_2$ and the corresponding quarterly percentage of extreme damages (right $y$-axis). The $x$-axis presents the various quarters.}}
\end{figure}

\vspace{-2ex}
In the previous subsection, we focused on the modelling of the damage severity, with a particular interest for the extreme damages. Now, we are looking for fully modelling those extremes, taking into account not only their magnitude but also their frequency. To do so, if the extreme observations constitute a stationary time series, we can introduce a Poisson-GPD model combining a one-dimensional Poisson process with parameter $\lambda(>0)$ for modelling the frequency at which exceedances over the threshold $u$ occur, with a GPD for representing their magnitude (see \cite{Smith2003} for details).  
The distribution of this model is expressed as:
\begin{equation}\label{eq:G-Pmodel}
H_u(x; \xi,\beta,\lambda):=exp\left\{-\lambda\left( 1+\xi \,\frac{x-u}{\beta}  \right)_+^{-1/\xi}\right\},\quad x>u.
\end{equation}
If there is some non-stationarity in the data, as, for instance, a change over time in the frequency of exceedances, or an increase of the severity of damages due to inflation, then time variability should be introduced in the scale parameter of the GPD, say $\beta(t)$, and in the Poisson intensity parameter, say $\lambda(t)$.\\[1ex]
Turning to our dataset, let us look at the frequency of extremes exceeding the threshold $u_2$ evaluated in the previous subsection (see Table~\ref{tab:param-PositiveDamages}). We consider the frequency with various time horizons from 1 to 4 months. Whatever the chosen horizon, we do not observe any clear trend, as illustrated in Figure~\ref{fig:freq-extremes} for quaterly frequency. In this figure, we present the quarterly frequency (left plot) and the quarterly percentage (right plot) of exceedances. The percentage allows to differentiate trend in the frequency of damages from trend in the exceedances. On the plots, there is no obvious trend in both cases (frequency and percentage). 

To assess such statement, we perform stationary tests on R, namely the Augmented Dickey-Fuller and the Phillips-Perron unit root tests, for the time series of exceedances and of exceedances monthly  frequency, respectively. For the exceedances time series, both tests strongly reject their null hypothesis, from which we conclude to the stationarity. For the frequency time series, although we considered a monthly frequency to have more observations than for the quaterly one, the number of observations is still small (46) to obtain a statistically significant and conclusive result. We display the obtained results in Table~\ref{tab:stationarity}.
%
\begin{table}[htbp]
  \centering
\parbox{330pt}{\caption{ \label{tab:stationarity}\sf\small Results of various stationarity tests on the exceedances dataset. 
\vspace{0.7ex}}}
\small
    \begin{tabular}{lrcrl}
    \hline
&&&&\\[-1.5ex]
    \multicolumn{1}{c}{\textbf{Stationarity Test}} & \multicolumn{1}{c}{\textbf{Test Value}} & \textbf{Lag} & \multicolumn{1}{c}{\textbf{p-value}} & \multicolumn{1}{l}{\textbf{Interpretation}} \\[0.5ex]
    \hline
    \hline
&&&&\\[-1.5ex]
  \textit{For Exceedances } {\footnotesize (2,994 observations)}&&&&\\[0.2ex]
    Augmented Dickey-Fuller & -10.42 & 14    &  < 0.01   & non-stationarity strongly rejected\\
    Phillips-Perron & -1,290.2 & 9     &  < 0.01   & non-stationarity strongly rejected\\
&&&&\\[-1.6ex]
    \multicolumn{2}{l}{\textit{For Monthly Exceedances Frequency}  {\footnotesize(46 observations)}} &&&\\[0.2ex]
    Augmented Dickey-Fuller & -2.58 &   3    &  0.34 & non-stationarity not rejected \\
    Phillips-Perron & -22.52 &  3     &  0.02 & non-stationarity rejected \\
    \hline
    \end{tabular}%
\end{table}%

\begin{figure}[h]
\center
\includegraphics[width=8cm,height=8cm]{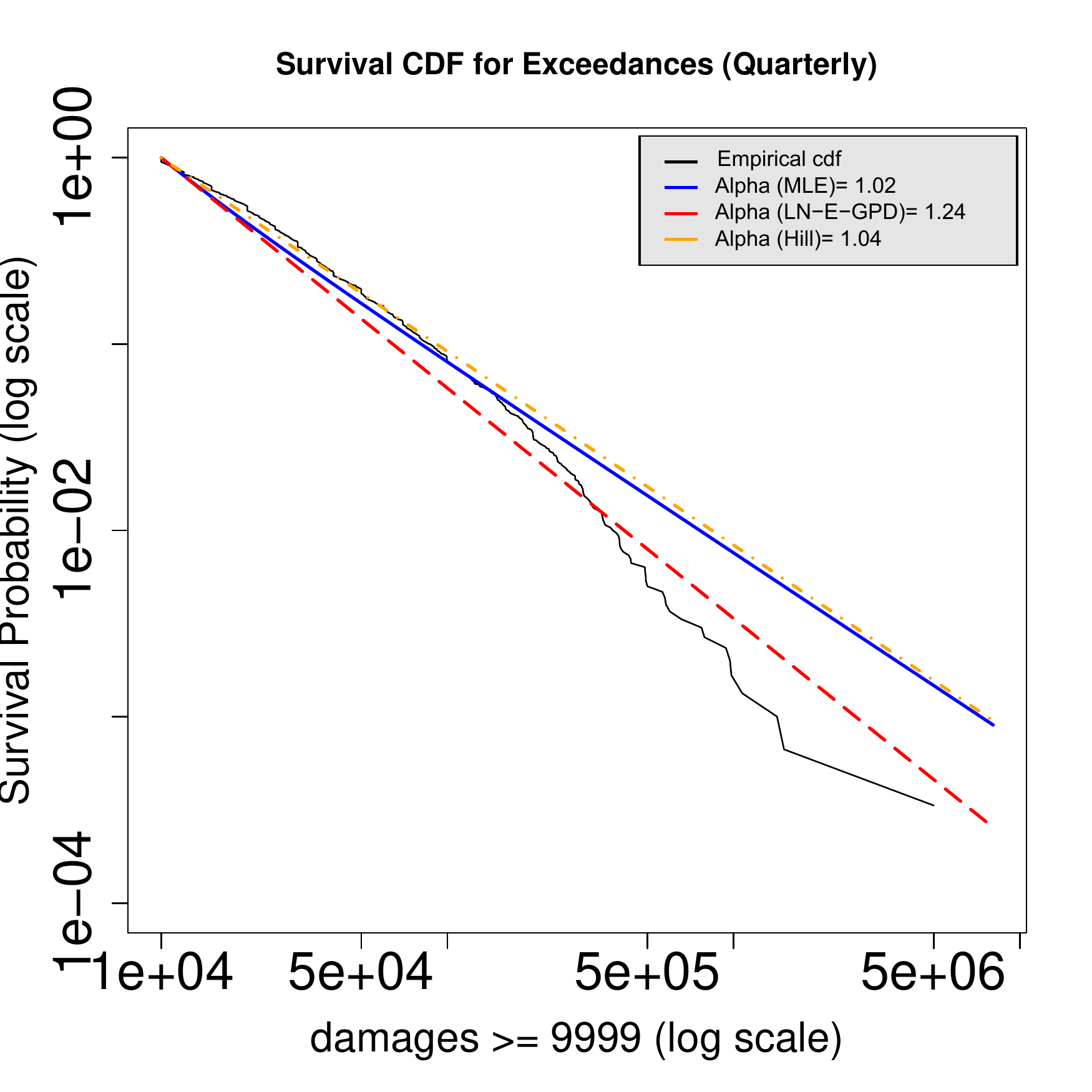}
\parbox{430pt}{\caption{\label{fig:surv-extremes} \small \sf Survival GPD for the extreme damages with parameters estimated from different methods: MLE for the Poisson-GPD model (dark/blue line), algorithmic method for the LN-E-GPD model (light/red dashed line), Hill estimator for the tail index in the LN-E-GPD model (light/yellow dotted line). In black, the empirical survival cdf.  Double log scale representation.}}
\end{figure}

Therefore, due to this stationarity, we consider the Poisson-GPD model with distribution \eqref{eq:G-Pmodel} and calibrate it on the exceedances above the threshold $u_2$. As time unit for the Poisson model, given that our dataset covers 45 months (ignoring the last month of our datset, namely April 2019, to keep full quarters), we choose the quarterly frequency of exceedances (above $u_2$) because it is the minimum interval size giving enough values.  In the considered sample, there are 2,994 exceedance observations. Then, we estimate the 4 parameters of the model on those observations, using the maximum likelihood method (run in R). The estimates are presented in Table~\ref{tab:PGP-MLEparam}; we observe that the estimate of the exceedance rate $\lambda$ is relatively close to the average quarterly frequency, namely 2,994/15=199.6 exceedances per quarter (45 months corresponding to 15 quarters), which lies in the 95\% confidence interval of $\lambda$. We also notice that the estimate of the tail index $\xi$ is higher than the one  ($0.81$) computed in the previous section (see Table~\ref{tab:param-PositiveDamages}) and that the latter lies outside of the Jackknife confidence range. 
\begin{table}[b]
\begin{center}
\parbox{450pt}{\caption{\label{tab:PGP-MLEparam}\sf\small Estimation of the parameters of the Poisson-GPD model for the $N=2,994$ exceedances above the threshold $u_2= 9,999.34$. The confidence range (CR) for $\xi$ and $\beta$ is obtained via the Jackknife method.\vspace{0.7ex}}}
\begin{tabular}{cccc}
\hline
&&&\\[-1.5ex]
Parameter & Exceedance rate & Scale parameter & Tail index \\
 & $\lambda$ & $\beta$ & $\xi$ \\
&&&\\[-1.5ex]
\hline \hline
&&&\\[-1.5ex]
ML estimates & 187.07 & 8,087.63  & 0.983\\
CR 95\% &  [171.84 ; 227.36]  & [8,087.53 ; 8,087.73] & [0.930 ; 1.036]\\ 
&&&\\[-1.5ex]
Hill estimate &  & &  0.962 \\
CI 95\% & & &  [0.55;1.37] \\
\hline
\end{tabular}
\end{center}
\end{table}
Let us plot in Figure~\ref{fig:surv-extremes}, using a double log-scale, the survival GPD of the Poisson-GPD model with parameters estimated via the ML method  (given in Table~\ref{tab:PGP-MLEparam}) (dark/blue line), and, for comparison, the survival GPD of the LN-E-GPD model calibrated via our algorithm (light/red dashed line), as well as the one calibrated when using the Hill estimate for the tail index (light/yellow dotted line).
We clearly observe that the LN-E-GPD model calibrated with our algorithm provides, among the three models, the best overall fit for the tail of the distribution, while in the two other cases, the fit is better at the start of the tail, but then overestimates the fatness of the tail.
Our method is especially designed to emphasize the tail observations, while the MLE method has to compromise between $\lambda$ for the Poisson and the GPD parameters, putting the same weights to all the points. 
This is another, {\em a posteriori}, justification of the use of the algorithmic method. Nevertheless, the Poisson-GPD model is giving a more complete view on the extremes. Now, if we use as Poisson-GPD parameters, on one hand the GPD parameters evaluated by the algorithmic method, on the other hand the Poisson intensity parameter $\lambda$ estimated by the empirical average quarterly frequency, the likelihood would decrease by only $0.1\%$ of the maximum likelihood computed on the initial Poisson-GPD model, which is already very close.

\vspace{-2ex}
\subsection{Some consequences for risk management}
\label{ss:RM}
\vspace{-2ex}
Recall our main focus on the tail of the distribution, as we want to characterize the cyber risk by how big is the probability of occurrence of extremes. This information is particularily relevant for (re)insurance, to know how much capital is required to cover such risk.
This is assessed with risk measures. Using regulatory ones and our model, we evaluate the standalone capital ({\it i.e.} without considering diversification benefits of the company risk portfolio), then compare our results to those obtained by standard EVT methods such as Hill. This is detailed in Section~\ref{ss:RM}.

The role of risk models in risk management practices is to help quantify both the liabilities of a (re)insurance company through the mathematical expectation (important for the computation of the risk premium), and the capital through risk measures. 
Our model takes into account both concerns, as it is built to provide a good modelling for both the bulk and the tail of the distribution. We estimated the mean and third quartile with our calibrated model; it  reproduces well the empirical values, with an error of 0.2\% for the mean and 0.4\% for the third quartile.
Moreover the knowledge of the tail also helps better understand descriptive features of the underlying distribution of the data. On a finite sample, any statistical quantity that we estimate is finite; it does not mean that theoretical moments, including expectation, exist. 
So, the first message we have drawn is that, most probably, the expectation of cyber risk exists since $\alpha$ is significantly larger than 1, as observed in Table~\ref{tab:Jack} (see Section~\ref{ss:EVT} for the relation between moments and tail fatness). Recall that the finiteness of the expectation is a necessary condition for the risk to be insurable. Hence cyber risk, as explored in this database, satisfies this condition.

Now, let us study the capital requirement for the insurance solvency and the risk capital required from banks. We do it using the two regulatory risk measures, value-at-risk (VaR) and expected shorfall (ES), and compute VaR(99.5\%), according to Solvency II, and ES(97.5\%), for Basel 4. Since we do not know the company portfolio, we evaluate the capital standalone (for cyber risk), which already gives an indication about the statistical nature of the considered risk.

Recall the advantage of using a model (rather than empirical values): We can compute probabilities beyond the data present in the database. It allows also to estimate risk measures (e.g. ES) in a sharper way via an analytical formula.

The risk measures being evaluated from the tail distribution, we consider the GPD estimated with our approach (hybrid model calibrated via our algorithm), and the 12 EVT-methods based on the Hill estimator for the tail index, proposed in the {\it tea} R-package that also contains the references to the methods we quote. We refer to Table~\ref{tab:tea-results} of the Appendix, where results obtained via these methods for the tail-threshold $u_2$ and the inverse of the tail index $\alpha=1/\xi$ are reported, then commented.

Recall (see e.g.~\cite{McNeil2016}) that for $G\sim$ GPD$(\xi,\sigma(u_2))$ (where $0<\xi<1$), we have, for $p\ge G(u_2)$, with $\beta\sim \xi\,u_2$ for high threshold $u_2$ (so that $p\to 1$),
\begin{equation}\label{eq:VaR-GPD}
VaR(p)= u_2 - \frac{\beta}{\xi}\left[1-\left( \frac{1-p}{1-G(u_2)}\right)^{-\xi} \right]  \underset{p\to 1}{\sim} 
u_2 \left(\frac{ 1-p}{1-G(u_2)}\right)^{-\xi}
\end{equation}
and 
\begin{equation}\label{eq:ES-GPD}
 ES(p)=\frac{VaR(p)}{1-\xi}+ \frac{\beta - \xi\,u_2}{1-\xi} \underset{p\to 1}{\sim} \frac{VaR(p)}{1-\xi}.
\end{equation}
We use those relations to estimate VaR and ES from the calibrated GPD (with each method), replacing the  parameters by their estimates and $G(u_2)$ by  $G_n(u_2)= 1 - N_{u_2}/n$ where $G_n$ denotes the empirical cdf of $G$, with $n$ the sample size and $N_{u_2}$ the number of observations above $u_2$. We denote those estimates by $\widehat{VaR}(p)$ and $\widehat{ES}(p)$, respectively.

To estimate $ES(p)$ directy from the data (without using the calibrated GPD), we proceed as in \cite{Kratz2018} (in the context of backtesting ES), simply averaging $k$ quantiles from $VaR(p)$:
\begin{equation}\label{eq:ES-numeric}
\widetilde{ES}_{n,k}(p):= \frac1k \sum_{i=1}^{k} VaR(p_i),\quad
\text{with}\quad p_j = p + \frac{j-1}{k}(1-p), \quad j=1,\ldots,k,  \; k \in \mathbb{N}.
\end{equation}
In our case, we take a large $k$ as we are interested in the strong precision of the numerical estimate and chose $k=20,000$. 

Considering our algorithmic approach, we evaluate $VaR(99.5\%)$, $ES(p)$ with $p=97.5\%$ and $99.77\%$, respectively, using \eqref{eq:ES-GPD} for VaR, and two possible estimates for ES, namely \eqref{eq:ES-GPD} and \eqref{eq:ES-numeric} (averaging the VaR's estimates from \eqref{eq:VaR-GPD}). The latter  is chosen to avoid resorting a second time to the asymptotic relation in \eqref{eq:ES-GPD} between ES and VaR.
We compare those estimates with the empirical values obtained directly from the data (using \eqref{eq:ES-numeric} for ES with empirical quantiles).

Then we select, on one hand, the two EVT methods (among the 12 of the {\it tea} R-package) that look the most stable accross samples with a threshold $u_2$ that remains below $q(99.90)$  (see Table~\ref{tab:tea-results} of the Appendix), namely the Reiss \& Thomas (2007) approach that is very stable and the Hall (1990) one, on the other hand, two other (less stable) EVT methods exhibiting a threshold $u_2$ the closest to ours, namely AMSE, performed on our whole dataset (of size $n=60,985$), and  Danielsson et al. (2001), performed on the 5,026 largest observations of the dataset (only case for this method where $u_2$ is below $q(97.5)$ and high enough (we cannot apply the Pickands-Balkema-de Haan theorem for a threshold close to the 3rd quartile).

The selection of the Reiss \& Thomas (2007) and Hall (1990) methods avoids making any arbitrary choice because of their stability. Nevertheless, the problem is that generally their threshold $u_2$ (corresponding to $q(99.6\%)$ and $q(99.77\%)$, respectively) is larger than ours and than VaR(99.5\%), meaning that comparisons of the regulatory quantities of interest obtained with the various methods become less direct.
So, we will also compute $ES(99.77\%)$ to make the comparison straightforward.
The two additional EVT methods, AMSE and Danielsson et al. (2001), provide the thresholds $u_2=q(97.47\%)$ and $u_2=q(97.4\%)$, respectively, which are closer to our threshold and allow for comparison between the various evaluations of $VaR(99.5\%)$, $ES(97.5\%)$, and, of course, also $ES(99.77\%)$. In addition, we added a specific case of the Hall method, where the threshold $u_2$ drops from about 160,000 to 80,000, such that we can compute all quantities of interest.

When evaluating the risk measures with the respective parameters estimates, we express them as a factor of the empirical mean, to ease the comparison with other risks. Then, we compute the relative variation $\Delta$ between the empirical quantity (obtained directly on the data) and the quantity evaluated via the estimated GPD (with the different approaches, respectively). We report the results of this analysis in Table~\ref{tab:risk-measures}. 

\begin{table}[H]
  \centering
  \parbox{500pt}{\caption{\label{tab:risk-measures}\sf\small Estimates of $VaR(99.5\%)$ (via \eqref{eq:VaR-GPD}) and $ES(p)$ (via \eqref{eq:ES-GPD} and \eqref{eq:ES-numeric}, respectively) for $p=97.5\%$ and $99.77\%$, expressed as a multiplying factor of the mean (which value is 3476~\EUR) for various models. Comparison with the empirical values $\widetilde{VaR}(99.5\%)$ and $\widetilde{ES}(p)$ by computing the relative variation $\Delta$ in $\%$. \vspace{0.7ex}}}
\small
\hspace*{-20pt}
    \begin{tabular}{|l|c|c|c|cc|c|cc|}
    \hline
		&&&&&&&&\\[-1.5ex]
                           & $\widehat{VaR}$ &    $\Delta$      & $p=97.5\%$ &     
\multicolumn{2}{c|}{$\Delta$ (in \%)}   & $p=99.77\%$ &  \multicolumn{2}{c|}{ $\Delta$ (in \%)} \\
& (99.5\%) & (in \%) & $\widehat{ES}(p)$ \, $\widetilde{ES}(p)$ & & & $\widehat{ES}(p)$ \, $\widetilde{ES}(p)$ &  & \\
    &&&&&&&&\\[-1.5ex]
    \hline
    \hline
		&&&&&&&&\\[-1.5ex]
  Empirical            &    $\widetilde{VaR}$= 25      &            &  23  &   &   & 114   &   &   \\
\hline
\footnotesize{Our model ($\alpha=1.24$)}    &     13      & -46.5 	&   19  $\quad$ 17  & -17.1  & -28.2 & 132  $\quad$ 104 &  15.9 &  -8.7 \\
\footnotesize{AMSE ($\alpha=1.17$)}  	& 24  & -3.7 &  43   $\quad$ 33 & 85.1 & 43.6 & 331 $\quad$ 227 &  190.6 &  99.1 \\
\footnotesize{Danielsson-al.(01) ($\alpha=1.15$)}  & 24 &  -2.9 &  47  $\quad$ 35 & 101.7 & 49.6 & 373 $\quad$ 242 & 227.1  &  112.1 \\
\footnotesize{Hall (1990)($u_2\!\!=\!\!q(99.45\%)$;$\alpha\!=\!1.37$)}   	&     25      & -2 &   28  $\quad$ 26 & 21.4 & 14.4 & 159 $\quad$ 142 &  39.9 &  24.5 \\
 \footnotesize{Hall (1990) ($\alpha=1.61$)}  &     $-$      & $-$ &   $-$   & $-$ & $-$ & 119 $\quad$ 114 &  4.2 &  -0.3 \\
\footnotesize{Reiss \&Thomas(07) ($\alpha=1.47$)}   & $-$ & $-$ &   $-$ & $-$ & $-$ & 130 $\quad$ 121 &  14.2 &  5.9 \\
  \hline
    \end{tabular}
\end{table}%
The results presented in this table, illustrate the difficulty in modelling our data due to its high noise content. This is well illustrated in the survival plot in Figure~\ref{fig:cdf-positiveDamages}. Moreover, 
an important difficulty with the extremes reported in our database is the strong biais towards round numbers in the filed complaints. This is particularly sensitive in extremes as we might have an accumulation of  large values on one round number and then one complaint with precise number introducing artificial discontinuities.

 Models will result from a compromise between various properties of the data (very extremes, moderately extremes, ...). For instance, it seems that our model reflects well the tail of the distribution whatever the chosen zone but not pointwise (via VaR). 

Concerning the VaR(99.5\%), we observe that our method provides a bad estimation, while the two other EVT methods (AMSE and Danielsson et al.) and the specific case for Hall's method give an accurate estimation (slight underestimation). The other two stable methods do not allow for the computation of the VaR at this threshold.

However, taking only one point (as we do here for the $99.5$\% quantile) does not reflect the tail of the distribution in a proper way. To understand further the phenomenon, we computed several quantiles and could see how it fluctuates a lot around the empirical value, with good (for instance, for VaR(90\%) with an empirical value of 4,300 \EUR and 4,155 estimated with our model; relative error of -3.4\%) and bad (under and over) estimations (as in this example of 99.5\%). This becomes more obvious when examining the right plot in Figure~\ref{fig:cdf-positiveDamages} where we see that the empirical values are underestimated for low quantiles and become well fitted with quantiles above 99.9\% ($1e-03$ on the graph).
This is why we turn to ES that gives a much better picture of the tail, as well recognized nowadays. It has already been a long debate in regulation; Basel 4 moved for market risk from VaR to ES. This might be even more needed for cyber risk, as we can experiment here.

Evaluating ES in the two described ways, we observe that our tail modelling reflects better the data, whatever $p$, while the AMSE and Danielsson et al. methods provide a gross over-evaluation and different results according to the way ES is estimated. For $p=99.77\%$, we can evaluate ES with the methods by Hall and Reiss \& Thomas, with which we obtain also good results, especially for the Hall estimate. This very good fit by Hall's method might be explained by the fact that the level 99.77\% is that of his estimated threshold. For this latter method, when considering the specific case where $u_2=q(99.45\%)$, while the quantile at $u_2$ is perfectly distributed, ES at 97.5\% gives similar results as with our method, but looses accuracy when $p$ increases. 

Finally, looking at the two ways ES is estimated based on the GPD model, we observe, as often in practice (see e.g. the discussion about it in \cite{McNeil2016}), that the numerical estimation based on averaging quantiles provides generally a smaller $\Delta$ (taking into account the signs) than when using the asymptotic relation in $\eqref{eq:ES-GPD}$ between VaR and ES. 

Clearly more research will be needed to produce credible values for the solvency risk measures like VaR, while the ES is better estimated as it concerns values beyond 97.5\% and the tail in this case is better captured by our model. We also see that the factors for the empirical risk measures are quite high (25 times the mean for the VaR(99.5\%) and 61 times the mean for the ES(97.5\%)), which is a sign that we are confronted here with very volatile risks; even the EVT models are not able to catch this for the VaR. In natural catastrophes like windstorms or flood, the factor for VaR(99.5\%) are usually around 20 times the mean. For earthquakes, values around 30 times the mean are found. 
Therefore, when underwriting cyber risks, approaches implemented for natural catastrophes could be borrowed, as, for instance, developing IT systems to control the accumulation of exposures and set limits to them. This will help diversifying the risks, which is key to successfully underwrite extreme risks.

This is also why we want to refine our understanding of cyber risk by differentiating it by type of attack, as presented in the next subsection.

\vspace{-2ex}
\subsection{Comparing the types of cyber attacks via their tail index}
\label{ss:classes}
\vspace{-2ex}
To conclude this section on modelling, we apply our method on three samples, the full sample, the sample related to the fraud only (but representing 87.3\% of the data) and the breach of trust one (representing only 4.9\% of the data). The idea here is to look at the possibility of finding significant differences in the statistics of the various types of attack. This is made possible as the Jackknife results show a relative robustness of the parameters estimation resulting from the application of the algorithmic method. Our assumption is that the tail index could be a discriminant between various forms of cyber complaints. Given the little number of qualified damages, this can only be a first attempt to see if this assumption can gain ground in our data. We provide, in Table~\ref{tab:tail-index},  only the tail index and associated shape parameter ({\it i.e.} the inverse of the tail index), as well as the threshold $u_2$ above which the extremes are modelled with a GPD. 

\begin{table}[h]
\begin{center}
\parbox{430pt}{\caption{\label{tab:tail-index}\sf\small Estimation of the tail index $\xi$ and shape parameter $1/\xi$, as well as of the tail-threshold (also expressed as a quantile) above which the GPD is fitted. Three samples are considered, the full one on the period July 2015-April 2019, a second one restricted to fraud-related damages (87.3\% of the full sample), and the last one restricted to breach-of-trust-related damages (4.9\% of the full sample). The confidence ranges are computed using the Jackknife method. \vspace{0.7ex}}}
\begin{tabular}{lccc}
\hline 
&&&\\[-1.5ex]
LN-E-GPD &  Full sample & Fraud sample & Breach of Trust sample  \\
&&&\\[-1.5ex]
\hline \hline
&&&\\[-1.5ex]
Number of observations& 60,985 & 53,260 & 3,004\\
Tail index            & 0.8088 &  0.8114 &  0.852 \\
Shape parameter       &   1.24 &   1.23 &  1.17\\
{\small with 95\% confidence range} &  [1.21 ; 1.26] & [1.21 ; 1.26] & [1.09 ; 1.27] \\[0.5ex]
\hline
&&&\\[-1.5ex]
Threshold (quantile)  &  9,999 (96.6\%) &  8,999 (96.3\%)&  14,999 (97.3\%)\\
95\% confidence range & [9,980 ; 10,018] & [8,826 ; 9,172] & [13,481 ; 16,517]\\[0.5ex]
\hline
\end{tabular}
\end{center}
\end{table}
\vspace{-2ex}

The results are in line with our expectations. Fraud, representing 87.3\% of the data, gives a shape parameter close to the full sample one. The interesting result is for breach of trust, where the shape parameter is about 6\% smaller than for the full sample. The Jackknife confidence ranges are in line with the fact that the numerical stability depends heavily on the number of observations. The confidence range is much wider for "breach of trust" than for "fraud", but still with a reasonable range. Nevertheless, the narrow confidence range for "fraud" points out to a useful discriminating method whenever the sample sizes are comparable: in such a case, if there is enough difference between tail indices, the algorithm will detect it. 

Given the small size of the data sample, it is not possible here to come up with a definitive conclusion, but it suggests a possible way for exploring further with more data (e.g. when we will have access to the complaints registered since April 2019) and a better characterization of the type of complaints in the database. 

Finally, the shape parameters confirm the common intuition that cyber risk is susceptible to systemic risk; indeed, the type of tails we observe here are extremely heavy. The shape parameters are close to those from earthquake or floods risk in insurance; both risks are characterized by the wide spread of the damages. 

\vspace{-2ex}
\section{Management and research perspectives}
\label{sec:concl}
\vspace{-2ex}
We  first want to emphasize the importance of understanding the data and making sure they are representative of the phenomenon under study. We spent a fair amount of time with our colleagues of SCRC to understand but also review manually the cleansing of their database (done automatically at C3N). We did it in general, through a preliminary statistical exploration of the database, and specifically, complaint by complaint, for the 1,100 largest declared damages. It gives us confidence that we are having here a very important source of information for studying cyber crimes and cyber risk in general, all the more since it is a large database, including various types of variables.  Cyber attacks are a massive phenomenon, especially when considering the iceberg effect. They reach every place in the country given the delocalised nature of Internet. 

Second, we observe that the GN dataset, quite different from those studied so far in the literature, shows similar perspectives in terms of very heavy-tailed distributions. Indeed, the results we obtain for the tail of the damage severity distribution confirm the presence of extremes, which is, quoting  \cite{Tang2019}, {\it a signature of common shocks or systemic risk}. Systemic risk originates from the combination of the existence of extremes, the interconnection between the various systems, and the weight of this set of systems in the general economy. Undoubtedly, these three properties are characteristic of cyber risk, as already discussed at the beginning of the paper. 

To reveal the presence of extremes, we adapted a recent algorithm for fitting heavy-tailed distribution to the case of positive asymmetric data. This tool is very important as it allows for an automatic fit of both the main and extreme behaviors of the empirical distribution. By EVT, we know that the extreme behavior follows a GPD. So, we compared our results with those obtained with other standard EVT methods and the shape parameter is of the same range across methods. Here, we introduce an additional way to judge the quality of the estimated model, by computing the standalone capital requirements with standard risk measures. We observe that our model evaluates well ES(97.5\%), with the closest value to the empirical estimate among tested methods.

We would like to point out the benefit of using this algorithm, not only for the study we carried out, but also for our next investigation on this dataset, when considering a multivariate setting and a dynamic view. Indeed, this method detects by itself the threshold above which observations are considered as extremes without resorting to heavy computations; this solves a practical issue encountered with standard methods of EVT that require separate treatment for the tail, or with other dynamic EVT methods resorting to an arbitrary high threshold. It should then lighten the procedure when introducing covariates and make it more accurate.  The OR field, where probability of extremes matters, 
would benefit from a method that integrates seamlessly the presence of extremes to the modelling of the whole distribution, as proposed in this study, as well as a straightforward way to assess its confidence range through the Jackknife method.

Moreover, to take into account not only the magnitude of the largest damages but also their frequency, we introduced a Poisson-GPD model. We studied the frequency of the extremes to see how it would evolve with time. Contrary to what is often assumed for cyber risk in general, for the extremes, we did not find a strong dynamic component. Namely, on the given period of observed data, it does not seem that, from a statistical point of view, the nature of the risk is changing. 
Nevertheless, this will be investigated again when having access to data recorded on a longer time period. We will use the same approach as described in Section~\ref{ss:Poisson}, introducing time variability in the parameters of the Poisson-GPD model, if needed.

Those statistical results, obtained for material damages, represent a solid basis for helping build resilience. They will now be interpreted by criminal intelligence analysts (from the Intelligence Division of the GN) in order to establish hypotheses in terms of explanation and anticipation by the SCRC. On the insurance side, the results obtained on the tail of the distribution confirm that cyber risk as a whole is insurable, and help evaluate how costly it can be to cover such risk. The existence of an expectation for the loss is crucial to compute the insurance technical premium, as it is its main component. Its second component is the risk-loading, which is related to the capital allocated to the risk. Once the probability distribution of the risk is known, the capital can be estimated in relation to the risk measure used for computing the solvency capital requirements. That is why, it is crucial for the insurability of a risk to have a good knowledge of its entire probability distribution, which is provided by our model. The heaviness of the (right) tail of the distribution of the damages has been estimated with a shape parameter of $1.24\ \pm \ 0.025$. As this parameter has a value significantly larger than 1, it indicates a finite expectation, which is a necessary (but not sufficient) condition for insurability of cyber risk. Nevertheless, the shape parameter with a value below 2 ({\it i.e.} infinite variance), classifies cyber as a very high risk, in the same range as natural catastrophes. But there are important differences between the two risks: the main characteristic of cyber risk is that cyber attacks are performed directly by humans, contrary to natural catastrophes. Also, the geographical location is crucial for the latter, while of much less importance for cyber attacks. For cyber risk, the self-hygiene of the IT system plays the most important role in terms of vulnerability. Other factors that should be studied for making the system more resilient are attackers' motivations, the possible targets in the system (databases, reputation, financials), and the security protocols of users. 

Finally, this study, performed on a novel and exhaustive database, establishes and measures the potential high intensity of cyber risk, a crucial information for the various actors involved in helping society to be more resilient, including the GN itself, insurance companies, strategic management, and policy makers. It also opens various interesting avenues of investigation. One of them is the automatic classification of the types of cyber crimes, according to their tail index, as started to be tackled in this paper. Comparison with existing classifications, such as those of the GN or the Ministry of Justice, will be made and discussed with SCRC and other experts. Given the rich, multi-fields GN database, we will also turn our attention to the modelling of cyber in a multivariate context, as already mentionned. It is an ongoing work. A further step will be to adapt our models and methods to fully account for systemic risk, building from studies on this topic developed in the aftermaths of the 2008/2009 financial crisis. Collaboration with experts of various disciplines will remain essential, taking into account the multiple factors playing a role,  for interpreting the results of the suggested models, and for developing adequate resilience management strategies. Indeed, cyber security and resilience are major challenges in our modern economies, and are top priorities on the agenda of governments, security forces, and management of companies and organizations. Therefore all those efforts are necessary steps for building an agreement on how to assess and manage this risk both quantitatively and qualitatively. 

\vspace{-2ex}
\section*{Acknowledgement(s)}
\vspace{-2ex}
The PJGN database we used for this study has been entrusted by the Gendarmerie under confidentiality agreement. Use and interpretation are the strict responsibility of the authors. As required by  Gendarmerie Nationale, any communication on this study should mention that the source is from "Gendarmerie Nationale – PJGN – treated by ESSEC-CREAR".
The authors are grateful to G\'en\'eral Daoust and Colonel Piat to have made possible this collaboration with the CREAR and its associated or invited members. Our warm thanks to Lieutenant Colonel J\'er\^ome Barlatier, Head of the Intelligence Division of SCRC, to have made this collaboration effective, through the partnership between CREAR and SCRC-PJGN, and to him and Commandant Edouard Klein (C3N) for hosting and helping us with the database. 
The CREAR also acknowledges with gratitude the financial support of Labex MME-DII (ANR-11-LBX-0023-01) for two research visits of Nehla Debbabi.

\bibliographystyle{apacite}
\bibliography{../Lit.bib}

\begin{thebibliography}{}

\bibitem [\protect \citeauthoryear {%
Accenture%
\ \BBA {} {Ponemon Institute LLC}%
}{%
Accenture%
\ \BBA {} {Ponemon Institute LLC}%
}{%
{\protect \APACyear {2019}}%
}]{%
Accenture2019}
\APACinsertmetastar {%
Accenture2019}%
\begin{APACrefauthors}%
Accenture%
\BCBT {}\ \BBA {} {Ponemon Institute LLC}.%
\end{APACrefauthors}%
\unskip\
\newblock
\APACrefYearMonthDay{2019}{}{}.
\newblock
\APACrefbtitle {The Cost of Cybercrime: Ninth Annual Cost of Cybercrime Study
  unlocking the value of improved cybersecurity protection.} {The cost of
  cybercrime: Ninth annual cost of cybercrime study unlocking the value of
  improved cybersecurity protection.}
\newblock
\APAChowpublished
  {\url{https://www.accenture.com/us-en/insights/security/cost-cybercrime-study}}.
\PrintBackRefs{\CurrentBib}

\bibitem [\protect \citeauthoryear {%
Advisen%
\ \BBA {} PartnerRe%
}{%
Advisen%
\ \BBA {} PartnerRe%
}{%
{\protect \APACyear {2018}}%
}]{%
Advisen2018}
\APACinsertmetastar {%
Advisen2018}%
\begin{APACrefauthors}%
Advisen%
\BCBT {}\ \BBA {} PartnerRe.%
\end{APACrefauthors}%
\unskip\
\newblock
\APACrefYearMonthDay{2018}{}{}.
\newblock
\APACrefbtitle {Survey of Cyber Insurance Market Trends.} {Survey of cyber
  insurance market trends.}
\newblock
\APAChowpublished
  {\url{https://partnerre.com/wp-content/uploads/2018/10/2018-Survey-of-Cyber-Insurance-Market-Trends.pdf}}.
\PrintBackRefs{\CurrentBib}

\bibitem [\protect \citeauthoryear {%
Agrafiotis%
, Nurse%
, Goldsmith%
, Creese%
\BCBL {}\ \BBA {} Upton%
}{%
Agrafiotis%
\ \protect \BOthers {.}}{%
{\protect \APACyear {2018}}%
}]{%
Agrafiotis2018}
\APACinsertmetastar {%
Agrafiotis2018}%
\begin{APACrefauthors}%
Agrafiotis, I.%
, Nurse, J.%
, Goldsmith, M.%
, Creese, S.%
\BCBL {}\ \BBA {} Upton, D.%
\end{APACrefauthors}%
\unskip\
\newblock
\APACrefYearMonthDay{2018}{}{}.
\newblock
{\BBOQ}\APACrefatitle {A taxonomy of cyber harms: Defining the impacts of
  cyber-attacks and understanding how they propagate} {A taxonomy of cyber
  harms: Defining the impacts of cyber-attacks and understanding how they
  propagate}.{\BBCQ}
\newblock
\APACjournalVolNumPages{Journal of Cybersecurity}{4}{1}{}.
\PrintBackRefs{\CurrentBib}

\bibitem [\protect \citeauthoryear {%
Aven%
}{%
Aven%
}{%
{\protect \APACyear {2016}}%
}]{%
Aven2016}
\APACinsertmetastar {%
Aven2016}%
\begin{APACrefauthors}%
Aven, T.%
\end{APACrefauthors}%
\unskip\
\newblock
\APACrefYearMonthDay{2016}{}{}.
\newblock
{\BBOQ}\APACrefatitle {Risk assessment and risk management: Review of recent
  advances on their foundation} {Risk assessment and risk management: Review of
  recent advances on their foundation}.{\BBCQ}
\newblock
\APACjournalVolNumPages{European Journal of Operational Research}{253}{}{1-13}.
\PrintBackRefs{\CurrentBib}

\bibitem [\protect \citeauthoryear {%
Aven%
}{%
Aven%
}{%
{\protect \APACyear {2019}}%
}]{%
Aven2019}
\APACinsertmetastar {%
Aven2019}%
\begin{APACrefauthors}%
Aven, T.%
\end{APACrefauthors}%
\unskip\
\newblock
\APACrefYearMonthDay{2019}{}{}.
\newblock
{\BBOQ}\APACrefatitle {The call for a shift from risk to resilience: What does
  it mean?} {The call for a shift from risk to resilience: What does it
  mean?}{\BBCQ}
\newblock
\APACjournalVolNumPages{Risk Analysis}{39}{6}{1196-1203}.
\PrintBackRefs{\CurrentBib}

\bibitem [\protect \citeauthoryear {%
Baldwin%
, Gheyas%
, Ioannidis%
, Pym%
\BCBL {}\ \BBA {} Williams%
}{%
Baldwin%
\ \protect \BOthers {.}}{%
{\protect \APACyear {2017}}%
}]{%
Baldwin2017}
\APACinsertmetastar {%
Baldwin2017}%
\begin{APACrefauthors}%
Baldwin, A.%
, Gheyas, I.%
, Ioannidis, C.%
, Pym, D.%
\BCBL {}\ \BBA {} Williams, J.%
\end{APACrefauthors}%
\unskip\
\newblock
\APACrefYearMonthDay{2017}{}{}.
\newblock
{\BBOQ}\APACrefatitle {Contagion in cyber security attacks} {Contagion in cyber
  security attacks}.{\BBCQ}
\newblock
\APACjournalVolNumPages{Journal of Cybersecurity}{68}{7}{780--791}.
\PrintBackRefs{\CurrentBib}

\bibitem [\protect \citeauthoryear {%
Beirlant%
, Goegebeur%
, Segers%
\BCBL {}\ \BBA {} Teugels%
}{%
Beirlant%
\ \protect \BOthers {.}}{%
{\protect \APACyear {2004}}%
}]{%
beirlant2004}
\APACinsertmetastar {%
beirlant2004}%
\begin{APACrefauthors}%
Beirlant, J.%
, Goegebeur, Y.%
, Segers, J.%
\BCBL {}\ \BBA {} Teugels, J.%
\end{APACrefauthors}%
\unskip\
\newblock
\APACrefYear{2004}.
\newblock
\APACrefbtitle {Statistics of Extremes: Theory and Applications} {Statistics of
  extremes: Theory and applications}.
\newblock
\APACaddressPublisher{West Essex, England}{John Wiley \& Sons}.
\PrintBackRefs{\CurrentBib}

\bibitem [\protect \citeauthoryear {%
B\"ohme%
, Laube%
\BCBL {}\ \BBA {} Riek%
}{%
B\"ohme%
\ \protect \BOthers {.}}{%
{\protect \APACyear {2018}}%
}]{%
Boehme2018}
\APACinsertmetastar {%
Boehme2018}%
\begin{APACrefauthors}%
B\"ohme, R.%
, Laube, S.%
\BCBL {}\ \BBA {} Riek, M.%
\end{APACrefauthors}%
\unskip\
\newblock
\APACrefYearMonthDay{2018}{}{}.
\newblock
{\BBOQ}\APACrefatitle {A fundamental approach to cyber risk analysis} {A
  fundamental approach to cyber risk analysis}.{\BBCQ}
\newblock
\APACjournalVolNumPages{Variance}{11}{2}{}.
\PrintBackRefs{\CurrentBib}

\bibitem [\protect \citeauthoryear {%
Bouveret%
}{%
Bouveret%
}{%
{\protect \APACyear {2018}}%
}]{%
Bouveret2018}
\APACinsertmetastar {%
Bouveret2018}%
\begin{APACrefauthors}%
Bouveret, A.%
\end{APACrefauthors}%
\unskip\
\newblock
\APACrefYearMonthDay{2018}{}{}.
\newblock
\APACrefbtitle {Cyber Risk for the Financial Sector: A Framework for
  Quantitative Assessment.} {Cyber risk for the financial sector: A framework
  for quantitative assessment.}
\newblock
\APAChowpublished {IMF Working Paper 18/143:
  \url{http://dx.doi.org/10.5089/9781484360750.001}}.
\PrintBackRefs{\CurrentBib}

\bibitem [\protect \citeauthoryear {%
Carfora%
, Martinelli%
, Mercaldo%
\BCBL {}\ \BBA {} Orlando%
}{%
Carfora%
\ \protect \BOthers {.}}{%
{\protect \APACyear {2019}}%
}]{%
Carfora2019}
\APACinsertmetastar {%
Carfora2019}%
\begin{APACrefauthors}%
Carfora, M.%
, Martinelli, F.%
, Mercaldo, F.%
\BCBL {}\ \BBA {} Orlando, A.%
\end{APACrefauthors}%
\unskip\
\newblock
\APACrefYearMonthDay{2019}{}{}.
\newblock
{\BBOQ}\APACrefatitle {Cyber risk management: an actuarial point of view}
  {Cyber risk management: an actuarial point of view}.{\BBCQ}
\newblock
\APACjournalVolNumPages{Journal of Operational Risk}{14}{4}{77--103}.
\PrintBackRefs{\CurrentBib}

\bibitem [\protect \citeauthoryear {%
Chavez-Demoulin%
, Embrechts%
\BCBL {}\ \BBA {} Hofert%
}{%
Chavez-Demoulin%
\ \protect \BOthers {.}}{%
{\protect \APACyear {2016}}%
}]{%
Chavez2016}
\APACinsertmetastar {%
Chavez2016}%
\begin{APACrefauthors}%
Chavez-Demoulin, V.%
, Embrechts, P.%
\BCBL {}\ \BBA {} Hofert, M.%
\end{APACrefauthors}%
\unskip\
\newblock
\APACrefYearMonthDay{2016}{}{}.
\newblock
{\BBOQ}\APACrefatitle {An extreme value approach for modeling operational risk
  losses depending on covariates} {An extreme value approach for modeling
  operational risk losses depending on covariates}.{\BBCQ}
\newblock
\APACjournalVolNumPages{Journal of Risk and Insurance}{83}{3}{735--776}.
\PrintBackRefs{\CurrentBib}

\bibitem [\protect \citeauthoryear {%
Cohen%
, Humphries%
, Veau%
\BCBL {}\ \BBA {} Francis%
}{%
Cohen%
\ \protect \BOthers {.}}{%
{\protect \APACyear {2019}}%
}]{%
Cohen2019}
\APACinsertmetastar {%
Cohen2019}%
\begin{APACrefauthors}%
Cohen, R.%
, Humphries, J.%
, Veau, S.%
\BCBL {}\ \BBA {} Francis, R.%
\end{APACrefauthors}%
\unskip\
\newblock
\APACrefYearMonthDay{2019}{}{}.
\newblock
{\BBOQ}\APACrefatitle {An investigation of cyber loss data and its links to
  operational risk} {An investigation of cyber loss data and its links to
  operational risk}.{\BBCQ}
\newblock
\APACjournalVolNumPages{Journal of Operational Risk}{14}{3}{1--25}.
\PrintBackRefs{\CurrentBib}

\bibitem [\protect \citeauthoryear {%
{CRO Forum}%
}{%
{CRO Forum}%
}{%
{\protect \APACyear {2016}}%
}]{%
CRO2016}
\APACinsertmetastar {%
CRO2016}%
\begin{APACrefauthors}%
{CRO Forum}.%
\end{APACrefauthors}%
\unskip\
\newblock
\APACrefYearMonthDay{2016}{}{}.
\newblock
\APACrefbtitle {{CRO Forum} Concept Paper on a proposed categorisation
  methodology for cyber risk.} {{CRO Forum} concept paper on a proposed
  categorisation methodology for cyber risk.}
\newblock
\APAChowpublished
  {\url{https://www.thecroforum.org/wp-content/uploads/2016/06/ZRH-16-09033-P1_CRO_Forum_Cyber-Risk_web.pdf}}.
\PrintBackRefs{\CurrentBib}

\bibitem [\protect \citeauthoryear {%
Dacorogna%
, Debbabi%
\BCBL {}\ \BBA {} Kratz%
}{%
Dacorogna%
\ \protect \BOthers {.}}{%
{\protect \APACyear {2018}}%
}]{%
Dacorogna2018}
\APACinsertmetastar {%
Dacorogna2018}%
\begin{APACrefauthors}%
Dacorogna, M.%
, Debbabi, N.%
\BCBL {}\ \BBA {} Kratz, M.%
\end{APACrefauthors}%
\unskip\
\newblock
\APACrefYearMonthDay{2018}{August}{}.
\newblock
{\BBOQ}\APACrefatitle {Analyse exploratoire des plaintes de crimes cyber
  renseign\'ees \`a la {C3N} de la {PJGN}} {Analyse exploratoire des plaintes
  de crimes cyber renseign\'ees \`a la {C3N} de la {PJGN}}.{\BBCQ}
\newblock
\APACjournalVolNumPages{Research report to the {PJGN}}{}{}{1-24}.
\PrintBackRefs{\CurrentBib}

\bibitem [\protect \citeauthoryear {%
Dacorogna%
\ \BBA {} Kratz%
}{%
Dacorogna%
\ \BBA {} Kratz%
}{%
{\protect \APACyear {2020}}%
}]{%
Ventre2020}
\APACinsertmetastar {%
Ventre2020}%
\begin{APACrefauthors}%
Dacorogna, M.%
\BCBT {}\ \BBA {} Kratz, M.%
\end{APACrefauthors}%
\unskip\
\newblock
\APACrefYearMonthDay{2020}{}{}.
\newblock
{\BBOQ}\APACrefatitle {Moving from Uncertainty to Risk: The Case of Cyber Risk}
  {Moving from uncertainty to risk: The case of cyber risk}.{\BBCQ}
\newblock
\BIn{} D.~Ventre, H.~Loiseau\BCBL {}\ \BBA {} H.~Aden\ (\BEDS), \APACrefbtitle
  {Cybersecurity in Humanities and Social Sciences: A Research Methods
  Approach} {Cybersecurity in humanities and social sciences: A research
  methods approach}\ (\BPGS\ 123--152).
\newblock
\APACaddressPublisher{Montreal and New York}{ISTE Scientific Publishing and
  Wiley}.
\PrintBackRefs{\CurrentBib}

\bibitem [\protect \citeauthoryear {%
Dacorogna%
\ \BBA {} Kratz%
}{%
Dacorogna%
\ \BBA {} Kratz%
}{%
{\protect \APACyear {2022}}%
}]{%
Dacorogna2022}
\APACinsertmetastar {%
Dacorogna2022}%
\begin{APACrefauthors}%
Dacorogna, M.%
\BCBT {}\ \BBA {} Kratz, M.%
\end{APACrefauthors}%
\unskip\
\newblock
\APACrefYearMonthDay{2022}{}{}.
\newblock
{\BBOQ}\APACrefatitle {Special {I}ssue “{C}yber {R}isk and {S}ecurity"}
  {Special {I}ssue “{C}yber {R}isk and {S}ecurity"}.{\BBCQ}
\newblock
\APACjournalVolNumPages{Risks}{10}{6}{112}.
\newblock
\begin{APACrefURL} \url{https:https://doi.org/10.3390/risks10060112}
  \end{APACrefURL}
\PrintBackRefs{\CurrentBib}

\bibitem [\protect \citeauthoryear {%
Das%
, Dhara%
\BCBL {}\ \BBA {} Natarajan%
}{%
Das%
\ \protect \BOthers {.}}{%
{\protect \APACyear {2021}}%
}]{%
DAS2021}
\APACinsertmetastar {%
DAS2021}%
\begin{APACrefauthors}%
Das, B.%
, Dhara, A.%
\BCBL {}\ \BBA {} Natarajan, K.%
\end{APACrefauthors}%
\unskip\
\newblock
\APACrefYearMonthDay{2021}{}{}.
\newblock
{\BBOQ}\APACrefatitle {On the Heavy-Tail Behavior of the Distributionally
  Robust Newsvendor} {On the heavy-tail behavior of the distributionally robust
  newsvendor}.{\BBCQ}
\newblock
\APACjournalVolNumPages{Operations Research}{}{}{}.
\newblock
\APAChowpublished {articles in advance:
  \url{https://doi.org/10.1287/opre.2020.2091}}.
\PrintBackRefs{\CurrentBib}

\bibitem [\protect \citeauthoryear {%
Debbabi%
\ \BBA {} Kratz%
}{%
Debbabi%
\ \BBA {} Kratz%
}{%
{\protect \APACyear {2014}}%
}]{%
Debbabi2014}
\APACinsertmetastar {%
Debbabi2014}%
\begin{APACrefauthors}%
Debbabi, N.%
\BCBT {}\ \BBA {} Kratz, M.%
\end{APACrefauthors}%
\unskip\
\newblock
\APACrefYearMonthDay{2014}{}{}.
\newblock
{\BBOQ}\APACrefatitle {A new unsupervised threshold determination for hybrid
  models} {A new unsupervised threshold determination for hybrid
  models}.{\BBCQ}
\newblock
\APACjournalVolNumPages{IEEE International Conference on Acoustics, Speech and
  Signal Processing (ICASSP)}{}{}{3440-3444}.
\PrintBackRefs{\CurrentBib}

\bibitem [\protect \citeauthoryear {%
Debbabi%
, Kratz%
\BCBL {}\ \BBA {} Mboup%
}{%
Debbabi%
\ \protect \BOthers {.}}{%
{\protect \APACyear {2017}}%
}]{%
Debbabi2017}
\APACinsertmetastar {%
Debbabi2017}%
\begin{APACrefauthors}%
Debbabi, N.%
, Kratz, M.%
\BCBL {}\ \BBA {} Mboup, M.%
\end{APACrefauthors}%
\unskip\
\newblock
\APACrefYearMonthDay{2017}{}{}.
\newblock
\APACrefbtitle {A self-calibrating method for heavy tailed data modelling.
  Application in neuroscience and finance.} {A self-calibrating method for
  heavy tailed data modelling. application in neuroscience and finance.}
\newblock
\APAChowpublished {Preprint, available at
  \url{https://arxiv.org/abs/1612.03974v2}}.
\PrintBackRefs{\CurrentBib}

\bibitem [\protect \citeauthoryear {%
{de}~Haan%
\ \BBA {} Ferreira%
}{%
{de}~Haan%
\ \BBA {} Ferreira%
}{%
{\protect \APACyear {2006}}%
}]{%
deHaan2006}
\APACinsertmetastar {%
deHaan2006}%
\begin{APACrefauthors}%
{de}~Haan, L.%
\BCBT {}\ \BBA {} Ferreira, A\BPBI J.%
\end{APACrefauthors}%
\unskip\
\newblock
\APACrefYear{2006}.
\newblock
\APACrefbtitle {Extreme Value Theory: An Introduction} {Extreme value theory:
  An introduction}.
\newblock
\APACaddressPublisher{Berlin}{Springer - Verlag}.
\PrintBackRefs{\CurrentBib}

\bibitem [\protect \citeauthoryear {%
Dr\'egoir%
}{%
Dr\'egoir%
}{%
{\protect \APACyear {2017}}%
}]{%
Dregoir2017}
\APACinsertmetastar {%
Dregoir2017}%
\begin{APACrefauthors}%
Dr\'egoir, M.%
\end{APACrefauthors}%
\unskip\
\newblock
\APACrefYearMonthDay{2017}{September}{}.
\newblock
{\BBOQ}\APACrefatitle {L'effet iceberg: d\'efinition, mesures et m\'ethodes de
  traitement et applications aux donn\'ees cybercriminelles} {L'effet iceberg:
  d\'efinition, mesures et m\'ethodes de traitement et applications aux
  donn\'ees cybercriminelles}.{\BBCQ}
\newblock
\APACjournalVolNumPages{PJGN report and ENSAI (Univ. Rennes 1) master
  thesis}{}{}{1-84}.
\PrintBackRefs{\CurrentBib}

\bibitem [\protect \citeauthoryear {%
Eling%
\ \BBA {} Schnell%
}{%
Eling%
\ \BBA {} Schnell%
}{%
{\protect \APACyear {2016}}%
}]{%
Eling2016}
\APACinsertmetastar {%
Eling2016}%
\begin{APACrefauthors}%
Eling, M.%
\BCBT {}\ \BBA {} Schnell, W.%
\end{APACrefauthors}%
\unskip\
\newblock
\APACrefYearMonthDay{2016}{}{}.
\newblock
{\BBOQ}\APACrefatitle {What do we know about cyber risk and cyber risk
  insurance?} {What do we know about cyber risk and cyber risk
  insurance?}{\BBCQ}
\newblock
\APACjournalVolNumPages{Journal of Risk Finance}{5}{17}{474-491}.
\PrintBackRefs{\CurrentBib}

\bibitem [\protect \citeauthoryear {%
Eling%
\ \BBA {} Wirfs%
}{%
Eling%
\ \BBA {} Wirfs%
}{%
{\protect \APACyear {2019}}%
}]{%
Eling2019}
\APACinsertmetastar {%
Eling2019}%
\begin{APACrefauthors}%
Eling, M.%
\BCBT {}\ \BBA {} Wirfs, J.%
\end{APACrefauthors}%
\unskip\
\newblock
\APACrefYearMonthDay{2019}{}{}.
\newblock
{\BBOQ}\APACrefatitle {What are the actual costs of cyber risk events?} {What
  are the actual costs of cyber risk events?}{\BBCQ}
\newblock
\APACjournalVolNumPages{European Journal of Operational
  Research}{272}{3}{1109--1119}.
\PrintBackRefs{\CurrentBib}

\bibitem [\protect \citeauthoryear {%
Embrechts%
, Kl\"uppelberg%
\BCBL {}\ \BBA {} Mikosch%
}{%
Embrechts%
\ \protect \BOthers {.}}{%
{\protect \APACyear {2011}}%
}]{%
Embrechts2011}
\APACinsertmetastar {%
Embrechts2011}%
\begin{APACrefauthors}%
Embrechts, P.%
, Kl\"uppelberg, C.%
\BCBL {}\ \BBA {} Mikosch, T.%
\end{APACrefauthors}%
\unskip\
\newblock
\APACrefYear{2011}.
\newblock
\APACrefbtitle {Modelling Extremal Events for Insurance and Finance} {Modelling
  extremal events for insurance and finance}\ (\PrintOrdinal{$1^{st}$ ed. 1997,
  $2^{nd}$}\ \BEd).
\newblock
\APACaddressPublisher{Berlin, Heidelberg and New York}{Springer - Verlag}.
\PrintBackRefs{\CurrentBib}

\bibitem [\protect \citeauthoryear {%
Embrechts%
, Mizgier%
\BCBL {}\ \BBA {} Chen%
}{%
Embrechts%
\ \protect \BOthers {.}}{%
{\protect \APACyear {2018}}%
}]{%
Embrechts2018}
\APACinsertmetastar {%
Embrechts2018}%
\begin{APACrefauthors}%
Embrechts, P.%
, Mizgier, K.%
\BCBL {}\ \BBA {} Chen, X.%
\end{APACrefauthors}%
\unskip\
\newblock
\APACrefYearMonthDay{2018}{}{}.
\newblock
{\BBOQ}\APACrefatitle {Modeling operational risk depending on covariates: an
  empirical investigation} {Modeling operational risk depending on covariates:
  an empirical investigation}.{\BBCQ}
\newblock
\APACjournalVolNumPages{Journal of Operational Risk}{13}{3}{17--46}.
\PrintBackRefs{\CurrentBib}

\bibitem [\protect \citeauthoryear {%
Fahrenwaldt%
, Weber%
\BCBL {}\ \BBA {} Weske%
}{%
Fahrenwaldt%
\ \protect \BOthers {.}}{%
{\protect \APACyear {2018}}%
}]{%
Fahrenwaldt2018}
\APACinsertmetastar {%
Fahrenwaldt2018}%
\begin{APACrefauthors}%
Fahrenwaldt, M\BPBI A.%
, Weber, S.%
\BCBL {}\ \BBA {} Weske, K.%
\end{APACrefauthors}%
\unskip\
\newblock
\APACrefYearMonthDay{2018}{}{}.
\newblock
{\BBOQ}\APACrefatitle {Pricing of Cyber Insurance Contracts In a Network Model}
  {Pricing of cyber insurance contracts in a network model}.{\BBCQ}
\newblock
\APACjournalVolNumPages{ASTIN Bulletin}{3}{48}{1175-1218}.
\PrintBackRefs{\CurrentBib}

\bibitem [\protect \citeauthoryear {%
Farkas%
, Lopez%
\BCBL {}\ \BBA {} Thomas%
}{%
Farkas%
\ \protect \BOthers {.}}{%
{\protect \APACyear {2021}}%
}]{%
Farkas2021}
\APACinsertmetastar {%
Farkas2021}%
\begin{APACrefauthors}%
Farkas, S.%
, Lopez, O.%
\BCBL {}\ \BBA {} Thomas, M.%
\end{APACrefauthors}%
\unskip\
\newblock
\APACrefYearMonthDay{2021}{}{}.
\newblock
{\BBOQ}\APACrefatitle {Cyber claim analysis using {G}eneralized {P}areto
  regression trees with applications to insurance} {Cyber claim analysis using
  {G}eneralized {P}areto regression trees with applications to
  insurance}.{\BBCQ}
\newblock
\APACjournalVolNumPages{Insurance Mathematics and Economics}{98}{}{92--105}.
\PrintBackRefs{\CurrentBib}

\bibitem [\protect \citeauthoryear {%
{Groupe de travail interminist{\'e}riel sur la lutte contre la
  cybercriminalit{\'e}}%
}{%
{Groupe de travail interminist{\'e}riel sur la lutte contre la
  cybercriminalit{\'e}}%
}{%
{\protect \APACyear {2020}}%
}]{%
natinf2014}
\APACinsertmetastar {%
natinf2014}%
\begin{APACrefauthors}%
{Groupe de travail interminist{\'e}riel sur la lutte contre la
  cybercriminalit{\'e}}.%
\end{APACrefauthors}%
\unskip\
\newblock
\APACrefYearMonthDay{2020}{}{}.
\newblock
\APACrefbtitle {Prot{\'e}ger les INTERNAUTES, Rapport sur la
  cybercriminalit{\'e}, Annexes.} {Prot{\'e}ger les internautes, rapport sur la
  cybercriminalit{\'e}, annexes.}
\newblock
\APAChowpublished
  {\url{http://www.justice.gouv.fr/include_htm/pub/rap_cybercriminalite_annexes.pdf}}.
\PrintBackRefs{\CurrentBib}

\bibitem [\protect \citeauthoryear {%
He%
, Zhuang%
\BCBL {}\ \BBA {} Rao%
}{%
He%
\ \protect \BOthers {.}}{%
{\protect \APACyear {2020}}%
}]{%
He2020}
\APACinsertmetastar {%
He2020}%
\begin{APACrefauthors}%
He, F.%
, Zhuang, J.%
\BCBL {}\ \BBA {} Rao, N\BPBI S\BPBI V.%
\end{APACrefauthors}%
\unskip\
\newblock
\APACrefYearMonthDay{2020}{}{}.
\newblock
{\BBOQ}\APACrefatitle {Discrete game-theoretic analysis of defense in
  correlated cyber-physical systems} {Discrete game-theoretic analysis of
  defense in correlated cyber-physical systems}.{\BBCQ}
\newblock
\APACjournalVolNumPages{Annals of Operations Research}{294}{}{741--767}.
\PrintBackRefs{\CurrentBib}

\bibitem [\protect \citeauthoryear {%
Hill%
}{%
Hill%
}{%
{\protect \APACyear {1975}}%
}]{%
Hill1975}
\APACinsertmetastar {%
Hill1975}%
\begin{APACrefauthors}%
Hill, B\BPBI M.%
\end{APACrefauthors}%
\unskip\
\newblock
\APACrefYearMonthDay{1975}{}{}.
\newblock
{\BBOQ}\APACrefatitle {A simple general approach to inference about the tail of
  a distributions} {A simple general approach to inference about the tail of a
  distributions}.{\BBCQ}
\newblock
\APACjournalVolNumPages{Annals of Statistics}{3}{}{1163--1174}.
\PrintBackRefs{\CurrentBib}

\bibitem [\protect \citeauthoryear {%
{Institut National de la Statistique et des Etudes Economiques}%
}{%
{Institut National de la Statistique et des Etudes Economiques}%
}{%
{\protect \APACyear {2018}}%
}]{%
INSEE2018}
\APACinsertmetastar {%
INSEE2018}%
\begin{APACrefauthors}%
{Institut National de la Statistique et des Etudes Economiques}.%
\end{APACrefauthors}%
\unskip\
\newblock
\APACrefYearMonthDay{2018}{}{}.
\newblock
\APACrefbtitle {Pyramide des {\^a}ges de la population fran{\c c}aise.}
  {Pyramide des {\^a}ges de la population fran{\c c}aise.}
\newblock
\APAChowpublished
  {\url{https://www.insee.fr/fr/statistiques?debut=0&theme=1&categorie=1}}.
\newblock
\APACrefnote{{C}onsulted on July 21, 2018}
\PrintBackRefs{\CurrentBib}

\bibitem [\protect \citeauthoryear {%
Knecht%
\ \BBA {} Kn{\"u}ttel%
}{%
Knecht%
\ \BBA {} Kn{\"u}ttel%
}{%
{\protect \APACyear {2003}}%
}]{%
Knecht2003}
\APACinsertmetastar {%
Knecht2003}%
\begin{APACrefauthors}%
Knecht, M.%
\BCBT {}\ \BBA {} Kn{\"u}ttel, S.%
\end{APACrefauthors}%
\unskip\
\newblock
\APACrefYearMonthDay{2003}{}{}.
\newblock
{\BBOQ}\APACrefatitle {The {C}zeledin distribution function} {The {C}zeledin
  distribution function}.{\BBCQ}
\newblock
\APACjournalVolNumPages{Proceedings XXXIV ASTIN Colloquium}{}{}{}.
\PrintBackRefs{\CurrentBib}

\bibitem [\protect \citeauthoryear {%
Kratz%
}{%
Kratz%
}{%
{\protect \APACyear {2019}}%
}]{%
Kratz2019}
\APACinsertmetastar {%
Kratz2019}%
\begin{APACrefauthors}%
Kratz, M.%
\end{APACrefauthors}%
\unskip\
\newblock
\APACrefYearMonthDay{2019}{}{}.
\newblock
{\BBOQ}\APACrefatitle {Introduction to {E}xtreme {V}alue {T}heory.
  {A}pplications to {R}isk {A}nalysis \& {M}anagement} {Introduction to
  {E}xtreme {V}alue {T}heory. {A}pplications to {R}isk {A}nalysis \&
  {M}anagement}.{\BBCQ}
\newblock
\BIn{} D.~Wood, J.~de Gier, C.~Praeger\BCBL {}\ \BBA {} T.~Tao\ (\BEDS),
  \APACrefbtitle {Matrix Book Series, vol. 2 - 2017 MATRIX Annals - Mathematics
  of Risk} {Matrix book series, vol. 2 - 2017 matrix annals - mathematics of
  risk}\ (\BPG~591-636).
\newblock
\APACaddressPublisher{Berlin, Heidelberg and New York}{Springer-Verlag}.
\PrintBackRefs{\CurrentBib}

\bibitem [\protect \citeauthoryear {%
Kratz%
, Lok%
\BCBL {}\ \BBA {} McNeil%
}{%
Kratz%
\ \protect \BOthers {.}}{%
{\protect \APACyear {2018}}%
}]{%
Kratz2018}
\APACinsertmetastar {%
Kratz2018}%
\begin{APACrefauthors}%
Kratz, M.%
, Lok, Y.%
\BCBL {}\ \BBA {} McNeil, A.%
\end{APACrefauthors}%
\unskip\
\newblock
\APACrefYearMonthDay{2018}{}{}.
\newblock
{\BBOQ}\APACrefatitle {Multinomial \uppercase{V}a\uppercase{R} Backtests: A
  simple implicit approach to backtesting expected shortfall} {Multinomial
  \uppercase{V}a\uppercase{R} backtests: A simple implicit approach to
  backtesting expected shortfall}.{\BBCQ}
\newblock
\APACjournalVolNumPages{Journal of Banking and Finance}{88}{}{393--407}.
\PrintBackRefs{\CurrentBib}

\bibitem [\protect \citeauthoryear {%
K\"unsch%
}{%
K\"unsch%
}{%
{\protect \APACyear {1989}}%
}]{%
Kunsch1989}
\APACinsertmetastar {%
Kunsch1989}%
\begin{APACrefauthors}%
K\"unsch, H\BPBI R.%
\end{APACrefauthors}%
\unskip\
\newblock
\APACrefYearMonthDay{1989}{}{}.
\newblock
{\BBOQ}\APACrefatitle {The Jackknife and the bootstrap for general stationary
  observations} {The jackknife and the bootstrap for general stationary
  observations}.{\BBCQ}
\newblock
\APACjournalVolNumPages{Annals of Statistics}{17}{}{1217--1241}.
\PrintBackRefs{\CurrentBib}

\bibitem [\protect \citeauthoryear {%
Leadbetter%
, Lindgren%
\BCBL {}\ \BBA {} Rootz\'en%
}{%
Leadbetter%
\ \protect \BOthers {.}}{%
{\protect \APACyear {2011}}%
}]{%
Leadbetter2011}
\APACinsertmetastar {%
Leadbetter2011}%
\begin{APACrefauthors}%
Leadbetter, M\BPBI R.%
, Lindgren, G.%
\BCBL {}\ \BBA {} Rootz\'en, H.%
\end{APACrefauthors}%
\unskip\
\newblock
\APACrefYear{2011}.
\newblock
\APACrefbtitle {Extremes and Related Properties of Random Sequences and
  Processes} {Extremes and related properties of random sequences and
  processes}\ (\PrintOrdinal{$1^{st}$ ed. 1983, $2^{nd}$}\ \BEd).
\newblock
\APACaddressPublisher{Berlin}{Springer - Verlag}.
\PrintBackRefs{\CurrentBib}

\bibitem [\protect \citeauthoryear {%
Levenberg%
}{%
Levenberg%
}{%
{\protect \APACyear {1944}}%
}]{%
Levenberg1944}
\APACinsertmetastar {%
Levenberg1944}%
\begin{APACrefauthors}%
Levenberg, K.%
\end{APACrefauthors}%
\unskip\
\newblock
\APACrefYearMonthDay{1944}{}{}.
\newblock
{\BBOQ}\APACrefatitle {A method for the solution of certain non-linear problems
  in least squares} {A method for the solution of certain non-linear problems
  in least squares}.{\BBCQ}
\newblock
\APACjournalVolNumPages{Quarterly of Applied Mathematics}{2}{2}{164--168}.
\PrintBackRefs{\CurrentBib}

\bibitem [\protect \citeauthoryear {%
Marotta%
, Martinelli%
, Nanni%
, Orlando%
\BCBL {}\ \BBA {} Yautsiukhin%
}{%
Marotta%
\ \protect \BOthers {.}}{%
{\protect \APACyear {2017}}%
}]{%
Marotta2017}
\APACinsertmetastar {%
Marotta2017}%
\begin{APACrefauthors}%
Marotta, A.%
, Martinelli, F.%
, Nanni, S.%
, Orlando, A.%
\BCBL {}\ \BBA {} Yautsiukhin, A.%
\end{APACrefauthors}%
\unskip\
\newblock
\APACrefYearMonthDay{2017}{}{}.
\newblock
{\BBOQ}\APACrefatitle {Cyber-insurance survey} {Cyber-insurance survey}.{\BBCQ}
\newblock
\APACjournalVolNumPages{Computer Science Review}{24}{}{35--61}.
\PrintBackRefs{\CurrentBib}

\bibitem [\protect \citeauthoryear {%
Marquardt%
}{%
Marquardt%
}{%
{\protect \APACyear {1963}}%
}]{%
Marquardt1963}
\APACinsertmetastar {%
Marquardt1963}%
\begin{APACrefauthors}%
Marquardt, D\BPBI W.%
\end{APACrefauthors}%
\unskip\
\newblock
\APACrefYearMonthDay{1963}{}{}.
\newblock
{\BBOQ}\APACrefatitle {An Algorithm for Least-Squares Estimation of Nonlinear
  Parameters} {An algorithm for least-squares estimation of nonlinear
  parameters}.{\BBCQ}
\newblock
\APACjournalVolNumPages{SIAM Journal on Applied Mathematics}{11}{2}{431--441}.
\PrintBackRefs{\CurrentBib}

\bibitem [\protect \citeauthoryear {%
Marsh%
\ \BBA {} Microsoft%
}{%
Marsh%
\ \BBA {} Microsoft%
}{%
{\protect \APACyear {2019}}%
}]{%
MarshMicrosoft2019}
\APACinsertmetastar {%
MarshMicrosoft2019}%
\begin{APACrefauthors}%
Marsh%
\BCBT {}\ \BBA {} Microsoft.%
\end{APACrefauthors}%
\unskip\
\newblock
\APACrefYearMonthDay{2019}{}{}.
\newblock
\APACrefbtitle {2019-Global cyber risk perception survey.} {2019-global cyber
  risk perception survey.}
\newblock
\APAChowpublished
  {\url{https://www.microsoft.com/security/blog/wp-content/uploads/2019/09/Marsh-Microsoft-2019-Global-Cyber-Risk-Perception-Survey.pdf}}.
\PrintBackRefs{\CurrentBib}

\bibitem [\protect \citeauthoryear {%
Mc~Neil%
, Frey%
\BCBL {}\ \BBA {} Embrechts%
}{%
Mc~Neil%
\ \protect \BOthers {.}}{%
{\protect \APACyear {2016}}%
}]{%
McNeil2016}
\APACinsertmetastar {%
McNeil2016}%
\begin{APACrefauthors}%
Mc~Neil, A\BPBI J.%
, Frey, R.%
\BCBL {}\ \BBA {} Embrechts, P.%
\end{APACrefauthors}%
\unskip\
\newblock
\APACrefYear{2016}.
\newblock
\APACrefbtitle {Quantitative Risk Management} {Quantitative risk management}\
  (\PrintOrdinal{$1^{st}$ ed. 2005, $2^{nd}$}\ \BEd).
\newblock
\APACaddressPublisher{Princeton}{Princeton Series in Finance}.
\PrintBackRefs{\CurrentBib}

\bibitem [\protect \citeauthoryear {%
Nagurney%
, Daniele%
\BCBL {}\ \BBA {} Shukla%
}{%
Nagurney%
\ \protect \BOthers {.}}{%
{\protect \APACyear {2017}}%
}]{%
Nagurney2017}
\APACinsertmetastar {%
Nagurney2017}%
\begin{APACrefauthors}%
Nagurney, A.%
, Daniele, P.%
\BCBL {}\ \BBA {} Shukla, S.%
\end{APACrefauthors}%
\unskip\
\newblock
\APACrefYearMonthDay{2017}{}{}.
\newblock
{\BBOQ}\APACrefatitle {A supply chain network game theory model of
  cybersecurity investments with nonlinear budget constraints} {A supply chain
  network game theory model of cybersecurity investments with nonlinear budget
  constraints}.{\BBCQ}
\newblock
\APACjournalVolNumPages{Annals of Operations Research}{248}{}{405--427}.
\PrintBackRefs{\CurrentBib}

\bibitem [\protect \citeauthoryear {%
NCSC%
}{%
NCSC%
}{%
{\protect \APACyear {2021}}%
}]{%
NCSC2020}
\APACinsertmetastar {%
NCSC2020}%
\begin{APACrefauthors}%
NCSC.%
\end{APACrefauthors}%
\unskip\
\newblock
\APACrefYearMonthDay{2021}{}{}.
\newblock
\APACrefbtitle {Current figures - Announcements per week (2020-2021).} {Current
  figures - announcements per week (2020-2021).}
\newblock
\APAChowpublished
  {\url{https://www.ncsc.admin.ch/ncsc/en/home/aktuell/aktuelle-zahlen.html}}.
\PrintBackRefs{\CurrentBib}

\bibitem [\protect \citeauthoryear {%
Pasculli%
}{%
Pasculli%
}{%
{\protect \APACyear {2020}}%
}]{%
Pasculli2020}
\APACinsertmetastar {%
Pasculli2020}%
\begin{APACrefauthors}%
Pasculli, L.%
\end{APACrefauthors}%
\unskip\
\newblock
\APACrefYearMonthDay{2020}{}{}.
\newblock
{\BBOQ}\APACrefatitle {The global cause of cyber-crimes and state
  responsibilities- Towards an integrated interdisciplinary theory} {The global
  cause of cyber-crimes and state responsibilities- towards an integrated
  interdisciplinary theory}.{\BBCQ}
\newblock
\APACjournalVolNumPages{Journal of Ethics and Legal
  Technologies}{2}{1}{48--74}.
\PrintBackRefs{\CurrentBib}

\bibitem [\protect \citeauthoryear {%
Paul%
\ \BBA {} Zhang%
}{%
Paul%
\ \BBA {} Zhang%
}{%
{\protect \APACyear {2021}}%
}]{%
Paul2021}
\APACinsertmetastar {%
Paul2021}%
\begin{APACrefauthors}%
Paul, J\BPBI A.%
\BCBT {}\ \BBA {} Zhang, M.%
\end{APACrefauthors}%
\unskip\
\newblock
\APACrefYearMonthDay{2021}{}{}.
\newblock
{\BBOQ}\APACrefatitle {Decision support model for cybersecurity risk planning:
  A two-stage stochastic programming framework featuring firms, government, and
  attacker} {Decision support model for cybersecurity risk planning: A
  two-stage stochastic programming framework featuring firms, government, and
  attacker}.{\BBCQ}
\newblock
\APACjournalVolNumPages{European Journal of Operational
  Research}{291}{}{349-364}.
\PrintBackRefs{\CurrentBib}

\bibitem [\protect \citeauthoryear {%
Peng%
, Xu%
, Xu%
\BCBL {}\ \BBA {} Hu%
}{%
Peng%
\ \protect \BOthers {.}}{%
{\protect \APACyear {2018}}%
}]{%
Peng2018}
\APACinsertmetastar {%
Peng2018}%
\begin{APACrefauthors}%
Peng, C.%
, Xu, M.%
, Xu, S.%
\BCBL {}\ \BBA {} Hu, T.%
\end{APACrefauthors}%
\unskip\
\newblock
\APACrefYearMonthDay{2018}{}{}.
\newblock
{\BBOQ}\APACrefatitle {Modeling multivariate cybersecurity risks} {Modeling
  multivariate cybersecurity risks}.{\BBCQ}
\newblock
\APACjournalVolNumPages{Journal of Applied Statistics}{45}{15}{2718--2740}.
\PrintBackRefs{\CurrentBib}

\bibitem [\protect \citeauthoryear {%
Reiss%
\ \BBA {} Thomas%
}{%
Reiss%
\ \BBA {} Thomas%
}{%
{\protect \APACyear {2007}}%
}]{%
Reiss2007}
\APACinsertmetastar {%
Reiss2007}%
\begin{APACrefauthors}%
Reiss, R\BPBI D.%
\BCBT {}\ \BBA {} Thomas, M.%
\end{APACrefauthors}%
\unskip\
\newblock
\APACrefYear{2007}.
\newblock
\APACrefbtitle {Statistical Analysis of Extreme Values: With Applications to
  Insurance, Finance, Hydrology and Other Fields} {Statistical analysis of
  extreme values: With applications to insurance, finance, hydrology and other
  fields}\ (\PrintOrdinal{$1^{st}$ ed. 1997, $2^{nd}$}\ \BEd).
\newblock
\APACaddressPublisher{Basel, Switzerland}{{Birkh\"auser} Verlag}.
\PrintBackRefs{\CurrentBib}

\bibitem [\protect \citeauthoryear {%
Resnick%
}{%
Resnick%
}{%
{\protect \APACyear {2007}}%
}]{%
Resnick2007}
\APACinsertmetastar {%
Resnick2007}%
\begin{APACrefauthors}%
Resnick, S\BPBI I.%
\end{APACrefauthors}%
\unskip\
\newblock
\APACrefYear{2007}.
\newblock
\APACrefbtitle {Heavy-Tail Phenomena: Probabilistic and Statistical Modeling}
  {Heavy-tail phenomena: Probabilistic and statistical modeling}.
\newblock
\APACaddressPublisher{Berlin}{Springer - Verlag}.
\PrintBackRefs{\CurrentBib}

\bibitem [\protect \citeauthoryear {%
Resnick%
}{%
Resnick%
}{%
{\protect \APACyear {2008}}%
}]{%
Resnick2008}
\APACinsertmetastar {%
Resnick2008}%
\begin{APACrefauthors}%
Resnick, S\BPBI I.%
\end{APACrefauthors}%
\unskip\
\newblock
\APACrefYear{2008}.
\newblock
\APACrefbtitle {Extreme Values, Regular Variation, and Point Processes}
  {Extreme values, regular variation, and point processes}\
  (\PrintOrdinal{$1^{st}$ ed. 1987, $2^{nd}$}\ \BEd).
\newblock
\APACaddressPublisher{Berlin}{Springer - Verlag}.
\PrintBackRefs{\CurrentBib}

\bibitem [\protect \citeauthoryear {%
Romanosky%
, Ablon%
, Kuehn%
\BCBL {}\ \BBA {} Jones%
}{%
Romanosky%
\ \protect \BOthers {.}}{%
{\protect \APACyear {2019}}%
}]{%
Romanosky2019}
\APACinsertmetastar {%
Romanosky2019}%
\begin{APACrefauthors}%
Romanosky, S.%
, Ablon, L.%
, Kuehn, A.%
\BCBL {}\ \BBA {} Jones, T.%
\end{APACrefauthors}%
\unskip\
\newblock
\APACrefYearMonthDay{2019}{}{}.
\newblock
{\BBOQ}\APACrefatitle {Content analysis of cyber insurance policies: how do
  carriers price cyber risk?} {Content analysis of cyber insurance policies:
  how do carriers price cyber risk?}{\BBCQ}
\newblock
\APACjournalVolNumPages{Journal of Cybersecurity}{5}{1}{1--19}.
\PrintBackRefs{\CurrentBib}

\bibitem [\protect \citeauthoryear {%
Smith%
}{%
Smith%
}{%
{\protect \APACyear {2003}}%
}]{%
Smith2003}
\APACinsertmetastar {%
Smith2003}%
\begin{APACrefauthors}%
Smith, R\BPBI L.%
\end{APACrefauthors}%
\unskip\
\newblock
\APACrefYear{2003}.
\newblock
\APACrefbtitle {Statistics of extremes, with applications in environment,
  insurance and finance} {Statistics of extremes, with applications in
  environment, insurance and finance}.
\newblock
\APACaddressPublisher{London}{in Extreme Values in Finance, Telecommunications
  and the Environment, insurance and finance, ed. by B. Finkenstadt and H.
  Rootz\`en, Chapman and Hall/CRC Press}.
\PrintBackRefs{\CurrentBib}

\bibitem [\protect \citeauthoryear {%
{Swiss Re}%
}{%
{Swiss Re}%
}{%
{\protect \APACyear {2017}}%
}]{%
SwissRE2017}
\APACinsertmetastar {%
SwissRE2017}%
\begin{APACrefauthors}%
{Swiss Re}.%
\end{APACrefauthors}%
\unskip\
\newblock
\APACrefYearMonthDay{2017}{}{}.
\newblock
\APACrefbtitle {Cyber: getting to grips with a complex risk.} {Cyber: getting
  to grips with a complex risk.}
\newblock
\APAChowpublished
  {\url{https://www.swissre.com/dam/jcr:995517ee-27cd-4aae-b4b1-44fb862af25e/sigma1_2017_en.pdf}}.
\PrintBackRefs{\CurrentBib}

\bibitem [\protect \citeauthoryear {%
Tang%
, Tang%
\BCBL {}\ \BBA {} Yang%
}{%
Tang%
\ \protect \BOthers {.}}{%
{\protect \APACyear {2019}}%
}]{%
Tang2019}
\APACinsertmetastar {%
Tang2019}%
\begin{APACrefauthors}%
Tang, Q.%
, Tang, Z.%
\BCBL {}\ \BBA {} Yang, Y.%
\end{APACrefauthors}%
\unskip\
\newblock
\APACrefYearMonthDay{2019}{}{}.
\newblock
{\BBOQ}\APACrefatitle {Sharp asymptotics for large portfolio losses under
  extreme risks} {Sharp asymptotics for large portfolio losses under extreme
  risks}.{\BBCQ}
\newblock
\APACjournalVolNumPages{European Journal of Operational
  Research}{276}{}{710-722}.
\PrintBackRefs{\CurrentBib}

\bibitem [\protect \citeauthoryear {%
Tencaliec%
, Favre%
, Naveau%
, Prieur%
\BCBL {}\ \BBA {} Nicolet%
}{%
Tencaliec%
\ \protect \BOthers {.}}{%
{\protect \APACyear {2020}}%
}]{%
Tencaliec2020}
\APACinsertmetastar {%
Tencaliec2020}%
\begin{APACrefauthors}%
Tencaliec, P.%
, Favre, A\BHBI C.%
, Naveau, P.%
, Prieur, C.%
\BCBL {}\ \BBA {} Nicolet, G.%
\end{APACrefauthors}%
\unskip\
\newblock
\APACrefYearMonthDay{2020}{}{}.
\newblock
{\BBOQ}\APACrefatitle {Flexible semiparametric Generalized Pareto modeling of
  the entire range of rainfall amount} {Flexible semiparametric generalized
  pareto modeling of the entire range of rainfall amount}.{\BBCQ}
\newblock
\APACjournalVolNumPages{Environmetrics}{31}{2}{}.
\PrintBackRefs{\CurrentBib}

\bibitem [\protect \citeauthoryear {%
Wang%
}{%
Wang%
}{%
{\protect \APACyear {2019}}%
}]{%
Wang2019}
\APACinsertmetastar {%
Wang2019}%
\begin{APACrefauthors}%
Wang, S.%
\end{APACrefauthors}%
\unskip\
\newblock
\APACrefYearMonthDay{2019}{}{}.
\newblock
{\BBOQ}\APACrefatitle {Integrated framework for information security investment
  and cyber insurance} {Integrated framework for information security
  investment and cyber insurance}.{\BBCQ}
\newblock
\APACjournalVolNumPages{Pacific-Basin Finance Journal}{57}{}{}.
\PrintBackRefs{\CurrentBib}

\bibitem [\protect \citeauthoryear {%
Xu%
, Hua%
\BCBL {}\ \BBA {} Xu%
}{%
Xu%
\ \protect \BOthers {.}}{%
{\protect \APACyear {2017}}%
}]{%
Xu2017}
\APACinsertmetastar {%
Xu2017}%
\begin{APACrefauthors}%
Xu, M.%
, Hua, L.%
\BCBL {}\ \BBA {} Xu, S.%
\end{APACrefauthors}%
\unskip\
\newblock
\APACrefYearMonthDay{2017}{}{}.
\newblock
{\BBOQ}\APACrefatitle {A Vine Copula Model for Predicting the Effectiveness of
  Cyber Defense Early-Warning} {A vine copula model for predicting the
  effectiveness of cyber defense early-warning}.{\BBCQ}
\newblock
\APACjournalVolNumPages{Technometrics}{4}{59}{508--520}.
\PrintBackRefs{\CurrentBib}

\bibitem [\protect \citeauthoryear {%
Yang%
\ \BBA {} Robinson%
}{%
Yang%
\ \BBA {} Robinson%
}{%
{\protect \APACyear {1986}}%
}]{%
Yang1986}
\APACinsertmetastar {%
Yang1986}%
\begin{APACrefauthors}%
Yang, M\BPBI C\BPBI K.%
\BCBT {}\ \BBA {} Robinson, D\BPBI H.%
\end{APACrefauthors}%
\unskip\
\newblock
\APACrefYear{1986}.
\newblock
\APACrefbtitle {Understanding and learning statistics by Computer}
  {Understanding and learning statistics by computer}.
\newblock
\APACaddressPublisher{Singapore}{volume 4 of Series in Computer Science, World
  Scientific Publishing}.
\PrintBackRefs{\CurrentBib}

\bibitem [\protect \citeauthoryear {%
Zeller%
\ \BBA {} Scherer%
}{%
Zeller%
\ \BBA {} Scherer%
}{%
{\protect \APACyear {2021}}%
}]{%
Zeller2020}
\APACinsertmetastar {%
Zeller2020}%
\begin{APACrefauthors}%
Zeller, G.%
\BCBT {}\ \BBA {} Scherer, M.%
\end{APACrefauthors}%
\unskip\
\newblock
\APACrefYearMonthDay{2021}{}{}.
\newblock
{\BBOQ}\APACrefatitle {A comprehensive model for cyber risk based on marked
  point processes and its application to insurance} {A comprehensive model for
  cyber risk based on marked point processes and its application to
  insurance}.{\BBCQ}
\newblock
\APACjournalVolNumPages{European Actuarial Journal}{}{}{}.
\newblock
\APAChowpublished {Published online at
  \url{https://doi.org/10.1007/s13385-021-00290-1}}.
\PrintBackRefs{\CurrentBib}

\end{thebibliography}

\newpage

\appendix

\section{With or without an exponential bridge?  A simulation study showing its benefits} 
\label{App-test2and3components}
\vspace{-3ex}
%
In order to test the relevance of introducing an exponential bridge in the hybrid model, we conduct a series of experiments based on simulated data. 
Indeed, it is an essential step to challenge the algorithm to find the true model that has been used for generating the data. Once conclusive, we can turn with confidence to applications on real data.
The experiments consist in using two self-calibrating algorithms with and without an exponential bridge,  to fit two datasets produced with the two models LN-E-GPD and LN-GPD, respectively. We test if the algorithms find out the proper generating models. In particular, are the algorithms able to give a reasonable estimate of the tail fatness?

In this simulation study, we consider heavy-tailed data coming from a two and three components distribution, namely a LN-GPD and LN-E-GPD, respectively. Then, we fit each of the two generated datasets using the two algorithms based on a LN-GPD and LN-E-GPD model, respectively. To generate the data, we perform 100 Monte Carlo (MC)  simulations considering various sample sizes, from 1,000 to 50,000 realizations ({\it i.e.} for each sample size, we change 100 times the seed of the random generator). We choose different sets of parameters to explore a wide spectrum.
For the GPD component (with tail index $0<\xi<1$, as we assumed heavy-tailed data with finite mean), we additionally consider two cases depending if we have a very heavy-tailed distribution (no variance), choosing $\xi=0.8$ (similar to our cyber data results), or a moderate heavy one when $\xi=1/3$ (meaning a finite variance). 
In total, we proceed to 40 experiments, as we consider 2 models, 2 tail indices, 5 samples, and 2 self-calibrating algorithms. 

While the LN-E-GPD model is explicited in \eqref{eq:LNEGPD} and \eqref{eq:relations-Param}, the LN-GPD one is defined by its pdf $\tilde{h}$ given as follows:
\begin{equation}\label{eq:LNGPD}
\tilde{h}(x;\boldsymbol{\tilde{\theta}})= \tilde{\gamma_1}\, f(x;\mu,\sigma)\,\1_{(x\leq u)} + \tilde{\gamma_2}\, g(x-u;\xi,\beta) \,\1_{( x \geq u)},  
\end{equation}
where, under the $C^1$ assumption, the parameters satisfy
\begin{equation}\label{eq:relations-Param2}
\left\{
  \begin{array}{ll}
  \xi=\displaystyle\frac{1}{\alpha}=\displaystyle\frac{\sigma^2}{\log(u)-\mu}; &   \tilde{\gamma_1}=\displaystyle \frac{1}{\beta f(u;\mu,\sigma)+F(u;\mu,\sigma)}; \\ 
\beta=\xi \, u; &\quad \tilde{\gamma_2}=1-\hat{\gamma_1} F(u;\mu,\sigma).
  \end{array}
\right.
\end{equation}
Notice that \eqref{eq:LNGPD} corresponds to a particular case of \eqref{eq:LNEGPD} without exponential bridge, \emph{i.e} $u_1=u_2$. In  \eqref{eq:LNGPD} the unique junction point is called $u$, $\boldsymbol{\tilde{\theta}}=[\mu,\sigma,u]$, and  $\tilde{\gamma_1}$ and $ \tilde{\gamma_2}$ denote the non-uniform weights associated to lognormal distribution and the GPD, respectively. 

Let us mention that the LN-GPD model \eqref{eq:LNGPD} corresponds to a generalized version of the Czeledin distribution (see \cite{Knecht2003}) that connects an unweighted lognormal ($\tilde{\gamma_1}=1$) to a weighted Pareto distribution ($\tilde{\gamma_2}=1-F(u;\mu,\sigma)$) . The LN-GPD gives clearly more flexibility with non-uniform weights.

In Table~\ref{tab:theo_param}, we display the parameters used to generate the various samples. Recall that the number of parameters are reduced (4 for the LN-E-GPD and 3 for the LN-GPD, instead of 10 and 7 respectively) by imposing continuity conditions at the junction points (leading to \eqref{eq:relations-Param} and \eqref{eq:relations-Param2}). Here we choose independently the tail indices $\xi$ for the two targeted cases, and consider various values of $\mu$ and $\sigma$ with different tail-thresholds, exploring then different settings.
\begin{table}[h]
\begin{center}
\parbox{350pt}{\caption{\label{tab:theo_param}\sf\small Parameters used to generate the data coming from the two hybrid models, respectively. The chosen thresholds are given as real numbers, with the corresponding  quantiles in parenthesis. In the case of two components, we display the threshold $u$ in the column of $u_2$ where the extremes start. \vspace{0.7ex} }}
\begin{tabular*}{355pt}{l>{\centering}p{30pt}>{\centering}p{30pt}>{\centering}p{60pt}>{\centering}p{80pt}c}
\hline 
&&&&&\\[-1.5ex]
~Model       & $\mu$ & $\sigma$ &   $u_1$   &   $u_2$   & $\xi$  \\
&&&&&\\[-1.5ex]
\hline \hline
&&&&&\\[-1.5ex]
~LN-E-GPD   & ~~1~~ &   ~~2~~  &  ~4 (80\%)   & 14.59~(98.28\%) & 1/3\\
            & ~~0~~ &   ~~5~~  & ~2 (85.91\%) & ~4.38~(94.37\%) & 0.8~~\\[0.7ex]
\hline
\\[-0.5ex]
~LN-GPD			& ~~2~~ &  ~~0.5~~ &       --     & 15.65~(91.89\%) & 1/3\\
            & ~~0~~ &   ~~1~~  &       --     & ~3.5~(86.49\%)  & 0.8~~\\[0.5ex]
\hline
\end{tabular*}
\end{center}
\end{table}

Turning to the fit of both models on the generated datasets, we use our algorithmic approach to  estimate the parameters of the LN-E-GPD and LN-GPD models, respectively. 
As already mentioned, we have adapted the unsupervised iterative algorithm introduced in \citep{Debbabi2017}. Besides the model and relations between parameters, another difference between modelling the bulk data by a Gaussian or a lognormal distribution, lies on the choice of initial parameters conducting the algorithm to the convergence. Indeed,  we recall that for  G-E-GPD, the Gaussian mean corresponds to the distribution mode, which gives a nice strategy to initialize the Gaussian parameters. Unfortunately, it is no longer the case for the lognormal distribution for which we tested several techniques to obtain initialization that holds up. This is of particular importance when fitting 100 times different datasets.

An advantage of MC techniques is to minimize the numerical noise and concentrates on the model performance. This is why we performed MC simulations (100 here) to provide the mean result for each estimate, and its standard deviation (that measures as well the numerical variation), which is a distinct notion from the confidence intervals (computed in our case with the Jackknife method).
To assess the performance of our algorithm, we compute the relative error between the mean MC estimate of the considered parameter and its theoretical value used to generate the data. We repeat this evaluation varying the sample size to study the speed of convergence of the algorithm in terms of the number of realizations, focusing on the tail index estimation. 

In Table~\ref{tab:bridge_res}, we display the MC fitting results over $100$ simulations with a sample size of $N=10,000$ for all the parameters, together with their relative error evaluated as explained above. 
In the first half of this table corresponding to the fit using the 3 components algorithm, there is no relative error above 5\%. For the tail indices, the error varies between $0.86\%$ to $3.8\%$, which is very reasonable in terms of performance. We notice that the error made on the tail index estimation slightly increases with the tail index. For the data generated with the two components distribution (LN-GPD), the three components algorithm is able to detect the absence of an exponential bridge, by setting $u_1=u_2$, demonstrating here the flexibility of the three components algorithm. Turning to the threshold estimates, they are quite accurate with errors ranging from $-0.99\%$ to $4.53\%$, hence well within the $5\%$ error range that we consider to judge about the performance of the algorithm. The LN parameters $(\mu,\sigma)$ are sharply evalauted, with errors between $-.29\%$ to $1.48\%$. Overall, the three components algorithm performs well.

Now let us analyse the second case where the fitting algorithm has only two components. We clearly notice that the algorithm is not able to estimate properly the tail index of models containing an exponential bridge, with errors over $26.29\%$ and $31.06\%$, pushing the threshold of the extremes to  99.99\% quantile. The parameters of the lognormal distribution are also badly estimated (with errors ranging from $-31.03\%$ to $-80.96\%$.
However, this fitting algorithm reproduces well the results for data generated by a two components model. The estimation accuracy is similar to that achieved with the three components model.
\begin{table}[t]
\centering
\parbox{472pt}{\caption{\label{tab:bridge_res}\sf\small Results for the Monte Carlo simulations. In the case of two components, only the junction point $u_2$ (where the extremes start) is displayed (since in this case, $u=u_1=u_2$). ND means {\it Not Defined} as the theoretical value is 0; in such a case, we look at the absolute error. Bold font indicates (relative) errors larger than 5\%. \vspace{0.7ex}}}
\footnotesize
\begin{tabular}{l l >{\centering}p{40pt} >{\centering}p{40pt} >{\centering}p{65pt} >{\centering}p{65pt} c}
\hline 
&&&&&&\\[-1.5ex]
~Fitting Algorithm & Generating Data &  $\mu$  & $\sigma$ &     $u_1$      &      $u_2$      & $\xi$  \\
&&&&&&\\[-1.5ex]
\hline \hline
&&&&&&\\[-1.5ex]
~Three components	& LN-E-GPD 1/3 	& ~1.01~  	&  ~2.00~  	& ~3.96~(75.67\%) & 15.01~(98.04\%)	& 0.336   \\
				& Relative Error		& 1.48\% 	&  0.18\%	&    -1.12\%     	   &   2.86\%		& 0.86\% \\[0.8ex]  
	          		& LN-GPD   1/3 		& ~1.99~  	&  ~0.49~	& 15.22~(90.59\%) & 15.49~(91.09\%)	& 0.337   \\
                   		& Relative Error 		& -0.05\%	&  -0.29\% &       --        	   &   -0.99 \% 		& 1.17\% \\[0.8ex]  
                   		& LN-E-GPD 0.8		& ~0.04~  	&  ~5.01~ 	& ~1.94~(85.01\%) & ~4.21~(93.37\%)	& 0.825   \\
                   		& Relative Error		& ND		&  0.22\%	&     -3.07\%          &   -3.93\% 		& 3.08\% \\ [0.8ex]                  
                   		& LN-GPD   0.8		& ~0.01~  	&  ~1.00~ 	& ~3.66~(86.75\%) & 3.66~(86.75\%)	& 0.782   \\
                   		& Relative Error		& ND		&  0.48\% 	&       --       	   &   4.53\%		& -1.99\%  \\[0.8ex]
\hline
\\[-0.5ex]         
~Two components	& LN-E-GPD 1/3 	& ~0.19~  		&  ~1.38~		&       -- 	&  105~(99.99\%)	& 0.437   \\
                   		& Relative Error		& {\bf -80.96\%}	&{\bf -31.03\%} &       -- 	&   {\bf 625\%}		& {\bf 31.06\%} \\ [0.8ex]            
                  		 & LN-GPD 1/3    	& ~1.99~  		& ~0.49~   		&       -- 	& 15.39~(91.01\%) 	& 0.337   \\
                   		& Relative Error 		& -0.07\% 		& -0.39\%  	&       -- 	&   -1.65\%       		& 1.41\% \\[0.8ex]   
                 	 		& LN-E-GPD   0.8  	& -2.72~  		& ~3.11~   		&       --  	& 1860~(99.99\%)  	& ~1.01   \\
                   		& Relative Error  		&  {\bf ND}		& {\bf -37.61\%}	&       -- 	&  {\bf 42,391\%}     	& {\bf 26.29\%} \\ [0.8ex]                
		 		& LN-GPD   0.8    	& -0.005  		& ~1.01~   		&       -- 	& ~3.55~(86.09\%) 	& 0.799   \\
                   		& Relative Error  		& ND			& 0.54\%  		&       -- 	&   1.52\% 			& 0.12\% \\[0.8ex]
\hline
\end{tabular}
\end{table}
The exponential bridge is thus a useful artifact to obtain a flexible self-calibrating method for various sorts of data, here for asymmetric non-negative ones (while for symmetric ones in \citep{Debbabi2017}).

This MC simulation study confirms the relevance of this algorithmic approach based on a general hybrid model which main and extreme behaviors are linked with an exponential bridge. It completes the overall method with a second algorithm for treating asymmetric non-negative heavy-tailed data.

An additional by-product of this MC simulation study is the exploration of the influence of sample sizes on the stability and quality of the tail index estimation. In Figure~\ref{tab:bridge_res}, we present mean and standard deviation of the tail index estimate, obtained with the three components algorithm, for various sample sizes, $N=1000; 5000; 10\,000; 20\,000$ and $50\,000$, and for each of the two tail indices. The data is generated from the LN-E-GPD model with the parameters of Table~\ref{tab:theo_param}, with $\xi=1/3$ and $\xi=0.8$. Note that the latter case and the fitting algorithm have been chosen to be in a similar situation as that encountered in our application on cyber data (the fit of our algorithm on 60,985 observations led to a tail index estimate $\hat\xi=0.81$).
\begin{figure}[t]
  \centering
  \includegraphics[scale=0.8]{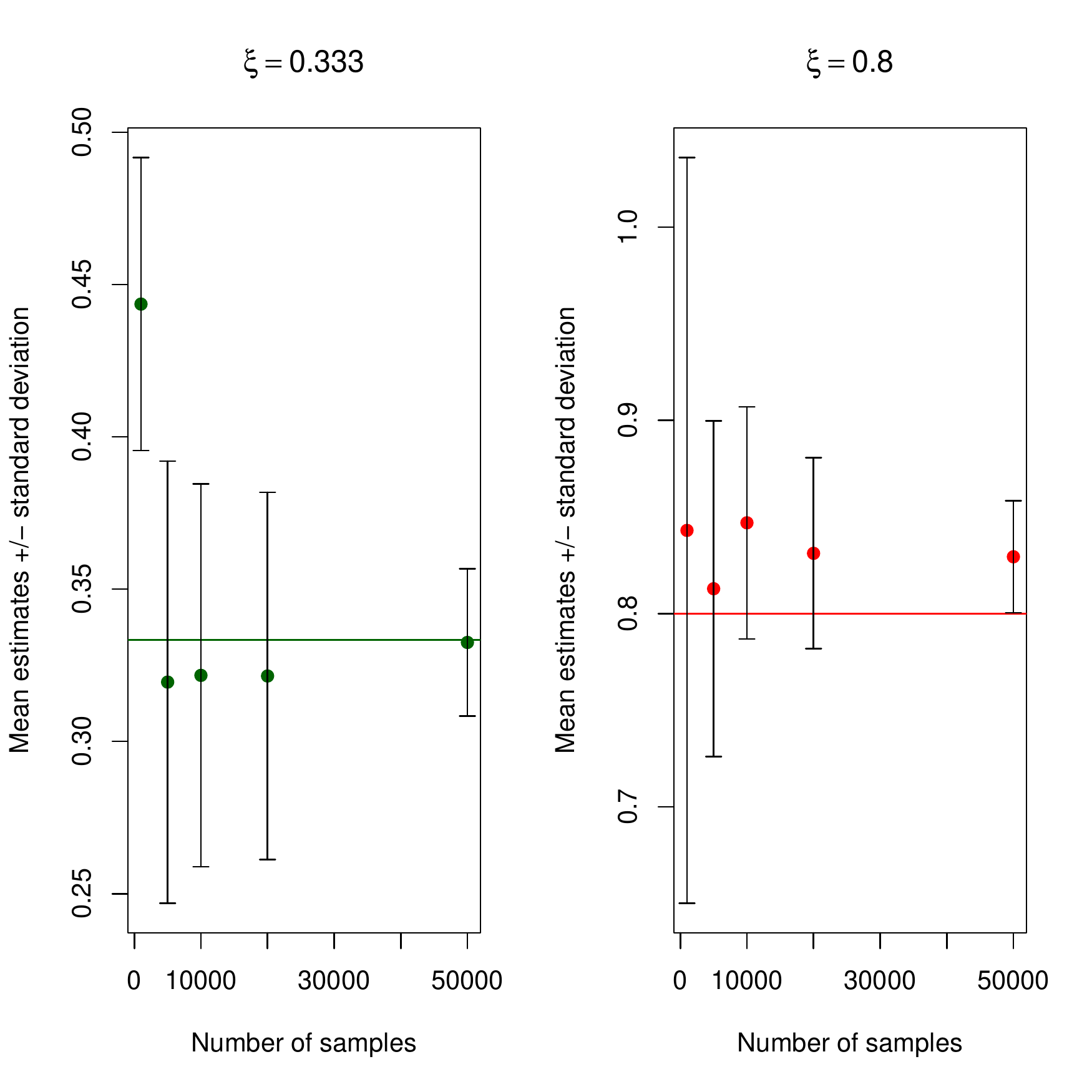}
\vspace{-2ex}
\parbox{400pt}{\caption{ \label{fig:Errors_and_obs} \sf\small We plot the relative error of the shape index $\alpha=1/\xi$ as a function of the number of observations (sample size) used for the fit. The data is generated from the LN-E-GPD model with the parameters of Table~\ref{tab:theo_param}, with $\xi=1/3$ ($\alpha=3$; dark green dots) in the left plot and $\xi=0.8$ (or $\alpha=1.25$; red dots) in the right plot.}}
\end{figure}
We observe that, as expected, the convergence is better for small tail index. This is a well known feature when using MC simulations for heavy-tailed data (see e.g. \cite{Dacorogna2018}). The numerical noise generally decreases when the sample size increases. A good convergence is reached with 50,000 observations for $\xi=1/3$, while the tail estimation remains within 5\% of the theoretical value, whatever the sample size and for both tail indices. We also see that a sample size of 1000 observations gives the worst result for both indices.

\vspace{-2ex}
\section{Estimation of the tail index: Comparison with EVT methods}
\label{App-estim-comparison-EVTmethods}
\begin{figure}[h]
  \centering
  \includegraphics[scale=1.1]{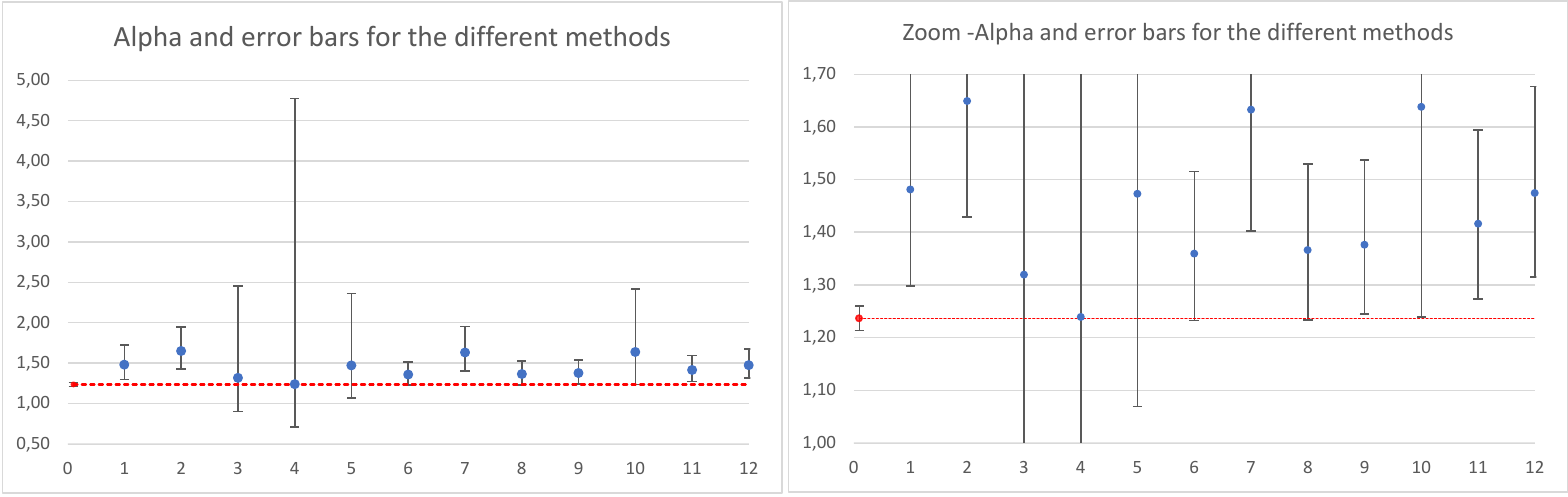}
  \caption{ \label{fig:tea-confidenceBands} \small\sf On the $y$-axis the shape parameter $\alpha$ is displayed, while on the $x$-axis the various methods are put in the same order as in the table (from 1 for AMSE to 12 for Reiss \& Thomas, and 0 for our model). The right plot is the same with a zoom on the $y$-axis to better visualize the confidence bands w.r.t. our LN-E-GPD model (dotted line).}
\end{figure}

\vspace{-2ex}
Many methods exist to evaluate the tail-threshold and the tail index (or, equivalently, its inverse), as well as software packages (e.g. on R or Python). To complete our study, we provide results obtained for the estimation of those two parameters, using the various methods included in the {\it tea} R-package (see \url{https://rdocumentation.org/packages/tea/versions/1.1} for details and the paper references therein on the considered method), and compare them with those obtained via our algorithm.
To judge  the robustess of the estimations, we order our sample and build subsamples keeping the $m$ largest observations (of our sample), varying $m$. Results are given in Table~\ref{tab:tea-results}. Confidence intervals are displayed in the plots for a given subsample and all methods; see Figure~\ref{fig:tea-confidenceBands}.

In Table~\ref{tab:tea-results}, columns correspond to subsamples ordered by increasing size. For each method (row),  three quantities are displayed: the tail-threshod value (in \EUR ), its associated quantile order in the full sample (of size $n=60,985$) for which the number of extreme observations can be deduced, and the inverse of the tail index $\alpha(=1/\xi)$. The last row shows the results for the entire dataset including zeroes.
'NaN' answers correspond either to the crash of the program or when stopping it after more than 5 days of run. This type of non-answer occurs for large sample sizes. For instance, the R-code for the Danielsson et al. (2001) method ran for 12 hours to give a result for the sample of size 50'039, while, for the sample of size 208,037, we did not obtain yet the result after 6 days of run, so we stopped it. We did not face such an issue for the Danielsson et al. (2016) method. Nevertheless, for very large sample sizes, some of the methods (mainly the non stable ones) needed a few hours to provide the results.  

All those methods have been successful in providing a good modelling of tail distributions in various contexts; that is why they have been included in the {\it tea} R-package. Nevertheless results provided in Table~\ref{tab:tea-results} are the testimony of the general challenge of modelling extremes, in particular on our large dataset. 
We observe that, whatever the method, the $\alpha$-parameter is mostly above 1 and below 1.5, confirming the finite mean and infinite variance. We can disregard the cases where $\alpha < 1$, ($\xi > 1$) as the tail-threshold (between $VaR(68.9\%)$ and $VaR(89.1\%)$) may be considered as too low for EVT methods. We also have to be careful not taking into account results obtained with too high thresholds for which estimation will be done on very few extreme values (for instance, for the rather stable eye-balling technique, the tail-threshold $q(99.99\%)$ corresponds only to 7 extreme observations!). We observed only once $\alpha=2$ (in Hall-Welsch (1985) for the first subsample), which must be discarded as the tail-threshold is such that only the maximum is taken as extreme observation. Some methods (among the less stable ones across samples) are subject to numerical noise and do not give the same results at each run (although close).

The Guillou and Hall (2001) and Reiss and Thomas (2007) methods are quite robust, given a constant value of $u_2$ and $\alpha$, whatever the sample size. We selected the latter one in our comparison study  given its reasonable number of extreme observations (263, compared with 27 for  Guillou and Hall). The approach by Drees and Kaufmann (1998) is also stable for small sample sizes, as well as for Hall (1990) where the variability is small. For the other methods, we observe much more variability.

\begin{landscape}
\begin{table}[h]
\small
  \centering
  \parbox{670pt}{\caption{  \label{tab:tea-results} \large\sf Estimates of the tail-threshold $u_2$ (expressed also as a quantile) and the inverse of the tail index $\alpha=1/\xi$ obtained with the 12 methods given in the {\it tea} R-package (presented in the same order as in the package). Quantiles levels are rounded up to 2 decimal digits. \vspace{0.7ex}}}
    \begin{tabular}{rlccccccccc}
    \toprule
    \multicolumn{1}{l}{\textbf{Initial sample}} & Nb. Obs. & \multicolumn{1}{r}{357} & 557   & 609   & 1006  & 5026  & 10001 & 50039 & 60985 & 208037 \\
    \midrule
    \midrule
    \multicolumn{1}{l}{\textbf{AMSE}} & \textit{ threshold } &      100,000  &   124,570  &   254,655  &   278,415  &   200,000  &   124,110  &     28,100  &     21,000  &  NaN  \\
          & \textit{ quantile } & 99.59\% & 99.69\% & 99.89\% & 99.90\% & 99.84\% & 99.68\% & 98.12\% & 97.47\% &  NaN  \\
          & \textit{ alpha } & 1.42  & 1.48  & 1.74  & 1.78  & 1.65  & 1.49  & 1.23  & 1.17  & NaN \\
    \multicolumn{1}{l}{\textbf{Danielsson et al. (2001)}} & \textit{ threshold } &      165,000  &   148,500  &   352,971  &   198,426  &     20,200  &     36,000  &        1,500  &        1,506  &  NaN  \\
          & \textit{ quantile } & 99.78\% & 99.74\% & 99.94\% & 99.83\% & 97.40\% & 98.64\% & 75.00\% & 75.49\% &  NaN  \\
          & \textit{ alpha } & 1.63  & 1.65  & 1.55  & 1.76  & 1.15  & 1.22  & 0.88  & 0.87  & NaN \\
    \multicolumn{1}{l}{\textbf{Drees \& Kaufmann (1998)}} & \textit{ threshold } &      493,623  &   493,623  &   493,623  &   493,623  &  NaN  &  NaN  &  NaN  &  NaN  &  NaN  \\
          & \textit{ quantile } & 99.97\% & 99.97\% & 99.97\% & 99.97\% &  NaN  &  NaN  &  NaN  &  NaN  &  NaN  \\
          & \textit{ alpha } & 1.32  & 1.32  & 1.32  & 1.32  & NaN   & NaN   & NaN   & NaN   & NaN \\
    \multicolumn{1}{l}{\textbf{Eye-balling Technique}} & \textit{ threshold } &      493,623  &   982,101  &   971,000  &   656,210  &   305,000  &   207,712  &     62,500  &  NaN  &  NaN  \\
          & \textit{ quantile } & 99.97\% & 99.99\% & 99.99\% & 99.98\% & 99.92\% & 99.85\% & 99.26\% &  NaN  &  NaN  \\
          & \textit{ alpha } & 1.32  & 1.22  & 1.34  & 1.12  & 1.64  & 1.64  & 1.33  & NaN   & NaN \\
    \multicolumn{1}{l}{\textbf{Guillou \& Hall (2001)}} & \textit{ threshold } &      400,000  &   400,000  &   400,000  &   400,000  &   400,000  &   400,000  &   400,000  &   400,000  &   400,000  \\
          & \textit{ quantile } & 99.96\% & 99.96\% & 99.96\% & 99.96\% & 99.96\% & 99.96\% & 99.96\% & 99.96\% & 99.96\% \\
          & \textit{ alpha } & 1.47  & 1.47  & 1.47  & 1.47  & 1.47  & 1.47  & 1.47  & 1.47  & 1.47 \\
    \multicolumn{1}{l}{\textbf{Gomes et al. (2012)}} & \textit{ threshold } &      197,000  &     75,000  &   342,700  &   251,000  &   120,000  &   147,000  &        2,800  &        2,400  &  NaN  \\
          & \textit{ quantile } & 99.82\% & 99.41\% & 99.93\% & 99.89\% & 99.67\% & 99.73\% & 85.50\% & 83.30\% &  NaN  \\
          & \textit{ alpha } & 1.75  & 1.36  & 1.63  & 1.75  & 1.52  & 1.66  & 0.90  & 0.90  & NaN \\
    \multicolumn{1}{l}{\textbf{Hall (1990)}} & \textit{ threshold } &      155,633  &   160,000  &   159,000  &   160,000  &   165,000  &   161,878  &   171,227  &     80,000  &   171,227  \\
          & \textit{ quantile } & 99.76\% & 99.77\% & 99.76\% & 99.77\% & 99.78\% & 99.78\% & 99.79\% & 99.45\% & 99.79\% \\
          & \textit{ alpha } & 1.63  & 1.63  & 1.62  & 1.61  & 1.62  & 1.60  & 1.65  & 1.37  & 1.65 \\
    \multicolumn{1}{l}{\textbf{Caeiro \& Gomes (2012)}} & \textit{ threshold } &      157,000  &     80,000  &   355,815  &   318,403  &   147,550  &   177,973  &        3,471  &        2,970  &  NaN  \\
          & \textit{ quantile } & 99.76\% & 99.45\% & 99.95\% & 99.93\% & 99.73\% & 99.79\% & 87.99\% & 85.98\% &  NaN  \\
          & \textit{ alpha } & 1.62  & 1.37  & 1.52  & 1.66  & 1.65  & 1.73  & 0.90  & 0.92  & NaN \\
    \multicolumn{1}{l}{\textbf{Hall \& Welsh (1985)}} & \textit{ threshold } &   8,069,984  &     77,900  &   313,913  &   200,000  &   245,860  &   251,000  &   105,000  &     96,000  &  NaN  \\
          & \textit{ quantile } & 100\% & 99.43\% & 99.93\% & 99.84\% & 99.88\% & 99.89\% & 99.62\% & 99.55\% &  NaN  \\
          & \textit{ alpha } & 2.09  & 1.36  & 1.66  & 1.65  & 1.72  & 1.75  & 1.40  & 1.43  & NaN \\
    \multicolumn{1}{l}{\textbf{Danielsson et al. (2016)}} & \textit{ threshold } &      411,028  &   350,000  &   300,000  &   300,000  &   371,000  &   400,000  &   489,656  &   489,656  &   939,950  \\
          & \textit{ quantile } & 99.96\% & 99.94\% & 99.92\% & 99.92\% & 99.95\% & 99.96\% & 99.97\% & 99.97\% & 99.99\% \\
          & \textit{ alpha } & 1.38  & 1.64  & 1.67  & 1.71  & 1.50  & 1.47  & 1.30  & 1.30  & 1.19 \\
    \multicolumn{1}{l}{\textbf{Caeiro \& Gomes (2016)}} & \textit{ threshold } &        87,500  &     87,500  &     87,500  &     31,000  &        7,478  &        3,975  &        1,199  &        1,199  &        1,199  \\
          & \textit{ quantile } & 99.50\% & 99.50\% & 99.50\% & \multicolumn{1}{r}{98.35\%} & 93.52\% & 89.12\% & 68.91\% & 68.91\% & 68.91\% \\
          & \textit{ alpha } & 1.41  & 1.41  & 1.41  & 1.21  & 1.00  & 0.92  & 0.90  & 0.90  & 0.90 \\
    \multicolumn{1}{l}{\textbf{Reiss \& Thomas (2007)}} & \textit{ threshold } &      100,000  &   100,000  &   100,000  &   100,000  &   100,000  &   100,000  &   100,000  &   100,000  &   100,000  \\
          & \textit{ quantile } & 99.60\% & \multicolumn{1}{r}{99.60\%} & \multicolumn{1}{r}{99.60\%} & \multicolumn{1}{r}{99.60\%} & \multicolumn{1}{r}{99.60\%} & \multicolumn{1}{r}{99.60\%} & \multicolumn{1}{r}{99.60\%} & \multicolumn{1}{r}{99.60\%} & \multicolumn{1}{r}{99.60\%} \\
          & \textit{ alpha } & 1.47  & 1.47  & 1.47  & 1.47  & 1.47  & 1.47  & 1.47  & 1.47  & 1.47 \\
    \bottomrule
    \end{tabular}%
  \label{tab:addlabel}%
\end{table}%
 \end{landscape}

There are some methods that cannot handle a big number of data like Dress and Kaufmann or Danielsson et al. (2001) where the execution time becomes prohibitive. In general, when using the full dataset of 208,037, half of the methods do not give results. It is probably related to the problem we also encountered, which is the presence of zeroes that might not be real zeroes but simply missing amounts. It is the reason why we restricted our reference sample to the non-zeros values (60,935).

In Figure~\ref{fig:tea-confidenceBands}, we show the results of the subsample with 557 observations as we have there results for all the methods. For each point, we also draw the asymptotic confidence band of the Hill estimator (see its properties recalled in Section~\ref{ss:jackknife}). As expected, the methods providing results on high quantiles (Drees \& Kaufmann (1998), Eye-balling Technique, Guillou \& Hall (2001), Danielsson et al. (2016)) present the widest confidence intervals. The values are consistently higher than ours (except for the eye-balling technique but with a wide confidence range), but fluctuate around 1.4. In 7 cases out of 12, the confidence ranges include our result.

\end{document}